\documentclass[aps,pra,reprint,amsmath,amssymb,superscriptaddress,onecolumn,longbibliography,nofootinbib,notitlepage]{revtex4-2}
\usepackage{mathtools}
\usepackage{siunitx}
\usepackage[hidelinks]{hyperref}
\usepackage{svg}
\usepackage{caption}
\usepackage{float}
\usepackage{subcaption}
\usepackage{ragged2e}
\makeatletter
\AtBeginDocument{%
  \let\rz@oldcaption\caption
  \renewcommand{\caption}[2][]{%
    \ifx\\#1\\%
      \rz@oldcaption{\justifying #2}%
    \else
      \rz@oldcaption[#1]{\justifying #2}%
    \fi}%
}
\makeatother
\usepackage[braket, qm]{qcircuit}
\usepackage{bbm}
\usepackage{lipsum} 
\usepackage{microtype} 
\usepackage{tikz}
\usetikzlibrary{positioning}
\usetikzlibrary{calc} 
\usepackage{booktabs,multirow,subcaption} 
\usepackage{xparse}
\usepackage{amsmath}
\usepackage{bm}
\usepackage[most]{tcolorbox}
\usepackage[T1]{fontenc}
\usepackage{textcomp}

\usepackage{tocloft}
\newlistof{appendixsections}{apc}{Appendix Table of Contents}
\newif\ifapponly      \apponlyfalse
\newif\ifmainonly     \mainonlyfalse

\DeclareMathOperator*{\argmin}{argmin}

\newcommand{\Ham}{\mathcal{H}}

\usepackage[ruled,vlined]{algorithm2e}  
\SetKw{KwInParallel}{in parallel}

\makeatletter
\renewcommand{\theequation}{\arabic{equation}}
\makeatother

\newcommand{\panel}[1]{{\textbf{#1}}\,, }
\newcommand{\fig}[2]{Fig.~\ref{fig:#1}\,{#2}}
\newcommand{\Fig}[2]{Figure~\ref{fig:#1}\,{#2}}

\newcommand{\eq}[1]{Eq.~(\ref{eq:#1})}
\newcommand{\secRef}[1]{Section~\ref{sec:#1}}

\makeatletter 
\let\origtitle\title
\renewcommand{\title}[1]{\def\savedtitle{#1}\origtitle{#1}}
\makeatother

\NewDocumentCommand{\runname}{s}{%
  trial\IfBooleanT{#1}{s}%
}

\NewDocumentCommand{\stepname}{s}{%
  update step\IfBooleanT{#1}{s}%
}

\NewDocumentCommand{\schedulename}{s}{%
  update schedule\IfBooleanT{#1}{s}%
}

\usepackage{xcolor}

\usepackage[normalem]{ulem}

\begin{document}

\title{A fully parallel densely connected probabilistic Ising machine with inertia}

\ifapponly
  \renewcommand{\title}[1]{\origtitle{Supplementary material for\\``\savedtitle''}}
  \title{}  
\fi

\author{Ruomin~Zhu}
\email{ruomin.zhu@cornell.edu}
\affiliation{School of Applied and Engineering Physics, Cornell University, Ithaca, NY 14853, USA}

\author{Abhishek~Kumar~Singh}
\affiliation{Department of Computer Science, Princeton University, Princeton, NJ, 08540, USA}

\author{Jérémie~Laydevant}
\affiliation{School of Applied and Engineering Physics, Cornell University, Ithaca, NY 14853, USA}

\author{Fan~O.~Wu}
\affiliation{School of Applied and Engineering Physics, Cornell University, Ithaca, NY 14853, USA}

\author{Ari~Kapelyan}
\affiliation{School of Applied and Engineering Physics, Cornell University, Ithaca, NY 14853, USA}

\author{Davide~Venturelli}
\affiliation{USRA Research Institute for Advanced Computer Science, Moffett Field, CA, 94035, USA}

\author{Kyle~Jamieson}
\affiliation{Department of Computer Science, Princeton University, Princeton, NJ, 08540, USA}

\author{Peter~L.~McMahon}
\email{pmcmahon@cornell.edu}
\affiliation{School of Applied and Engineering Physics, Cornell University, Ithaca, NY 14853, USA}
\affiliation{Kavli Institute at Cornell for Nanoscale Science, Cornell University, Ithaca, NY, 14853, USA}

\ifapponly\else

\begin{abstract}

Ising machines---special-purpose hardware for heuristically solving Ising optimization problems---based on probabilistic bits (p-bits) have been established as a promising alternative to heuristic optimization algorithms run on conventional computers. However, it has---until now---been thought that Ising spins that are connected in probabilistic Ising machines (PIMs) cannot be updated in parallel without ruining the machine's solving ability. This has presented a major challenge to realizing the potential for probabilistic Ising machines to act as fast solvers for densely connected Ising problems. In this paper, we show that it is possible to circumvent this conventional wisdom. We introduce a modified form of Ising spin dynamics for probabilistic Ising machines, adding an inertia term, and verify in algorithm simulations, field-programmable gate array (FPGA) hardware emulation, and in FPGA hardware experiments that the modified dynamics enables fully parallel, synchronous updates and at the same time improves rather than degrades the achieved success probability.

Our evaluations were performed with various types of abstract (Max-Cut and Sherrington-Kirkpatrick model) and application-derived (multiple-input and multiple-output, MIMO detection) dense Ising benchmark instances. 
Performing fully parallel updates results in a speed advantage that grows \textit{superlinearly} with the number of spins, giving rise to large time-to-solution reductions for practical problem sizes.
For both Max-Cut and the Sherrington--Kirkpatrick model at a problem size of 200, our approach achieved an average speedup of ${\approx}34\times$, with the best single-instance speedup reaching $150\times$.

As an example of the practical utility of our approach in an application where speed is critical, we co-design the algorithm dynamics and hardware implementation for MIMO detection, achieving improved detection accuracy relative to the standard linear minimum mean-square error (MMSE) detector and higher throughput than the conventional sequential-update PIM.
\vspace{10pt}

\end{abstract}
\fi
\maketitle

\ifapponly\else

\section{Introduction}
\label{sec:intro}

Combinatorial optimization problems arise in a wide range of application domains, but many problems of interest are expensive, in energy and time, to solve---even heuristically---for practical problem sizes. The Ising optimization problem is a canonical combinatorial optimization problem to which many other problems can be efficiently mapped~\cite{Lucas2014}, motivating the investigation of \textit{Ising machines}~\cite{Mohseni2022}: special-purpose computers for heuristically solving the Ising problem.

Many approaches to constructing Ising machines have been and are being investigated. 
Among them, probabilistic Ising machines (PIMs) based on networks of probabilistic bits (p-bits) are particularly attractive because they combine simple binary-state dynamics, intrinsic stochasticity, and broad implementation flexibility in a hardware-friendly framework \cite{Camsari2017, sutton2017intrinsic, Borders2019, Sutton2020, Aadit2022, Chowdhury2023, Singh2024cmos, Niazi2024, Patel2024, Duffee2025, Iftakher2025, Garg2025}.
A p-bit is a binary stochastic unit that randomly takes on a value +1 or -1, with the probability bias towards one value or the other determined by an input signal \cite{Chowdhury2023}. One direct physical realization of p-bits is using stochastic nanomagnets whose probabilistic switching behavior arises from thermal fluctuations and spin-charge transduction \cite{Debashis2018, Camsari2020}. However, p-bits can also be emulated in conventional digital electronics using pseudorandom number generation, and various system-level demonstrations of PIMs have taken this approach, often on field-programmable gate arrays (FPGAs) \cite{Pervaiz2019, Aadit2022, Niazi2024}.

Conventional probabilistic Ising machines (PIMs) update spins sequentially, computing the local field and stochastic transition for one spin before proceeding to the next~\cite{Camsari2017, Pervaiz2019, Duffee2025, Iftakher2025}. 
As Ref.~\cite{Aadit2022} explains: ``\textit{A practical difficulty lies in the serial nature of this evolution. Connected nodes need to be updated one after the other since parallel updating leads to repeated oscillations in the network state, preventing the network from converging to the Boltzmann distribution. The need for sequential updating inherently serializes the network evolution}.''
In the noise-free limit, synchronous updates can produce a period-2 cycle in the system, where the full spin configuration repeats every two update steps but changes between successive steps: $\mathbf{s}(t+2)=\mathbf{s}(t)$ and $\mathbf{s}(t+1)\neq\mathbf{s}(t)$~\cite{Goles1981, Goles1985}.
Including stochastic noise breaks this exact recurrence, although the sampled spin configurations can remain concentrated near the underlying period-2 cycle. 

Although the sequential update rule avoids the oscillatory behavior,
it imposes a scalability bottleneck: updating each of the $N$ spins once (i.e., a full \textit{sweep}, in the language of simulated annealing~\cite{isakov2015optimised}) requires $\mathcal{O}(N)$ steps.
As a result, using sequential updates directly limits the achievable throughput and hardware efficiency of PIM implementations.
Aadit~et~al.~\cite{Aadit2022} addressed this by using graph coloring to partition the spins into independent sets, so that parallel updates were applied only to spins that are not directly connected.
However, for dense graphs, proposed solutions have either involved running much more slowly (since graph coloring produces only small independent sets and thus limits the number of spins that can be updated in parallel at each step,) or using much more hardware resources (e.g., serially running duplicated copies of the PIM \cite{Duffee2025}).

To overcome this limitation, \textbf{we introduce a variant of PIM: a \underline{P}robabilistic \underline{I}sing \underline{M}achine with \underline{I}nertia (PIMI)}.
A PIMI employs an inertia-like self-alignment term that biases each spin toward its previous state. As shown empirically in later sections, this algorithmic variant of PIM allows fully parallel updates while suppressing oscillations and stabilizing convergence to low-energy states \cite{Caccioli2008}.
Complementary approaches based on continuous variables include mean-field annealing, which evolves the average value of spin states, and the Free-Energy Machine, which combines variational free-energy minimization with gradient-based optimization over marginal probabilities to provide a general and highly parallel framework \cite{Bilbro1988, Shen2025}.
PIMI addresses a distinct setting by retaining stochastic binary spins and operating directly on the current configuration of the native $N$-spin Ising graph. 
While conceptually related to momentum-based ideas \cite{Qian1999, goh2017why}, PIMI differs in implementation: Momentum Annealing uses a $2N$-spin bipartite mapping together with per-instance preprocessing, including eigenvalue computation, dropout, and momentum scaling \cite{Okuyama2019}, whereas momentum-accelerated Coherent Ising Machines evolve continuous variables \cite{Brown2024}.

In addition, we showed how to optimize PIMI's hardware implementations through hardware--software co-design that jointly shapes the update rule and its architectural mapping.
This approach ensures that the resulting parallel dynamics can be efficiently realized on hardware platforms such as field-programmable gate arrays (FPGAs), with potential for implementation as application-specific integrated circuits (ASICs).
As a proof of concept, we implemented both a conventional PIM and a PIMI on a Xilinx U55C FPGA and evaluated their performance on standard Ising benchmarks including the Max-Cut problem \cite{Karp1972} and the Sherrington-Kirkpatrick (SK) model \cite{Sherrington1975}, demonstrating higher success probabilities and significantly reduced time-to-solution relative to conventional PIMs.
We further demonstrated the co-design of a PIMI for multiple-input multiple-output (MIMO) detection, a key signal-processing task in modern wireless communication systems \cite{Tse2005, Proakis2008, Larsson2014}.  
Motivated by prior studies of Ising-machine-based MIMO detection \cite{Kim2019, Singh2022ri, Singh2022cim_ri, DeLunaDucoing2022, Singh2022}, we developed an application-oriented FPGA demonstration in which PIMI achieves higher detection accuracy than the minimum mean-square error (MMSE) detector, a standard linear baseline, and higher throughput than the conventional sequential-update PIM.

\section{PIMI Algorithm and Hardware–Software Co-Design Framework}
\label{sec:PIMI}
Conventional Probabilistic Ising Machines (PIMs) aim to heuristically solve optimization problems $\argmin_{\mathbf{s}}\Ham(\mathbf{s})$ defined by an Ising Hamiltonian,
\begin{equation}
\Ham(\mathbf{s}) = -\sum_{i<j} J_{ij} s_i s_j - \sum_i h_i s_i,
\label{eq:ising_hamiltonian}
\end{equation}
where $s_i \in \{-1,+1\}$ are the binary spin variables being optimized, and the problem instance being solved is defined by $J_{ij}$, specifying spin-spin couplings, and $h_i$, specifying local bias terms (sometimes called external-field terms).
PIMs explore this energy landscape through sequential stochastic spin updates, as illustrated in \fig{algo_schematic}{a} (orange).
We discretize the dynamics into update steps (discrete time steps) indexed by $t$; in a conventional PIM, each update step updates a single selected spin. 
For a system of $N$ spins, the updated spin is chosen by cycling through the spins in a fixed order (equivalently, $i = t \bmod N$), and its effective local field is computed as
\begin{align}
I_i(t) &= \sum_j J_{ij} s_j(t) + h_i.
\label{eq:local_field}
\end{align}
The spin is then stochastically updated according to
\begin{align}
s_i(t+1) &= \text{sign}\Big[\tanh \big(\beta(t) I_i(t)\big) + \eta(t) U(-1,1)\Big],
\label{eq:PIM_conv}
\end{align}
where $J_{ij}$ are the couplings, $h_i$ the local biases, $\beta(t)$ is an inverse temperature, $U(-1,1)$ denotes a uniform random variable in $[-1,1]$ and $\eta(t)$ determines the noise amplitude. The spins are updated one at a time. As we have highlighted in the previous section,  this sequential approach results in a latency that increases with the problem size, since each update \textit{sweep} (updating each of the $N$ spins once) requires $\mathcal{O}(N)$ iterations.  

As we have also already noted, applying this update rule to each of the spins in parallel (\fig{algo_schematic}{a} (green), all spins are updated simultaneously at each update step) tends to cause groups of mutually coupled spins to flip together, resulting in oscillations of the spin configuration.
With the absence of noise, Goles~et~al.\cite{Goles1981, Goles1985} have shown that the dynamics settle into a \emph{period-2} oscillation between two configurations, which for symmetric couplings is the only non-trivial attractor of synchronous threshold dynamics.
Once noise is present, the alternation is no longer exact, but the dynamics remain trapped near these two configurations, leaving the PIM far from minimum-energy spin configurations~\cite{Aadit2022}.

\begin{figure}[htbp]
    \centering
    \includegraphics[width=\textwidth]{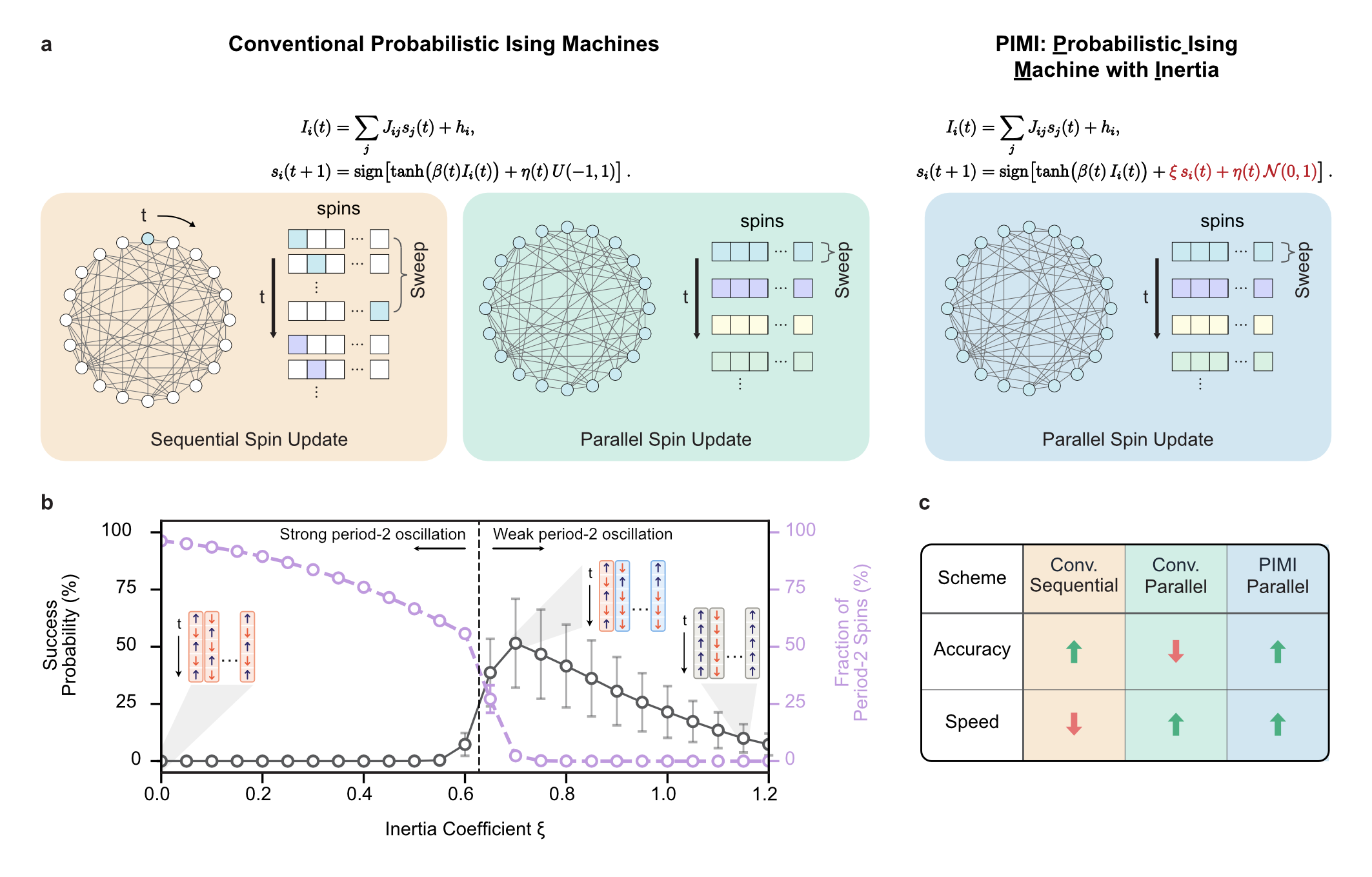}
    \vspace{-20pt}
    \caption{
    \textbf{\underline{P}robabilistic \underline{I}sing \underline{M}achine with \underline{I}nertia (PIMI) enables fully parallel probabilistic Ising dynamics with high solution quality and favorable hardware scaling.}
    \panel{a} Comparison between a conventional Probabilistic Ising Machine (PIM) and our proposed Probabilistic Ising Machine with Inertia (PIMI). 
    In conventional PIMs, spins ($s_i$) are updated one at a time (orange), as updating all spins simultaneously (green) repeated oscillations in the local field and prevents the system from converging \cite{Aadit2022}.   
    PIMI introduces a self-spin (inertia) term ($\xi s_i(t)$) that suppresses these oscillations and enables stable fully parallel updates, together with a stochastic term ($\eta(t)\mathcal{N}(0,1)$) that injects randomness for exploration.
    \panel{b} Introducing the inertia term suppresses period-2 oscillations and improves solution quality. 
    Success probability (black, left axis) and the fraction of spins in period-2 oscillation (purple, right axis) versus the inertia coefficient $\xi$, for Max-Cut instances at $N=150$ (random Erd\H{o}s--R\'enyi graphs with 50\% edge density~\cite{Erdos1959}; 100 instances, 256 \runname{s} each, $100N$ steps).
    The period-2 fraction is the fraction of spins in the alternating pattern $s_i(t)=-s_i(t-1)=s_i(t-2)$, averaged over the final 1000 steps of each anneal and across all \runname{s} and instances.
    The period-2 fraction drops as $\xi$ increases: at small $\xi$ (strong period-2
    oscillation) the success probability is near zero, whereas at larger $\xi$ (weak
    period-2 oscillation) it rises to a peak at $\xi^\star\approx0.70$. When $\xi$ is too large ($\xi>1$), the inertia term dominates and the spins tend to be frozen, which prevents exploration of the solution space --- resulting in a low success probability again.
    \panel{c} Accuracy and speed of the three schemes: sequential updates are accurate but limited in speed, parallel updates are fast but inaccurate, and PIMI is both accurate and fast. Arrows indicate favorable (up) or unfavorable (down) behavior.
    }
    \label{fig:algo_schematic}
\end{figure}

As shown in \fig{algo_schematic}{a} (blue), to enable fast and reliable convergence to low-energy states in probabilistic Ising systems, we introduce a \underline{P}robabilistic \underline{I}sing \underline{M}achine with \underline{I}nertia (PIMI), which allows fully parallel spin updates by including a self-spin term in the update rule.
In a PIMI, the effective local field for each spin is calculated in the same way as in conventional PIMs (\eq{local_field}). 
However, instead of updating spins one at a time, all spins are updated simultaneously at each update step $t$, according to
\begin{align}
s_i(t+1) &= \text{sign}\Big[\tanh\big(\beta(t) I_i(t)\big) + \xi s_i(t) + \eta(t) \mathcal{N}(0,1)\Big],
\label{eq:PIMI}
\end{align}
where $I_i(t)$ is the effective local field of \eq{local_field}, $\beta(t)$ is the inverse temperature, $\xi$ is the inertia coefficient setting the strength of the self-spin term, $\mathcal{N}(0,1)$ denotes a gaussian random variable, and $\eta(t)$ determines the noise amplitude. The inverse temperature $\beta(t)$ and the noise amplitude $\eta(t)$ follow problem-instance-specific update schedules.
For our benchmarking of Max-Cut and SK instances, we increased both quantities according to a coupled schedule that progressively strengthens the local-field response while retaining sufficient stochasticity to avoid premature locking.
The specific schedules and their interpretation are provided in \ref{sec:PIMI_MC_SK}, \ref{sec:CPIM_MC_SK}, \ref{sec:PIMI_MIMO}, and \ref{sec:CPIM_MIMO}.
Gaussian noise is used because its unbounded tails improve stochastic exploration relative to bounded uniform noise; see \ref{sec:noise_distribution} for analysis and numerical comparison.

The inertia term $\xi s_i(t)$ biases each spin toward its current value, making spin flips less likely and thereby suppressing the period-2 instability described above.
\fig{algo_schematic}{b} shows how the success probability and the fraction of spins remaining in period-2 oscillations ($s_i(t)=-s_i(t-1)=s_i(t-2)$, measured over the final 1000 update steps of each \runname) vary with the inertia coefficient $\xi$, for Max-Cut instances at $N=150$.
Without inertia ($\xi=0$), almost all spins oscillate, and fully parallel updates alone fail to solve these instances.
Increasing $\xi$ steadily reduces the fraction of spins in period-2 oscillation, and the success probability increases significantly once the oscillation is largely suppressed, reaching a maximum of $51.5\%$ at $\xi = 0.7$. 
For $\xi>0.7$, the success probability declines steadily as the inertia grows further: the self-bias becomes strong enough that spins tend to stick to their current values, making it hard to explore the solution space.
The inertia coefficient therefore balances two competing requirements --- enough self-bias to suppress oscillation, but not so much that exploration is lost. 
A detailed analysis of why the inertia term removes the period-2 attractor, together with the behavior of the underlying noise-free dynamics, is given in \ref{sec:period-2}.
\fig{algo_schematic}{c} shows how conventional PIMs with sequential and parallel updates compare with PIMI using parallel updates: sequential updates favor accuracy, parallel updates with conventional dynamics favor speed, and PIMI combines both. 
The accuracy and speed of the update schemes are characterized quantitatively in the following sections.

\definecolor{seqbg}{HTML}{F8E9D4}
\definecolor{parbg}{HTML}{D1EDE6}
\definecolor{pimibg}{HTML}{D1E6F1}

\begin{tcolorbox}[colback=white,colframe=gray,boxrule=2pt,arc=2pt,
                  title={Box 1. Definition of trials, update steps, and sweeps in this work},
                  fonttitle=\bfseries]
\textbf{A trial:} start from an initial spin configuration and execute the corresponding update rule for a fixed number of update steps, with each \textbf{update step} corresponding to one iteration of the loop below.

\begin{tcolorbox}[colback=seqbg,colframe=seqbg,boxrule=0pt,arc=2pt,left=4pt,right=4pt,top=4pt,bottom=4pt]
\raggedright
\textbf{Conventional PIM with sequential updates:}\\
\For{$u = 1$ \KwTo $T_{\mathrm{steps}}/N$}{
    \For{$i = 1$ \KwTo $N$}{
        $t \gets (u-1)N + i$\;\\
        $I_i(t) \gets \sum_j J_{ij}s_j(t) + h_i$\;\\
         $s_i(t+1) \gets \operatorname{sign}\!\Big[\tanh\!\big(\beta(t)\,I_i(t)\big) + \eta(t)\,U(-1,1)\Big]$\;
        update spin $s_i$ only\;
    }
}
\textit{Here, $t$ denotes the discrete update-step index. Each increment of $t$ corresponds to one single-spin update. One \textbf{sweep} corresponds to $N$ such update steps.}
\end{tcolorbox}

\vspace{-0.8em}
\begin{tcolorbox}[colback=parbg,colframe=parbg,boxrule=0pt,arc=2pt,left=4pt,right=4pt,top=4pt,bottom=4pt]
\raggedright
\textbf{Conventional PIM with parallel updates:}\\
\For{$t = 1$ \KwTo $T_{\mathrm{steps}}$}{
    $\bm{I}(t) \gets J\,\bm{s}(t) + \bm{h}$\;\\
    $\bm{s}(t+1) \gets \operatorname{sign}\!\Big[\tanh\!\big(\beta(t)\,\bm{I}(t)\big) + \eta(t)\,U(-1,1)^N\Big]$\;
    update all spins $\bm{s}$ simultaneously;
}
\textit{One \textbf{sweep} corresponds to a single update step.}
\end{tcolorbox}

\vspace{-0.8em}
\begin{tcolorbox}[colback=pimibg,colframe=pimibg,boxrule=0pt,arc=2pt,left=4pt,right=4pt,top=4pt,bottom=4pt]
\raggedright
\textbf{Probabilistic Ising Machine with Inertia (PIMI):}\\
\For{$t = 1$ \KwTo $T_{\mathrm{steps}}$}{
    $\bm{I}(t) \gets J\,\bm{s}(t) + \bm{h}$\;\\
    $\bm{s}(t+1) \gets \operatorname{sign}\!\Big[\tanh\!\big(\beta(t)\,\bm{I}(t)\big) + \xi\,\bm{s}(t) + \eta(t)\,\mathcal{N}(\bm{0},\bm{1})\Big]$\;
    update all spins $\bm{s}$ simultaneously;
}
\textit{One \textbf{sweep} corresponds to a single update step.}
\end{tcolorbox}
\end{tcolorbox}

We now briefly define several terms related to the PIMI and PIM algorithms, which we will use throughout this paper (see also Box~1). A \emph{\runname} is an independent run composed of multiple \emph{sweeps}, producing a single candidate solution. 
We discretize the dynamics into \emph{\stepname{s}} (discrete time steps), indexed by $t$.
Each sweep corresponds to one complete round of updates over all $N$ spins and can be viewed as $N$ single-spin updates. 
In conventional PIMs, these $N$ single-spin updates are typically performed serially (one spin per \stepname), whereas in a PIMI they are performed in parallel (all $N$ spins updated simultaneously in one \stepname). 
An \emph{\schedulename} specifies how the update parameters (the inverse temperature $\beta(t)$ and noise level $\eta(t)$) vary across \stepname{s} within a \runname.

\begin{figure}[htbp]
    \centering
    \includegraphics[width=\textwidth]{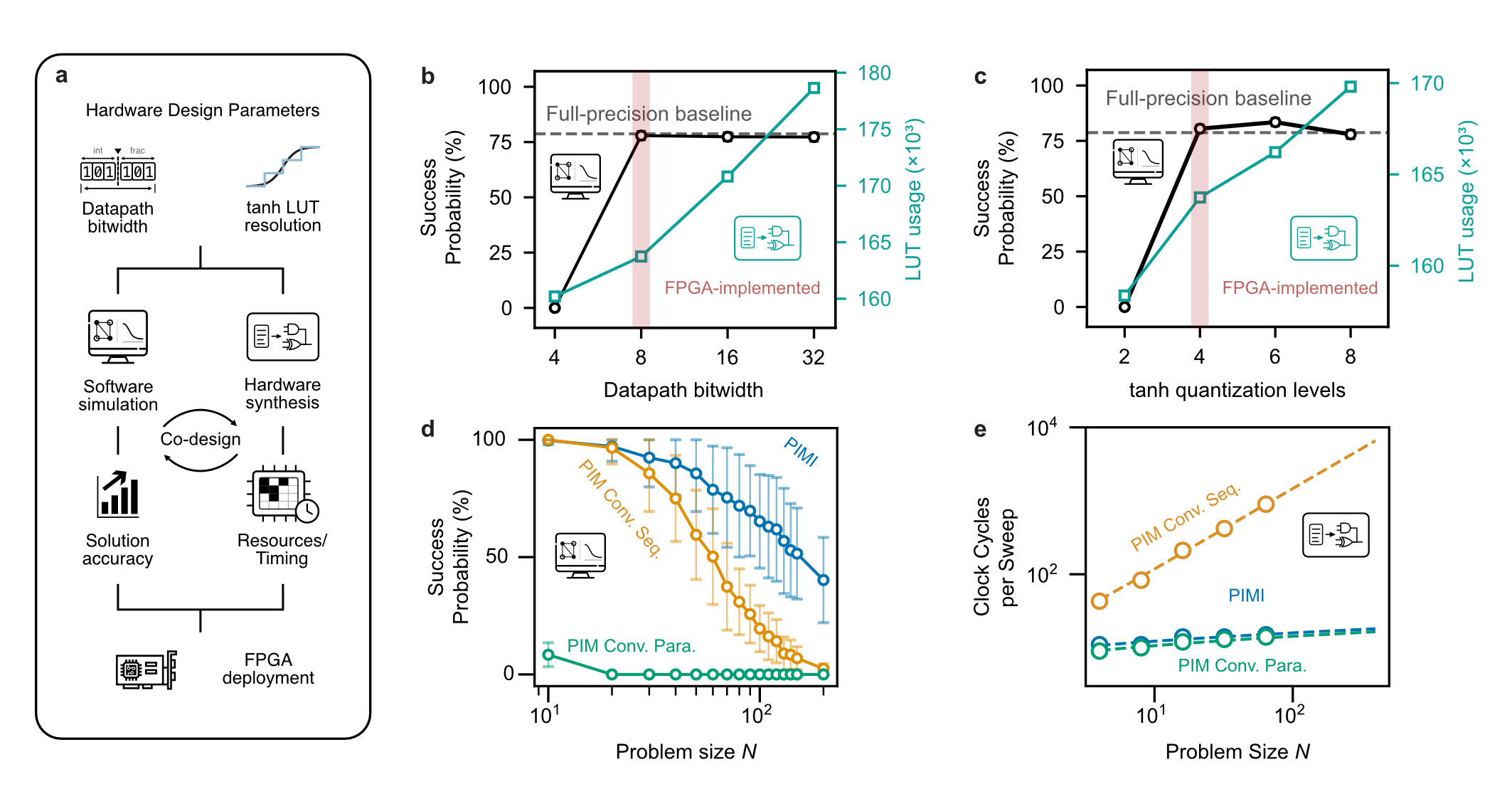}
    \caption{
    \textbf{Hardware--software co-design yields a Probabilistic Ising Machine with Inertia (PIMI) that achieves high solution quality with low field-programmable gate array (FPGA) resource usage and favorable clock-cycle scaling.}
    \panel{a} Hardware--software co-design workflow for implementing a PIMI on an  FPGA.
    Tunable FPGA hardware parameters (data bitwidth and quantized $\tanh$ look-up-table resolution) are jointly optimized using software simulations in Python for solution accuracy and hardware synthesis in C++ with Vitis high-level synthesis (HLS) for resource usage and latency before FPGA deployment.
    \panel{b} Success probability (black, left axis) and FPGA look-up table (LUT) usage (teal, right axis) versus datapath bitwidth, with the quantized $\tanh$  fixed at 4 levels, for Max-Cut instances at $N=60$ (random Erd\H{o}s--R\'enyi graphs with 50\% edge density, 100 instances, 256 \runname{s} each, $100N$ steps). 
    The fixed-point format allocates 1 sign bit and 2 integer bits, with the remaining bits assigned to the fractional part.
    The success probability reaches the full-precision baseline (dashed) at a bitwidth of 8 and is used in the FPGA implementation (shaded red).
    \panel{c} Success probability (black, left axis) and LUT usage (teal, right axis) versus the number of $\tanh$ quantization levels, with the datapath bitwidth fixed at 8 bits, for the same instances and protocol as in \textbf{b}. 
    Four levels are sufficient to reach the full-precision baseline (dashed) and are used in the FPGA implementation (shaded red).
    Full success-probability results for all datapath-bitwidth and $\tanh$-level combinations are provided in \ref{sec:mc_quant_tanh_map}.
    \panel{d} Success probability versus problem size $N$ for Max-Cut instances with the same protocol as in \textbf{b}. 
    Solid lines indicate the mean across the 100 instances at each size, and error bars denote the standard deviation. 
    PIMI (blue) outperforms conventional probabilistic Ising Machines (PIMs) with sequential updates, whereas conventional PIM with fully parallel updates fails to converge. See Section~\ref{sec:mc_sk_benchmark} and \ref{sec:mc_gen} for additional details.
    \panel{e} FPGA clock cycles required to update all spins once (one full sweep) as a function of problem size.
    PIMI exhibits logarithmic scaling, $C_{\mathrm{PIMI}} \simeq 1.1 \log_2(N) + 8.6$, in contrast to conventional PIMs with parallel updates ($C_{\mathrm{para}} \simeq 1.1 \log_2(N) + 7$) and sequential updates ($C_{\mathrm{seq}} \simeq N \log_2(N) + 8N + 4.67$).  See \ref{sec:CC_per_sweep} for more details.
    }
    \label{fig:co_design}
\end{figure}

As illustrated in \fig{co_design}{a}, we adopted a hardware-software co-design approach to realize a PIMI on an FPGA.
To minimize hardware resource usage , we aimed to use as few bits as possible in the datapath and to replace the exact $\tanh$ nonlinearity with a coarsely quantized look-up table (see \ref{sec:quant_sch} for details).
Algorithm development and parameter tuning were first performed in Python using a quantized simulation framework designed to match FPGA arithmetic behavior, where the $\tanh$ look-up table resolution and data quantization bit-widths (e.g. the coupling matrix $J$) were optimized for solution accuracy under fixed resource constraints.
The finalized design was then translated into C++, after which Vitis high-level synthesis (HLS)\cite{vitis_hls_2023_1} was used to generate the hardware implementation, with loop pipelining, unrolling, and memory partitioning applied to meet throughput targets while minimizing FPGA resource usage.
This workflow enabled direct mapping of PIMI's update rule to a fully parallel hardware architecture that maintains solution quality under practical FPGA resource constraints.
Throughout this work, the computer icon denotes software-level algorithm development and simulation, the code-to-gate icon denotes hardware-aware synthesis and emulation using HLS, and the FPGA icon denotes the deployed hardware implementation (more details about FPGA quantization and software-hardware equivalence calibration can be found in \ref{sec:quant_sch} and \ref{sec:hw_equiv}).

\fig{co_design}{b} and~c present our co-design decisions for the PIMI kernels used in the Max-Cut and SK solvers. 
Both panels report the success probability together with the corresponding FPGA Look-up-table usage (since it is the dominant resource used in our implementation) for Max-Cut instances at $N=60$.
The success probability is compared against a baseline obtained with full floating-point precision and an unquantized $\tanh$ evaluation (dashed lines).
\fig{co_design}{b} shows the success probability (left axis) and LUT usage (right axis) as a function of the datapath bitwidth, with the $\tanh$ look-up table held at 4 levels. 
The fixed-point format allocates 1 sign bit and 2 integer bits, with the remaining bits assigned to the fractional part. The success probability improves with increasing precision and reaches the full-precision baseline at a bitwidth of 8, beyond which additional precision yields no further gain in solution quality while FPGA LUT usage continues to grow. 
We therefore adopt an 8-bit fixed-point representation, the lowest precision that matches the full-precision baseline.
\fig{co_design}{c} shows the corresponding dependence on the number of quantization levels in the $\tanh$ look-up table, with the datapath bitwidth held at 8 bits. 
The success probability increases with LUT resolution and reaches the full-precision baseline at 4 levels, whereas LUT usage grows with the number of levels. 
Balancing solution quality against hardware cost, we therefore employ a 4-level $\tanh$ look-up table to minimize per-kernel resource usage. 
Results for all combinations of datapath bitwidth and $\tanh$ level are provided in \ref{sec:mc_quant_tanh_map}.

\fig{co_design}{d} compares the success probability of PIMI and conventional PIMs with sequential or parallel spin updates on Erd\H{o}s--R\'enyi random Max-Cut instances with 50\% edge density across a range of problem sizes \cite{Erdos1959}. For each problem size \(N\), we generated 100 instances. For each instance, the success probability was estimated from 256 independent trials, each run for \(100N\) update steps. 
A trial was counted as successful if, at any point during the run, it reached the reference-state energy for that instance, and the success probability was defined as the fraction of successful trials.
The reported value at each problem size is the mean across the 100 instances.
Our results suggest that a conventional PIM with sequential updates performs well at small sizes, while its success probability gradually decreases as the problem size grows.
Directly applying fully parallel updates to a conventional PIM, however, leads to unstable dynamics and very low success probabilities even for small system sizes. 
In contrast, PIMI consistently achieves higher success probability than conventional PIMs across all problem sizes tested here, showing that fully parallel updates can be effective without sacrificing performance.
To evaluate execution speed, we implemented a PIMI, together with conventional PIMs using sequential and parallel update schemes on an FPGA.
In these implementations, matrix–vector and vector–vector multiplications were fully unrolled to maximize parallelism and throughput.
\fig{co_design}{e} shows how the number of FPGA clock cycles required for one full sweep scales with problem size for PIMI and conventional PIMs.
In a conventional PIM with sequential updates, the latency scales as $\mathcal{O}(N \log_2 N)$ since each spin needs to wait for the previous one to be updated before its local field can be calculated. 
In contrast, PIMI performs all spin
updates in parallel, resulting in a near-constant sweep cost with only $\mathcal{O}(\log_2 N)$ overhead due to accumulation within matrix–vector multiplications and loop management (see \ref{sec:CC_per_sweep} for a detailed explanation).

As a result, PIMI reduces the clock cycles used per sweep by a factor of $\mathcal{O}(N)$ relative to sequential PIM updates, while avoiding the convergence issues associated with directly applying parallel updates to the conventional dynamics.
However, the speed advantage of PIMI over conventional PIM goes beyond this $\mathcal{O}(N)$ advantage in clock cycles per sweep: as we will see in the empirical results that are presented in the next section, the modifications to the dynamics in PIMI result in an improvement in clock-cycles-to-solution (i.e., time-to-solution) that scales even more rapidly than the $\mathcal{O}(N)$ improvement that arises from the parallelization of the spin updates.

\section{Results}

\subsection{Benchmarking PIMI on standard Ising problems}
\label{sec:mc_sk_benchmark}
\begin{figure}[htbp]
    \centering
    \includegraphics[width=\textwidth]{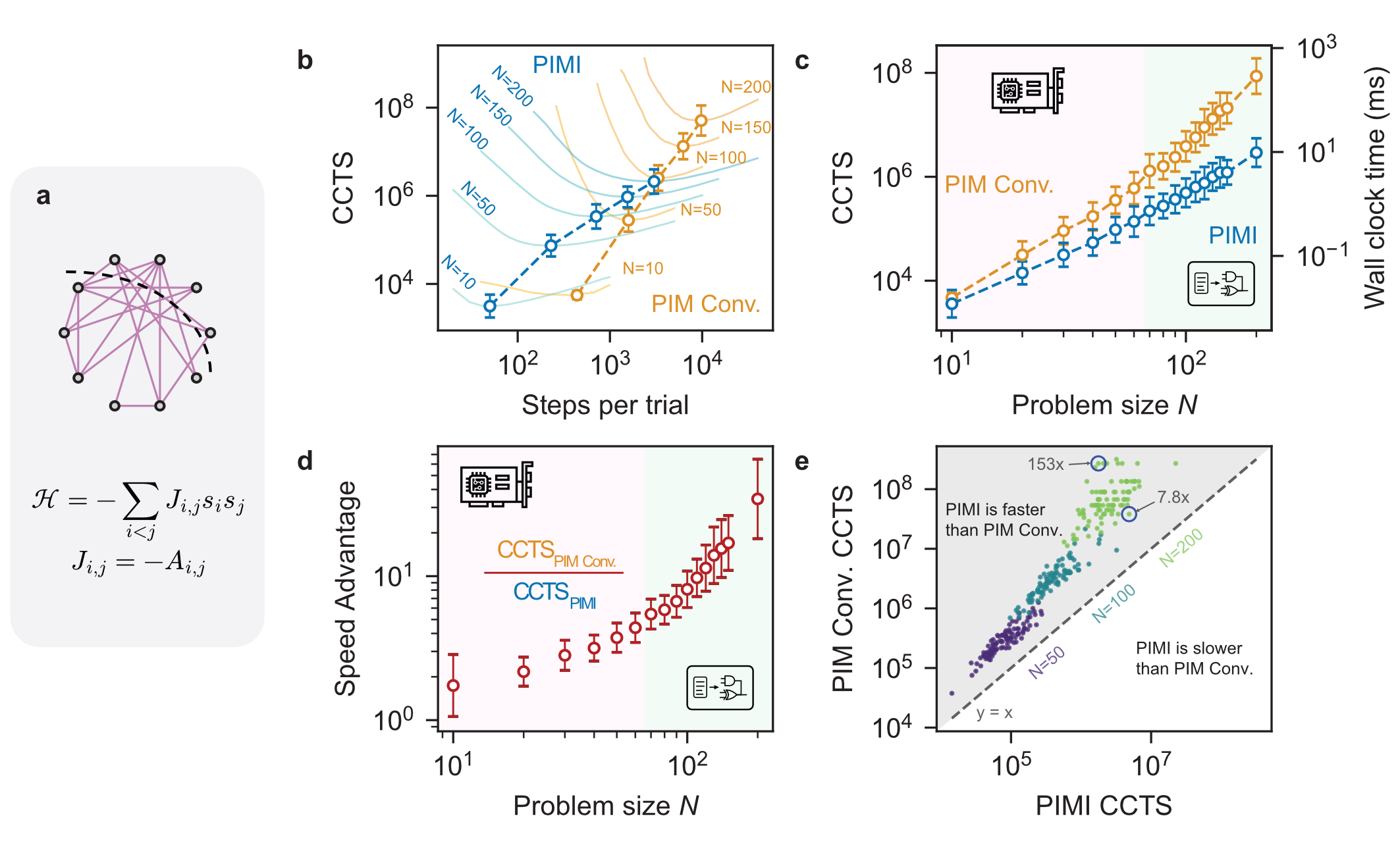}
    \caption{
    \textbf{Probabilistic Ising Machine with Inertia (PIMI) reduces the number of  clock cycles to solution compared to conventional probabilistic Ising machines on Max-Cut problems.}
    \panel{a} Mapping of a Max-Cut instance to an Ising Hamiltonian for unweighted Erd\H{o}s–R\'enyi graphs with edge probability 0.5, where $A_{ij}=1$ if vertices $i$ and $j$ are connected and $A_{ij}=0$ otherwise; the Ising coupling matrix is given by $J=-A$.
    \panel{b} Clock cycles to solution (CCTS) as a function of the step budget per trial for representative problem sizes ($N = 10, 50, 100, 150, 200$). 
    Blue curves correspond to probabilistic Ising machine with inertia (PIMI), whereas orange curves correspond to the conventional probabilistic Ising machine with sequential updates (PIM Conv.).
    For each problem size, the results were collected over 100 instances. For every instance and number of steps per trial, the success probability was estimated from 256 independent trials, and these values were then averaged across instances.
    The ensemble-averaged success probability was converted into the number of independent trials required to reach a cumulative success probability of $99.9\%$ using Eq.~\ref{eq:n_trials}; multiplying this quantity by the steps per trial and the clock-cycle cost per step yielded the CCTS (Eq.~\ref{eq:CCTS}).
    For each problem size, CCTS exhibits a minimum at an intermediate step budget, reflecting the trade-off between steps per trial and success probability. 
    Dashed lines connect the measured optima, and faint curves show the full landscapes for different $N$.
    \panel{c} Optimal CCTS versus problem size $N$, obtained by minimizing over the step budget in \textbf{b}. PIMI consistently requires fewer clock cycles to reach a solution than the conventional PIM. The right axis shows the corresponding wall-clock time estimated from the hardware clock rates of the two implementations.
    \panel{d} Speed advantage as a function of problem size, computed as the ratio of the optimal CCTS of the conventional PIM to that of PIMI. The advantage increases with N, indicating that the benefit of PIMI becomes more significant for larger instances.
    \panel{e} Instance-by-instance comparison of the optimal CCTS obtained with PIMI and with the conventional PIM for $N=50, 100 \text{ and } 200$, with 100 instances shown for each problem size. 
    Each point represents one instance and is colored by problem size. Points above the dashed $y=x$ line correspond to instances for which PIMI required fewer clock cycles to solution than the conventional PIM.
    For $N=200$, PIMI achieved an average speedup of $34\times$, with a maximum observed advantage of $153\times$ and a minimum of $7.8\times$.
    Error bars in \textbf{b–d} denote the standard deviation across instances of every problem size, evaluated in log space.
    Results for smaller problem sizes ($N \leq 60$) are obtained from FPGA runs, while results for larger problem sizes ($N > 60$) are obtained from hardware emulation.
    Details of the estimation procedure and extended figures are provided in \ref{sec:CCTS_estimation}.
    }
    \label{fig:max_cut}
\end{figure}

We evaluated the performance and speed of PIMI on two standard Ising benchmarks: the Max-Cut problem and the Sherrington-Kirkpatrick (SK) model.
The Max-Cut problem aims to partition a graph's nodes into two sets such that the number of edges crossing between them is maximized.  
As shown in \Fig{max_cut}{a}, the Ising coupling matrix $J$ is obtained from the graph's adjacency matrix $A$ by setting $J_{ij} = -A_{ij}$, where $A_{ij} = 1$ for connected nodes and $0$ otherwise (\ref{sec:mc_gen} provides more details about how the problems are generated).
To compare PIMI and conventional PIMs in terms of hardware runtime, we next introduce the clock cycles to solution (CCTS), which quantifies the total number of FPGA clock cycles required to solve a problem instance.
This metric depends on the number of update steps per trial (denoted by $T_{\mathrm{steps}}$): increasing $T_{\mathrm{steps}}$ generally improves the success probability of a trial, but also increases its hardware runtime cost, which makes the choice of $T_{\mathrm{steps}}$ critical for minimizing the overall CCTS \cite{Albash2018, Hamerly2019}.

Following established time-to-solution benchmarking conventions, we characterize each problem size $N$ using the ensemble-averaged single-trial success probability $\bar{p}(T_{\mathrm{steps}};N)$, obtained by averaging the empirical success probabilities across problem instances, with each trial run for $T_{\mathrm{steps}}$ update steps \cite{Ronnow2014,McMahon2016,Hamerly2019}.
The number of independent trials required for the probability of obtaining at least one successful outcome to reach $99.9\%$ is
\begin{equation}
n_{\mathrm{trials}}(T_{\mathrm{steps}};N)
=
\frac{\log(0.001)}
{\log\!\bigl(1-\bar{p}(T_{\mathrm{steps}};N)\bigr)}.
\label{eq:n_trials}
\end{equation}
The resulting number of clock cycles to solution is then
\begin{equation}
\mathrm{CCTS}(T_{\mathrm{steps}};N)=n_{\mathrm{trials}}(T_{\mathrm{steps}};N)\,T_{\mathrm{steps}}\,C_{\mathrm{step}}(N),
\label{eq:CCTS}
\end{equation}
where \(C_{\mathrm{step}}(N)\) is the FPGA clock-cycle cost per full update step (one sweep), as reported in \fig{co_design}{e} (see \ref{sec:CCTS_estimation} for details about the estimation procedure).
\Fig{max_cut}{b} plots CCTS versus $T_{\mathrm{steps}}$ for representative problem sizes (\(N=10,50,100,150,200\)).
For both PIMI and conventional PIM, the trade-off between hardware runtime and success probability is clearly observed in the data, with the minimum CCTS achieved at an intermediate number of steps per trial. The optimal number of steps per trial and the corresponding clock cycles to solution for each problem size are highlighted and connected by dashed lines in the plot.
\fig{max_cut}{c} further summarizes the optimal CCTS across all problem sizes studied ($N = 10, 20, \ldots, 150, 200$). PIMI consistently requires fewer clock cycles to solution than the conventional PIM.
The right axis shows the corresponding wall-clock times estimated from the hardware clock rates.

To quantify the relative speedup provided by PIMI, \fig{max_cut}{d} plots the ratio between optimal CCTS for the conventional PIM and for PIMI.
The speed advantage increases with problem size, indicating that the benefit of PIMI becomes more significant for larger instances.
This advantage is also evident at the level of individual instances. 
\fig{max_cut}{e} compares the optimal CCTS of PIMI and conventional PIM on individual instances for $N=50$, 100, and 200, with 100 instances shown for each problem size.
Points above the dashed $y=x$ line correspond to instances for which PIMI required fewer clock cycles to solution than conventional PIM. 
Our results show that PIMI consistently outperforms conventional PIM across all tested instances for practical problem sizes.
Specifically, for $N=200$, PIMI achieved an average advantage of $34\times$, with a maximum observed advantage of $153\times$ and a minimum observed advantage of $7.8\times$.
(See \ref{sec:CCTS_estimation} for extended figures.) 
We emphasize that the observed advantage is empirical and is limited to the range of problem sizes studied here.


The Sherrington--Kirkpatrick (SK) model provides a complementary benchmark to Max-Cut for evaluating performance on random spin-glass instances. Here, we use the SK-1 model, a fully connected variant in which every pair of spins is coupled and each coupling $J_{ij}$ independently takes the value $-1$ or $+1$ with equal probability (see \fig{sk1}{a} and \ref{sec:sk_gen}). 
The resulting dense mixture of positive and negative couplings makes this problem particularly challenging for parallel update rules.
The CCTS analysis for SK-1 was performed in the same way as for Max-Cut. 
\Fig{sk1}{b} shows CCTS versus the number of steps per trial $T_{\mathrm{steps}}$ for representative problem sizes, with both PIMI and conventional PIM exhibiting a minimum at an intermediate value of $T_{\mathrm{steps}}$.
The optimal number of steps per trial and corresponding minimum CCTS for each size are highlighted and connected by dashed lines. 
\Fig{sk1}{c} summarizes the optimal CCTS across all problem sizes studied and shows that PIMI consistently reaches a solution in fewer clock cycles than the conventional PIM; the right axis shows the corresponding wall-clock times. 
\Fig{sk1}{d} quantifies the relative speedup by plotting the ratio between optimal CCTS for the conventional PIM and for PIMI, which again increases with problem size. Finally, \Fig{sk1}{e} compares the optimal CCTS of PIMI and conventional PIM on individual instances for \(N=50\), 100, and 200, with 100 instances shown for each problem size. 
For \(N=200\), PIMI achieved an average advantage of $35\times$, with the greatest observed advantage of $161\times$ and the smallest of $7.5 \times$. Extended figures with all problem sizes are provided in \ref{sec:CCTS_estimation}.

\begin{figure}[htbp]
    \centering
    \includegraphics[width=\textwidth]{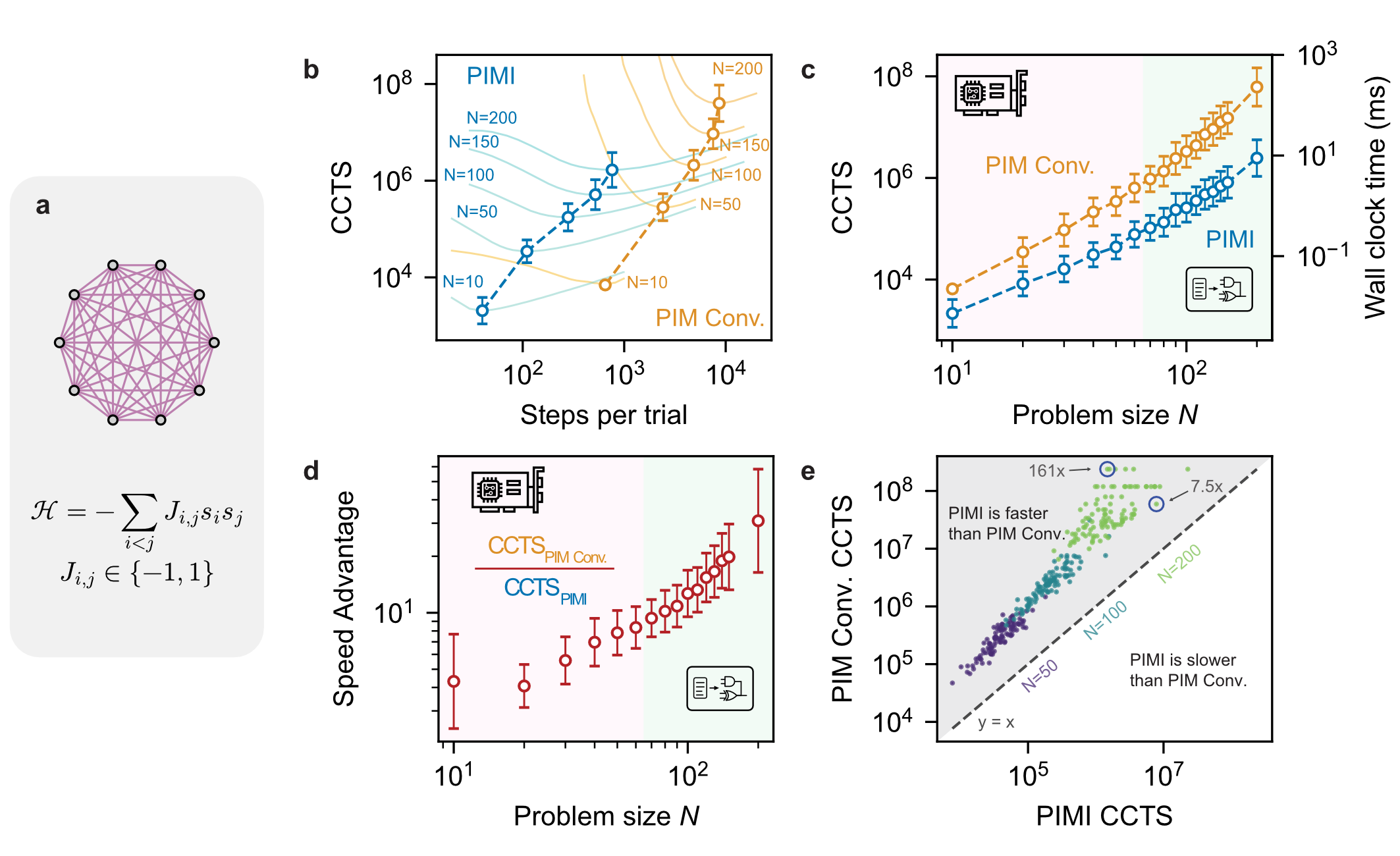}
    \caption{
    \textbf{Probabilistic Ising Machine with Inertia (PIMI) reduces the number of clock cycles to solution compared to conventional probabilistic Ising machines for solving the Sherrington-Kirkpatrick-1 (SK-1) model.}
    \panel{a} Mapping of an SK-1 spin-glass instance to the corresponding Ising Hamiltonian, where all spin pairs are connected and each coupling term $J_{ij}$ is independently set to +1 or -1 with equal probability.
    \panel{b} Clock cycles to solution (CCTS) as a function of the number of steps per trial for representative problem sizes (\(N = 10, 50, 100, 150, 200\)). Blue curves correspond to probabilistic Ising machine with inertia (PIMI), whereas orange curves correspond to the conventional probabilistic Ising machine with sequential updates (PIM Conv.). 
    CCTS was derived in the same way as in \fig{max_cut}{b} from success probabilities estimated over 100 instances with 256 independent trials per instance, together with the clock-cycle cost per step. Dashed lines connect the measured optima, and faint curves show the full landscapes for different $N$.
    \panel{c} Optimal CCTS versus problem size $N$, obtained by minimizing over the step budget in \textbf{b}. PIMI consistently requires fewer clock cycles to reach a solution than the conventional PIM. The right axis shows the corresponding wall-clock time estimated from the hardware clock rates of the two implementations.
    \panel{d} Speed advantage as a function of problem size, computed as the ratio of the optimal CCTS of the conventional PIM to that of PIMI. The advantage increases with $N$, indicating that the benefit of PIMI becomes more pronounced for larger instances.
    \panel{e} Instance-by-instance comparison of the optimal CCTS obtained with PIMI and with the conventional PIM for $N=50$, 100 and 200, with 100 instances shown for each problem size. 
    Each point represents one instance and is colored by problem size. Points above the dashed $y=x$ line correspond to instances for which PIMI required fewer clock cycles to solution than the conventional PIM. 
    For $N=200$, PIMI achieved an average speedup of $35\times$, with a maximum observed advantage of $161\times$ and a minimum of $7.5\times$.
    Error bars in \textbf{b--d} denote the standard deviation across instances of the same problem size, evaluated in log space. Results for smaller problem sizes (\(N \leq 60\)) are obtained from FPGA runs, while results for larger problem sizes (\(N > 60\)) are obtained from hardware emulation.
    Extended figures are provided in \ref{sec:CCTS_estimation}.
    }
    \label{fig:sk1}
\end{figure}

\subsection{FPGA-Based MIMO detection implementation}

\begin{figure}[htbp]
    \centering
    \includegraphics[width=\textwidth]{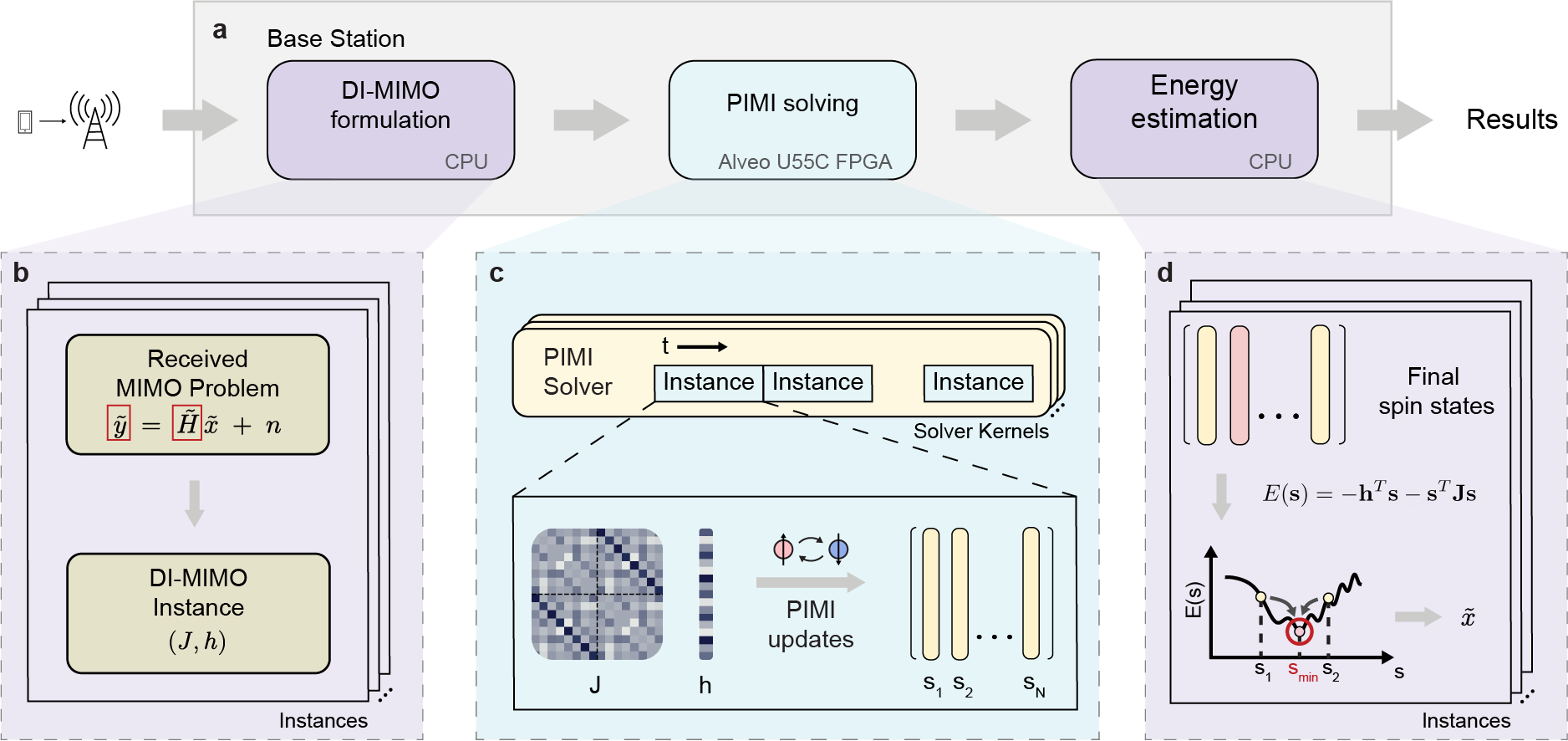}
    \caption{
    \textbf{Field-programmable gate array (FPGA) implementation of the Probabilistic Ising Machine with Inertia (PIMI) for multiple-input multiple-output (MIMO) detection.}
    \panel{a} System-level overview of PIMI-based MIMO detection in a base-station setting.
    DI-MIMO instance formulation and final energy evaluation are performed on the host computer, while dedicated solver kernels on the FPGA execute the PIMI update steps.
    \panel{b} Host-side DI-MIMO instance formulation.
    Each MIMO realization $(\tilde{y},\tilde{H})$ is mapped to a corresponding Delta-Ising MIMO (DI-MIMO) instance $(J,h)$.
    \panel{c} PIMI solvers implemented on an Alveo U55C FPGA.
    Multiple PIMI kernels run continuously in parallel, each loading successive DI-MIMO instances $(J,h)$, executing a fixed number of \runname* and \stepname*, while returning the final spin states.
    \panel{d} Host-side energy evaluation, solution selection, and signal reconstruction.
    For each DI-MIMO instance, spin configurations returned from the FPGA are evaluated using the Ising energy function, the minimum-energy state is selected, and the corresponding transmitted signal $\tilde{x}$ is reconstructed.
    }
    \label{fig:MIMO_workflow}
\end{figure}

Multiple-Input Multiple-Output (MIMO) is widely used in modern wireless systems to increase communication capacity and reliability by transmitting and receiving signals across multiple antennas \cite{Tse2005, Proakis2008, Larsson2014}.
In this work, we focus on the uplink MIMO detection problem at the base station, where the objective is to recover the transmitted symbol vector $\tilde{x} \in \Phi^{N_t}$ from $N_t$ transmitters, based on the received signal
\begin{align}
\tilde{y} = \tilde{H}\tilde{x} + n,    
\end{align}
where $\tilde{y} \in \mathbb{C}^{N_r}$ is the received vector at the base station with $N_r$ antennas, $\tilde{H} \in \mathbb{C}^{N_r \times N_t}$ is the channel matrix, and $n \sim \mathcal{CN}(0,\sigma^2 I_{N_r})$ is additive Gaussian noise.
Previous studies have shown that Maximum-likelihood (ML) detection provides the optimal solution by minimizing
$\|\tilde{y} - \tilde{H}u\|^2, \quad u \in \Phi^{N_t},$    
but the search space contains $|\Phi|^{N_t}$ candidate symbol vectors, making exact ML detection computationally prohibitive for large systems or high-order modulations \cite{Yang2015}.
As an alternative, linear detectors such as Minimum Mean Square Error (MMSE) achieve efficient polynomial-time complexity but exhibit reduced accuracy in challenging detection scenarios \cite{Yang2015} (see \ref{sec:MIMO_def} for details).
More recently, Ising-machine–based approaches have been proposed for MIMO detection by mapping the maximum-likelihood detection problem onto an Ising Hamiltonian and solving it via energy minimization, achieving near-ML performance with polynomial complexity  \cite{Kim2019, Singh2022ri,Singh2022cim_ri,DeLunaDucoing2022}.
Among these, the Delta-Ising MIMO (DI-MIMO) method \cite{Singh2022} combines an initial MMSE estimate with an Ising-based refinement of the residual detection problem through couplings $(J,h)$, which provides a structured and scalable mapping suitable for hardware implementation and enables up to $2\times$ higher spectral efficiency (transmitted symbols per unit bandwidth) than conventional linear detectors.
In this work, we adopt DI-MIMO as the basis for our implementation, enabling PIMI to achieve near-ML detection even for large systems.

\fig{MIMO_workflow}{a} summarizes the FPGA-based implementation of the PIMI solvers for MIMO detection at the base station.
The host CPU performs DI-MIMO problem formulation, after which PIMI \stepname* are executed on dedicated solver kernels implemented on the FPGA. The resulting spin states are then transferred back to the host for energy evaluation and signal reconstruction.
As shown in panel~b, the host computer maps each received MIMO problem $(\tilde{y},\tilde{H})$ to a corresponding DI-MIMO instance, generating coupling matrices $J$ and local fields $h$ (see \ref{sec:DI_MIMO} for details).
The runtime of this host-side preprocessing is negligible compared with the FPGA solver execution \cite{singh2025thesis}.
The resulting instances are scheduled and transferred to the FPGA, which hosts multiple parallel PIMI solver kernels (see \ref{sec:host_scheduler} for details about the scheduler).
As shown in \fig{MIMO_workflow}{c}, these kernels process DI-MIMO instances continuously by loading $(J,h)$, executing a fixed number of \runname* and \stepname*, while returning the resulting spin configuration before immediately accepting the next instance.
Their numerical precision was selected using the same co-design procedure applied to the Max-Cut and SK solvers (\fig{co_design}), with the datapath bitwidth and $\tanh$ look-up-table resolution chosen as the smallest values meeting the target bit error rate, thereby minimizing the resource cost per kernel and maximizing the number of kernels that fit on the device (see \ref{sec:MIMO_co_design} for figures).
\fig{MIMO_workflow}{d} demonstrates the final stage of the workflow: for every DI-MIMO instance, after all update steps, the final spin configurations from each \runname\ are returned to the host, where the minimum-energy state is selected and the transmitted signal is reconstructed (details are listed in \ref{sec:DI_MIMO}, subsection 3).
This post-processing consists of a lightweight, deterministic energy evaluation over a small candidate set and adds negligible overhead to the total latency.
As a result, the end-to-end throughput and latency of the workflow are dominated by the PIMI solver execution on the FPGA, while host-side preprocessing, scheduling, and energy evaluation result in negligible overhead, which was profiled in prior Ising-Machine-based MIMO detection studies \cite{singh2025thesis} (see details in \ref{sec:MIMO_cost_breakdown}).

\begin{figure}[htbp]
    \centering
    \includegraphics[width=\textwidth]{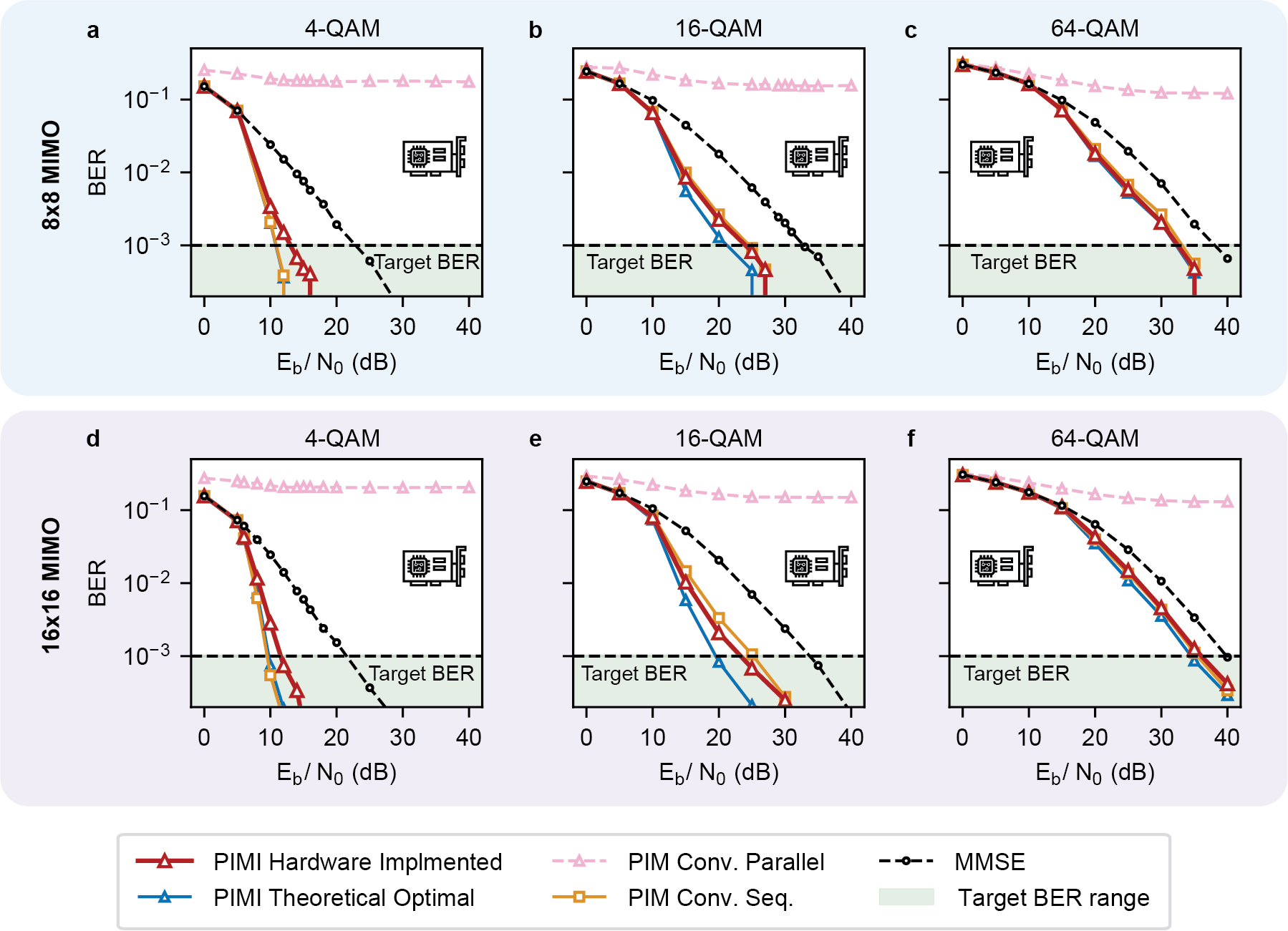}
    \caption{
    \textbf{Probabilistic Ising Machine with Inertia (PIMI) achieves low and robust bit error rate (BER) for Multiple-Input Multiple-Output (MIMO) detection across modulation orders and system sizes.}
    The BER is shown as a function of the signal-to-noise ratio per bit ($E_b/N_0$) for the PIMI solver, with comparisons to conventional probabilistic Ising machine (PIM) methods and a linear minimum mean-square error (MMSE) detector.
    Results are shown for multiple MIMO dimensions and Quadrature Amplitude Modulation (QAM) orders.
    \panel{a--c} BER performance for an $8\times8$ MIMO system using 4-QAM, 16-QAM, and 64-QAM modulation, respectively.
    \panel{d--f} Corresponding BER performance for a larger $16\times16$ MIMO system under the same modulation schemes.
    All results are obtained from FPGA implementations except for the theoretical optimal.
    Each BER point is estimated from $10,000$ independent MIMO instances.
    Error bars are omitted for clarity since the resulting statistical uncertainty is negligible relative to the line/marker width across the plotted range.
    }
    \label{fig:MIMO_BER_SNR}
\end{figure}

\begin{table}[h!]
\centering
\caption{
\textbf{Number of \stepname* for different solvers, modulation formats, and problem sizes in \fig{MIMO_BER_SNR}.}
For PIMI (Hardware Implemented), the number of \stepname* per \runname\ is determined by the trade-off between BER performance and real-time latency constraints.
For PIMI (Theoretical Optimal), additional \stepname* are used ($4\times$ the hardware-implemented value) to push the dynamics closer to convergence, providing an empirical near-convergence reference that is less constrained by the hardware latency target.
For conventional PIM with sequential updates, the numbers of steps are selected such that the total number of spin updates matches those of PIMI.
Details about parameter selection can be found in \ref{sec:slt_anneal_params}.
}

\label{tab:BER_SNR_steps}
\setlength{\tabcolsep}{6pt}
\small
\begin{tabular}{llccc}
\toprule
Problem size & Solver & 4\,-QAM & 16\,-QAM & 64\,-QAM \\
\cmidrule(lr){3-5}
 &  & \multicolumn{3}{c}{\emph{(All results use 32 \runname*)}} \\
\midrule
\multirow{4}{*}{8$\times$8} 
 & PIMI (Hardware Implemented)   & 16   & 32    & 64    \\
 & PIMI (Theoretical Optimal)    & 64   & 128   & 256   \\
 & PIM Conventional (Sequential) & 512  & 1024  & 2048  \\
 & PIM Conventional (Parallel)   & 16   & 32    & 64    \\
\midrule
\multirow{4}{*}{16$\times$16} 
 & PIMI (Hardware Implemented)   & 32   & 64    & 64    \\
 & PIMI (Theoretical Optimal)    & 128  & 256   & 256   \\
 & PIM Conventional (Sequential) & 2048 & 4096  & 4096  \\
 & PIM Conventional (Parallel)   & 32   & 64    & 64   \\
\bottomrule
\end{tabular}
\end{table}

\fig{MIMO_BER_SNR} reports the bit error rate (BER), defined as the fraction of incorrectly detected information bits (see \ref{sec:MIMO_def}, subsection 6 for details), as a function of the signal-to-noise ratio (SNR) per bit ($E_b/N_0$) for $8\times8$ and $16\times16$ ($N_t\times N_r$) uplink MIMO systems with 4-QAM (panels~a\&d), 16-QAM (panels~b\&e), and 64-QAM (panels~c\&f) constellations.
The x-axes show the SNR level, and the y-axes correspond to the resulting BER (lower is better).
Table~\ref{tab:BER_SNR_steps} lists the number of \stepname* per \runname\ used for FPGA implementation; the number of independent \runname* is fixed to 32 for all methods.
For the PIMI solvers implemented on FPGAs, these parameters are determined by the trade-off between detection performance and latency constraints, enabling a clear BER improvement over MMSE while remaining compatible with low-latency operation (see \ref{sec:slt_anneal_params} for details about parameter selection).
Across all tested cases, PIMI consistently solves the MIMO detection problem with significantly lower BER than MMSE.  
At the $10^{-3}$ target BER, which lies within the uncoded-BER range relevant to practical MIMO detection~\cite{Singh2022ri}, PIMI's hardware implementation reduces the required $E_b/N_0$ relative to MMSE by $\sim10$~dB for 4- and 16-QAM and by $\sim5$~dB for 64-QAM.
To provide a performance reference, we include a PIMI curve obtained by extending each \runname{} to $4\times$ the number of \stepname{s} used in the hardware implementation, yielding an empirical near-convergence reference for the given PIMI dynamics.
This curve indicates that an additional $\sim5$ dB gain over the hardware-implemented PIMI can be achieved at the target BER for 4- and 16-QAM, while the hardware results still remain close to this near-convergence reference overall.
In addition, the comparison between hardware-implemented PIMI and conventional PIM with sequential updates suggests that for 16- and 64-QAM, hardware PIMI achieves better accuracy than sequential PIM, although sequential PIM can still outperform hardware PIMI in the simpler 4-QAM case. 
Nevertheless, the near-convergence-reference PIMI curve consistently delivers better BER than sequential PIM across all modulations.
Lastly, conventional PIM with parallel updates fails to converge and does not yield usable results.

Beyond detection accuracy, we measured the batched solver throughput of the FPGA implementation as the number of parallel solver kernels was varied. 
To examine architectural scaling beyond the available FPGA resources, we additionally performed a synthesis-based comparison of the projected aggregate logic area of PIMI and conventional-PIM implementations at representative throughput levels defined in \ref{sec:ASIC_area_est}. \cite{Wolf2013, Clark2016} 
Under the stated logic-only projection assumptions, PIMI requires approximately $8\times$ and $7\times$ reduction in aggregate logic area than the conventional-PIM baseline for the $8\times8$ and $16\times16$ configurations, respectively. These values are synthesis-based architectural projections rather than measurements of a fabricated device or an end-to-end wireless receiver. The measurement procedure, scaling assumptions, and full throughput--area results are provided in \ref{sec:MIMO_throughput_measure} and \ref{sec:ASIC_area_est}.

\section{Discussion and Outlook}
\label{sec:discussion}

\subsection{Summary of the results}

By introducing an inertia-like update rule, we overcome the long-standing limitation of sequential updates in probabilistic Ising machines, enabling fully parallel spin updates while retaining the ability to reach the reference-state solutions used in our benchmarks. 
Our analysis and simulations show that the inertia term suppresses the period-2 behavior associated with synchronous updates, allowing the dynamics to reach low-energy configurations.
Compared with conventional probabilistic Ising machines (PIMs), the \underline{P}robabilistic \underline{I}sing \underline{M}achine with \underline{I}nertia (PIMI) achieves higher success probabilities and fewer clock cycles to solution (CCTS) on standard Ising benchmarks including the Max-Cut problem and the Sherrington-Kirkpatrick (SK) model. 
In both benchmarks, the speed advantage of PIMI increases with problem size, with the largest observed gain on a single instance reaching more than $150\times$ at $N=200$.
As a real-world proof of concept, we applied this co-design framework to implement a PIMI on a field-programmable gate array (FPGA) for uplink multiple-input multiple-output (MIMO) detection.
Our results show that PIMI achieves a lower bit-error rate (BER) than the standard linear minimum mean-square error (MMSE) detector. 
The fully parallel FPGA implementation of PIMI also delivers higher measured solver throughput than a conventional PIM with sequential updates.


\subsection{Relation to existing works}

PIMI builds on and complements a broad range of prior efforts aimed at enhancing the stability, parallelism, and hardware efficiency of Ising-machine dynamics.
Several studies have explored memory- or momentum-based modifications of Ising dynamics, in which update rules depend explicitly on spin histories.
In digital implementations, this dependence introduces per-spin state variables that require additional hardware resources for update and storage \cite{Caccioli2008, Okuyama2019, Brown2024}.
Another line of work considers continuous-variable formulations, including mean-field annealing \cite{Bilbro1988}, which replaces discrete spins with their average values, and the Free-Energy Machine \cite{Shen2025}, which combines variational free-energy minimization with gradient-based optimization. 
Complementary to these approaches, PIMI implements binary stochastic spins and relies only on the current spin configuration, providing an inertia-like stabilization mechanism that supports fully synchronous updates.
In Coherent Ising Machines (CIMs), spin states evolve according to coupled continuous-time dynamics, enabling intrinsically parallel updates through the underlying differential equations \cite{McMahon2016}.
When mapped to digital platforms, however, CIM-inspired approaches typically represent spins using high-precision continuous variables, leading to increased computational cost \cite{Singh2022cim_ri, Singh2022, Singh2022ri, Singh2024, Singh2025mmgap}.
Another p-bit-based approach, HOPbit \cite{Zeng2025}, addresses uplink MIMO detection using a higher-order unconstrained binary optimization (HUBO) formulation within a probabilistic Ising-machine framework. 
In contrast, PIMI preserves a pairwise Ising representation.

\subsection{Outlook}

At the algorithmic level, this work demonstrates that PIMI removes the fundamental bottleneck associated with sequential spin updates and enables fully parallel update dynamics.
In the current digital hardware realization, however, the effective level of parallelism is still constrained by available on-chip memory and logic resources, which limit the system size and throughput achievable on a single FPGA.
Our implementation further relies on pre-generated pseudo-random numbers to sustain high update throughput.
While suitable for digital prototyping, this approach introduces additional memory and data-movement overhead and does not exploit intrinsic physical sources of stochasticity.
Lastly, the \schedulename* and parameters used in this study were optimized empirically for the specific problem classes considered here and may require retuning when applied to different optimization tasks.


Looking beyond the present FPGA prototype, our results demonstrate that PIMI can map efficiently onto parallel hardware architectures.
Together with prior studies of Ising-machine-based MIMO detection \cite{Kim2019, Singh2022, Singh2022cim_ri, Singh2022ri, DeLunaDucoing2022, Sreedhara2023, Singh2024, Zeng2025, Seah2025, Kim2025, singh2025thesis}, these results motivate further investigation of PIMI-based detection across broader channel models, system configurations, and receiver operating conditions.

At a more physical level, PIMI naturally extends to mixed-signal and analog probabilistic computing platforms.
For instance, resistive crossbar architectures realize matrix–vector multiplications through Ohm’s and Kirchhoff’s laws.
The inherent parallelism of PIMI can be naturally mapped onto these architectures, enabling parallel computation of the local fields required for spin updates \cite{Sebastian2020, Cai2020, Jiang2023, Iftakher2025, He2025}.
Furthermore, in physical implementations of probabilistic Ising machines, intrinsic stochasticity arising from thermal or ionic fluctuations in stochastic magnetic tunnel junctions (MTJs) \cite{Camsari2017, Borders2019, Singh2024cmos, Daniel2024, Duffee2025, Iftakher2025}, 
low-barrier stochastic nanomagnets \cite{Debashis2018},
phase-transition nano-oscillators \cite{Dutta2021}, as well as shot noise in complementary metal--oxide--semiconductor (CMOS) circuits \cite{Patel2024}, can be directly harnessed to realize physical p-bits, rather than emulated stochastic units driven by pseudo-random number generators.
In such systems, the stochastic device itself acts as the probabilistic spin, with its switching behavior biased by the local field, enabling intrinsically parallel and energy-efficient updates.
Beyond eliminating digital random-number generation, the non-ideal and often non-Gaussian noise statistics of these devices may enhance sampling and help escape shallow local minima, motivating hardware realizations of probabilistic Ising machines that exploit physical stochasticity for massively parallel operation.
A recent hybrid implementation by Iftakher~et~al.~\cite{Iftakher2025} showed that these ingredients can be realized fully in analog hardware, combining memristor crossbars for coupling computation with stochastic magnetic tunnel junctions as physical p-bits.
While their prototype used the conventional sequential update rule, it nevertheless provides an important proof of concept for possible future physical implementations of PIMI using parallel analog and stochastic hardware.

\section{Conclusions}

This work addresses a central scalability bottleneck of probabilistic Ising machines: the reliance on sequential spin updates to obtain stable algorithmic convergence. 
By introducing an inertia-like update rule that suppresses the period-2 behavior associated with synchronous dynamics, a probabilistic Ising machine with inertia (PIMI) enables fully parallel spin updates while preserving robust convergence toward low-energy configurations.
On standard Ising benchmarks such as the Max-Cut problem and the Sherrington--Kirkpatrick (SK) model, this leads to higher success probabilities and substantially fewer clock cycles to solution than conventional probabilistic Ising machines, with the speed advantage growing superlinearly with the number of spins and the largest observed gain on a single $N=200$-spin instance reaching more than $150\times$.

Our hardware--software co-design framework enables efficient mapping of PIMI onto FPGA platforms.
As a proof of principle for applications where speed is critical, we apply this framework to uplink multiple-input multiple-output (MIMO) detection, achieving improved detection accuracy compared to standard linear detectors such as minimum mean-square error (MMSE) and higher solver throughput than the conventional sequential-update probabilistic Ising machine (PIM) baseline.
More broadly, PIMI's parallel dynamics align naturally with emerging analog and mixed-signal architectures, with the potential to enable fully parallel probabilistic computing machines with improved energy efficiency and throughput.

\vspace{1cm}
\noindent\textit{Note added:} During the preparation of our manuscript, we became aware of Ref.~\cite{Sajeeb2026}, which applies FPGA-based p-bit Ising solving with parallel tempering to MIMO detection. In contrast to that work, which retains sequential single-spin updates for connected spins within each replica (one copy of the Ising system evolved at a specific effective temperature), and approaches parallelization by mapping problems to larger sparse hardware graphs, our work introduces and applies to MIMO problems a complementary approach: modified Ising-solving dynamics that supports parallel spin updates directly.

\section*{Data and code availability}

The code and data underlying this study are publicly available via Zenodo: 
\url{https://doi.org/10.5281/zenodo.18215709}.

\section*{Author contributions}

P.L.M., K.J. and D.V. initiated the project. 
P.L.M. and R.Z. developed the algorithm. 
R.Z. performed the simulations, developed and tested the FPGA implementations, led the data analysis and wrote the manuscript with feedback from all authors. 
A.K.S. and K.J. contributed to the system evaluation and interpretation of results for MIMO detection. 
A.K.S. provided code to generate Ising problem instances for MIMO detection and to evaluate candidate solutions.
J.L. and F.O.W. contributed to the setup of the Max-Cut and SK evaluations, and to theoretical analysis of the algorithm.
A.K. performed initial investigations on Ising solver implementation on FPGAs, and assisted with FPGA system setup. 
D.V. advised on system evaluation. 
P.L.M. supervised the project.

\section*{Acknowledgements}
We gratefully acknowledge the Office of Naval Research Multidisciplinary University Research Initiative (ONR--MURI) funding under award number N000142312708. 
This material is also based upon work supported by the National Science Foundation under grants AST-2232457, OAC-2429485, CCF-1918549. We gratefully acknowledge a gift from Princeton NextG industrial affiliate program member Qualcomm Corporation.
USRA acknowledges the National Research Platform (NRP) for providing access to the Nautilus cluster and for their support of the computational infrastructure used in this work. 
A.K. acknowledges support of the USRA Feynman Academy Internship Program.
F.O.W. acknowledges support from the Eric and Wendy Schmidt AI in Science Postdoctoral Fellowship, a program of Schmidt Sciences, LLC. We would like to thank Maxwell~Anderson, Francesco~Caravelli, Kerem~Çamsarı and P.~Aaron~Lott for helpful discussions.

\bibliographystyle{mcmahonlab}
\bibliography{references}
\fi

\ifmainonly\else
    \appendix
    \renewcommand{\thefigure}{S\arabic{figure}}
\renewcommand{\thetable}{S\arabic{table}}
\renewcommand{\theequation}{S\arabic{equation}}
\renewcommand{\thesection}{Appendix Section~\arabic{section}}

\makeatletter
\gdef\appendixname{}
\@removefromreset{equation}{section}
\makeatother

\setcounter{figure}{0}
\setcounter{table}{0}
\setcounter{equation}{0}
\setcounter{section}{0}

\let\oldsection\section
\renewcommand{\section}[1]{%
  \oldsection{#1}%
  \addcontentsline{apc}{appendixsections}{\thesection\hspace{1em}#1}%
}

\listofappendixsections
\markboth{}{}
\newpage

\section{Summary of experimental procedures, data Preparation, and data analysis}

\subsection{Problem formulation and mappings}

\paragraph{Max-Cut.}
Max-Cut instances were generated as random unweighted graphs with 50\% edge density. Each graph was mapped to an Ising Hamiltonian using couplings $J_{ij} = -A_{ij}$ and zero biases, where $A$ is the adjacency matrix. 
The reference state of this Hamiltonian corresponded directly to a maximum cut, and was determined by executing an extended Breakout Local Search (BLS) run. 
Details of instance generation and reference state determination are shown in \ref{sec:mc_gen}.

\paragraph{Sherrington--Kirkpatrick Spin Glass.}
Sherrington--Kirkpatrick (SK) instances were generated as fully connected Ising models with i.i.d.\ couplings $J_{ij} \in \{\pm 1\}$ sampled with equal probability. 
These couplings were used directly without additional preprocessing, so the embedding into the PIPE framework was simply the Ising Hamiltonian with $h_i = 0$. The reference-state energy for each instance was determined by an extended simulated annealing run. 
More details can be found in \ref{sec:sk_gen}.

\paragraph{MIMO Detection.}
MIMO instances followed the DI-MIMO formulation: we drew i.i.d.\ Rayleigh fading channels, symbol vectors, and noise levels. 
These defined the real-valued model $\tilde{y} = \tilde{H}\tilde{x} + n$, from which the DI-MIMO mapping produced the corresponding Ising couplings $(J, h)$. 
Linear MMSE detection was used as the baseline for performance comparison.
Details about derivation and instance generation are provided in \ref{sec:MIMO_def} and \ref{sec:DI_MIMO}.

\subsection{PIMI update rule}
All experiments in this work used the fully parallel PIMI (Probabilistic Ising Machine with Inertia) update rule introduced in \secRef{PIMI}. 
As given in Eq.~(\ref{eq:PIMI}), all spins in $\mathbf{s} \in \{-1,+1\}^N$ were updated in parallel at time $t$ following the PIMI rule. 
Each update combined the instantaneous local field $I_i(t)$ scaled by the inverse-temperature schedule $\beta(t)$, a stabilizing self-alignment term $\xi\, s_i(t)$, and injected Gaussian noise whose amplitude was controlled by $\eta(t)$.

The schedules $\beta(t)$ and $\eta(t)$ controlled the relative strengths of the local-field and stochastic contributions during each trial, while the self-alignment term $\xi$ suppressed synchronous oscillations.
The schedule form was problem-dependent and is specified separately for each benchmark and implementation.
In practice, the shapes of the \schedulename\ were chosen empirically, and the associated schedule parameters were determined through software parameter search to achieve best results. 
A constant self-alignment strength was used for each problem, while its value differed across problem classes and sizes to account for their differing dynamic ranges.
All baseline experiments using traditional sequential PIM dynamics were tuned using the same procedure to ensure fair comparison. 

\subsection{Software-hardware co-design}
The PIMI kernels implemented on FPGAs were developed through a software--hardware co-design process.
A full software prototype of PIMI was implemented in Python and calibrated to match the exact quantization and look-up-table-based $\tanh$ function used in the FPGA kernels. 
This software model was employed to determine \schedulename*, numerical precision, as well as the appropriate number of \stepname* per run and the number of \runname* required for each problem class.

Once the solver configuration was established in software, the corresponding PIMI kernels were implemented on the FPGA using high-level synthesis (HLS) through AMD Vitis.
This co-design workflow ensured that the hardware kernels faithfully reproduced the intended PIMI dynamics while achieving high throughput and efficient resource utilization. 
This procedure was carried out independently for each problem class: the datapath bitwidth and $\tanh$ look-up-table resolution were swept for the Max-Cut and SK solvers (\ref{sec:mc_quant_tanh_map}) as well as the MIMO detection kernels
(\ref{sec:MIMO_co_design}), in each case selecting the smallest values that preserve
solution quality. 
Further details of the quantization scheme are given in \ref{sec:quant_sch}, and the corresponding FPGA resource usage is reported in \ref{sec:resource_v_bw_mc_sk}, \ref{sec:resource_v_tanh_mc_sk}, \ref{sec:resource_v_bw} and \ref{sec:resource_v_tanh}.

\subsection{FPGA implementation}

We implemented PIMI on AMD Alveo U55C FPGAs. 
PIMI solver kernels were instantiated on the FPGA board, with each kernel responsible for solving one Ising instance. 
Problem data---the coupling matrix $J$, local fields $h$, and initial spin states---were transferred from the host to the FPGA card over PCIe (Peripheral Component Interconnect Express) and stored in on-card HBM accessed through AXI-based 
HBM controllers. 
Once initialized, each kernel performed a prescribed number of PIMI update steps and returned either the full spin trajectories (for Max-Cut and SK-1) or only the final spin configurations (for MIMO detection). 
The host distributed problem instances across the available kernels, collected the returned solutions, and computed the corresponding Ising energies to identify the optimal spin configuration.

Each iteration of PIMI consisted of two computational steps. The FPGA first computed the effective fields $I_i(t) = \sum_j J_{ij} s_j(t) + h_i$ for all spins in parallel. 
It then applied the PIMI update rule to obtain the next spin vector, using the \schedulename\ $\beta(t)$ and $\eta(t)$ together with pre-generated Gaussian noise stored in on-board memory. 
For Max-Cut and SK-1, each kernel processed one \runname\ at a time, while for MIMO detection the same kernel structure evaluated multiple parallel \runname* of the same instance to improve throughput.
Implementation details for the different problem sets are described in \ref{sec:PIMI_MC_SK} and \ref{sec:PIMI_MIMO} for PIMI, and in \ref{sec:CPIM_MC_SK} and \ref{sec:CPIM_MIMO} for the conventional PIM, respectively.

\subsection{Evaluation metrics}

For Max-Cut and SK-1, we quantified performance using the clock cycles to solution (CCTS), defined as the expected number of FPGA clock cycles required to obtain a successful outcome with \(99.9\%\) probability. For each problem size \(N\), we measured the empirical single-\runname{} success probability as a function of the update-step budget \(T_{\mathrm{steps}}\), averaged this quantity over instances to obtain \(\bar{p}(T_{\mathrm{steps}};N)\), and estimated the required number of independent \runname{s} as
\begin{equation}
n_{\mathrm{\runname}}(T_{\mathrm{steps}};N)=\frac{\log(0.001)}{\log\!\bigl(1-\bar{p}(T_{\mathrm{steps}};N)\bigr)}.
\end{equation}
We then computed
\begin{equation}
\mathrm{CCTS}(T_{\mathrm{steps}};N)=n_{\mathrm{\runname}}(T_{\mathrm{steps}};N)\,T_{\mathrm{steps}}\,C_{\mathrm{step}}(N),
\end{equation}
where \(C_{\mathrm{step}}(N)\) is the FPGA clock-cycle cost per update step, and minimized this quantity over \(T_{\mathrm{steps}}\) to obtain the optimal CCTS. 
Details about CCTS estimation are provided in \ref{sec:CCTS_estimation}.

For MIMO detection, we evaluated performance using bit-error rate (BER) over 10,000 channel realizations at each SNR level. For each realization, multiple parallel \runname* were performed on the FPGA, and the solution vector was chosen as the configuration with the lowest Ising energy.
All baseline algorithms were evaluated under identical channel realizations and SNR settings.
We also reported the throughput for MIMO detection, defined as the number of completed \runname* produced per second. 
This metric reflected only the device-side running time and did not include the host-side energy computation used to select the final detected symbol vector.
This focus was motivated by prior work showing that the Ising solving time dominated the overall runtime in Ising-based MIMO detection \cite{singh2025thesis} (Figure 6.18).
See \ref{sec:MIMO_def} for BER estimation and \ref{sec:MIMO_throughput_measure} for details about MIMO throughput measurements.
\clearpage

\section{Max-Cut instance generation}
\label{sec:mc_gen}

\paragraph{Graph construction.}
Max–Cut benchmark instances are generated as Erd\H{o}s–R'enyi random graphs with N vertices and unweighted edges.
For each problem size N, each pair of vertices is connected independently with probability 0.5, and every edge is assigned unit weight $w_{ij}=1$, corresponding to the standard unweighted Max–Cut formulation.
We generate $100$ independent instances for each $N \in \{10,20,\ldots,150,200\}$.

\paragraph{Ising mapping.}
Each Max--Cut instance is converted into an Ising Hamiltonian of the form
\[
    \Ham(\mathbf{s}) = -\sum_{i<j} J_{ij} s_i s_j,
    \qquad s_i \in \{-1,+1\},
\]
using the standard transformation
\[
    J_{ij} = -A_{ij},
    \qquad h_i = 0,
\]
where $A$ is the adjacency matrix of the graph. Because all edge weights equal $1$, this produces an Ising model with
$J_{ij} \in \{0,-1\}$ and no local fields.  
Minimizing $E(s)$ is equivalent to computing a maximum cut.

\paragraph{Reference energy computation.}
To obtain reference energies, each instance is solved using a state-of-the-art Breakout Local Search (BLS) implementation. Following common practice, we run BLS with different restarts per instance with independent random seeds and retain the best solution found.
For problems with $N \leq 150$, BLS was run with 100 restarts, each consisting of 500 search cycles. For $N = 200$, BLS was run with 200 restarts, each consisting of 1000 search cycles.

\paragraph{Preprocessing}
During solver execution, the coupling matrix is normalized by \(2/\sqrt{N}\) so that the typical local-field magnitude remains approximately independent of problem size and remains compatible with fixed-point arithmetic.
For the Max-Cut mapping used here, the unnormalized coupling matrix is \(J^{\mathrm{raw}}=-A\), with \(J^{\mathrm{raw}}_{ij}\in\{0,-1\}\).
The corresponding unnormalized local field is
\[
I_i^{\mathrm{raw}}
=
\sum_{j\ne i}J^{\mathrm{raw}}_{ij}s_j.
\]
For a randomly oriented spin configuration, each nonzero contribution \(J^{\mathrm{raw}}_{ij}s_j\) takes the value \(+1\) or \(-1\) with equal probability and therefore has zero mean and unit variance.
If node \(i\) has degree \(d_i\), then
\[
\operatorname{Var}\!\left(I_i^{\mathrm{raw}}\right)
=
\sum_{j\in\mathcal{N}(i)}
\operatorname{Var}\!\left(J^{\mathrm{raw}}_{ij}s_j\right)
=
d_i.
\]
For the dense Erd\H{o}s--R\'enyi graphs considered here, whose edge probability is \(1/2\), \(d_i\approx N/2\), so the typical unnormalized local-field magnitude scales as
\[
\left|I_i^{\mathrm{raw}}\right|_{\mathrm{typ}}
\sim
\sqrt{\frac{N}{2}}.
\]
During solver execution, we apply the positive scaling factor \(2/\sqrt{N}\),
\[
I_i
=
\frac{2}{\sqrt{N}}I_i^{\mathrm{raw}},
\]
which gives
\[
\operatorname{Var}(I_i)
=
\frac{4d_i}{N}
\approx 2.
\]
The normalized local-field magnitude is therefore \(O(1)\) and approximately independent of problem size.
Because this scaling multiplies all couplings by the same positive constant, it does not change the minimizing spin configurations of the Max-Cut Hamiltonian.
No additional preprocessing or rescaling is applied.

\section{SK-1 instance generation}
\label{sec:sk_gen}

\paragraph{Coupling construction.}
SK-1 benchmark instances are fully connected Ising spin glasses with random $\pm 1$ couplings.  
For each system size $N$, we generate a symmetric coupling matrix $J \in \{-1,+1\}^{N\times N}$ by drawing each off-diagonal entry independently with equal probability and setting $J_{ii}=0$.  
We generate $100$ independent instances for each $N \in \{10,20,\ldots,150,200\}$.

\paragraph{Ising mapping.}
Each instance corresponds to the standard SK Hamiltonian
\[
    \Ham(\mathbf{s}) = -\sum_{i<j} J_{ij} s_i s_j, 
    \qquad s_i \in \{-1,+1\},
\]
with no local fields.  
This yields a fully dense Ising model with random $\pm 1$ interactions on every edge.

\paragraph{Reference energy computation.}
Reference energies are obtained using single-spin-flip simulated annealing.  
Starting from a random spin configuration, the algorithm performs Metropolis updates with a temperature schedule decreasing geometrically from $T_{\mathrm{init}}=5.0$ to $T_{\mathrm{final}}=0.01$ using a cooling factor $\alpha=0.995$.  
For instances with $N<70$, we used $10N$ proposed spin flips at each temperature. 
For larger instances, we increased the per-temperature budget to improve solution quality: for $N=70$,80,90, and 100, we use 10{,}000 proposed flips per temperature; for $N=110$,120,130,140, and 150, we use 20{,}000; and for $N=200$, we use 50{,}000. 
In all cases, 10 independent annealing runs are performed for each instance, and the lowest energy encountered across all runs is recorded as the reference energy.

\paragraph{Preprocessing.}
Coupling matrices are normalized by \(1/\sqrt{N}\) so that the typical scale of the local field remains approximately independent of problem size. For the SK model, the local field at spin \(i\) is
\[
I_i=\sum_{j\ne i} J_{ij}s_j,
\]
where \(s_j\in\{-1,+1\}\) and \(J_{ij}\in\{-1,+1\}\) are random couplings drawn symmetrically about zero. Each term \(J_{ij}s_j\) therefore has zero mean and unit variance, giving
\[
\mathrm{Var}(I_i)=\sum_{j\ne i}\mathrm{Var}(J_{ij}s_j)\sim N.
\]
Thus, the typical magnitude of the local field scales as
\[
|I_i|_{\mathrm{typ}} \sim \sqrt{N}.
\]
Normalizing \(J\) by \(1/\sqrt{N}\) compensates for this problem-size-dependent growth and keeps the local-field scale at \(O(1)\). No additional preprocessing is applied.

\clearpage
\section{Period-2 oscillations}
\label{sec:period-2}

\subsection{Period-2 oscillations under fully parallel updates}

Fully parallel updates in a conventional probabilistic Ising machine are prone to a collective instability in which the spins fail to converge \cite{Aadit2022}.
Each spin $s_i$ is updated to match the sign of its local field $I_i(t)=\sum_j J_{ij}s_j(t)$. Since all spins are updated simultaneously, a group of
mutually coupled spins can flip together in a single step; this collective flip reverses the fields acting on those spins, so that at the next step they are again misaligned and flip back. The dynamics therefore settle into a period-2 oscillation (Fig.~\ref{fig:period2_landscape}a), alternating between two configurations $\mathbf{s}^{A}$ and $\mathbf{s}^{B}$ that each recur every second step \cite{Goles1981}. For symmetric couplings, such period-2 cycles are in fact the only non-trivial attractor of synchronous threshold dynamics, as established by Goles and
co-workers using a decreasing energy functional \cite{Goles1985}. In the schematic landscape of Fig.~\ref{fig:period2_landscape}(a), a two-dimensional projection of the spin configuration space colored by the normalized Ising energy $\Ham/|\Ham_{\mathrm{ref}}|$, the conventional parallel machine fails to reach the low-energy basin and remains oscillating between two high-energy states(red points), whereas the conventional machine with sequential updates and PIMI relax into the basin and converge to the solution.

Panel (b) demonstrates such behavior in a full simulation of a Max-Cut instance at $N=150$ with 50\% edge density. The two axes are a two-dimensional projection of the sampled spin configurations, obtained from a principal-component analysis and transformed so that the solution is located at the center; the colour encodes the normalized Ising energy $\Ham$.
In the absence of noise, the conventional PIM with parallel updates converges onto the two period-2 configurations (large red circles) at high energy levels.
Under the noisy dynamics of the actual machine, the visited states
scatter around these configurations (small red points) and fail to reach the low-energy solution at the center.
The conventional machine with sequential updates (orange) and PIMI ($\xi=0.7$, blue), by contrast, converge to the low-energy solution at the centre.

\begin{figure}[htbp]
    \centering
    \includegraphics[width=\textwidth]{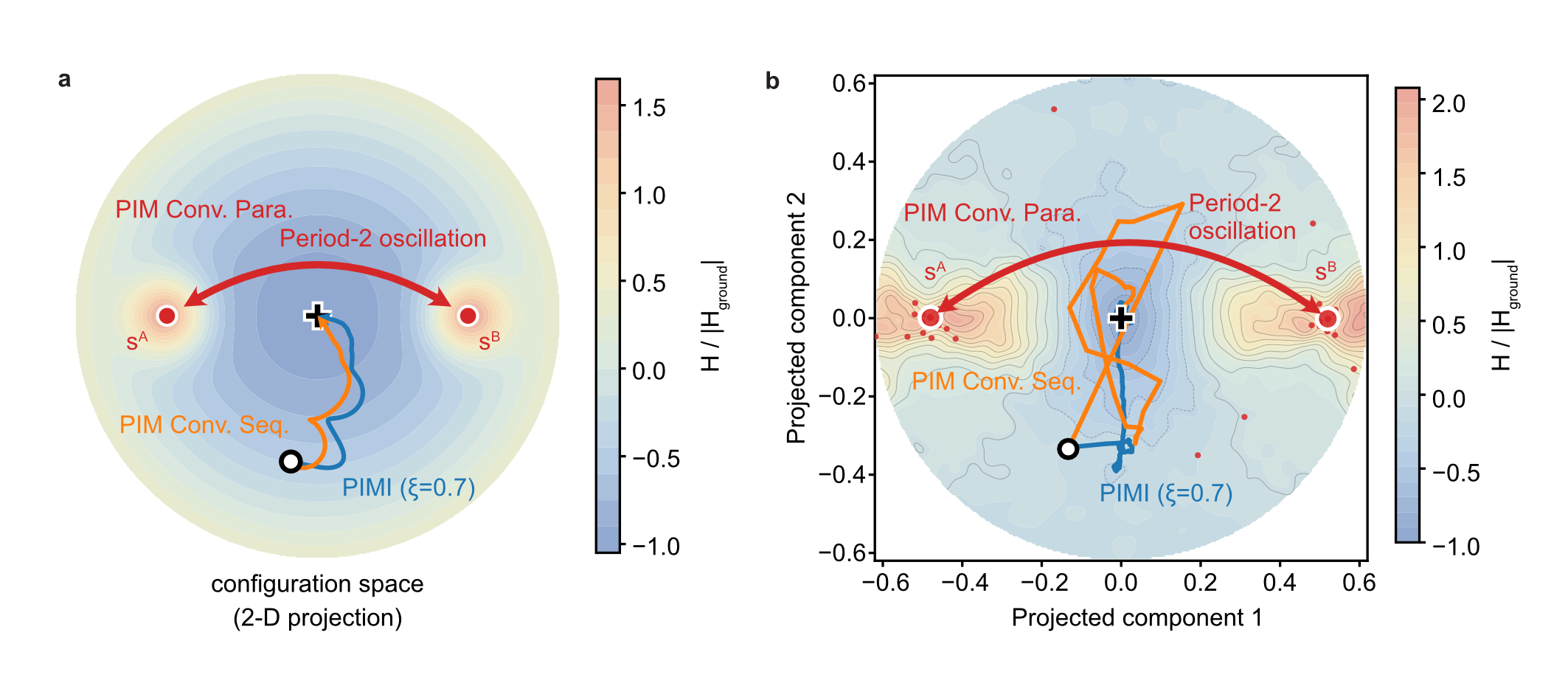}
    \caption{\textbf{Period-2 oscillation  between two high-energy configurations.} 
    Both panels employ two-dimensional projections of the spin configuration space, colored by the normalized Ising energy $\Ham/|\Ham_{\mathrm{ref}}|$ (equal to $-1$ at the reference energy state), so that the solution lies at the center of a low-energy (blue) basin. All trajectories start from a common initial state (open circle).
    \panel{a} Schematic landscape. Without inertia, conventional PIM with parallel updates (red) fails to reach the basin and instead oscillates between two high-energy configurations $\mathbf{s}^{A}$ and $\mathbf{s}^{B}$ (period-2 oscillation), whereas conventional PIM with sequential updates (orange) and PIMI (blue) descend into the basin and reach the solution ($+$).
    \panel{b} The same behavior for a Max-Cut instance at $N=150$ with $50\%$ edge density, projected onto its two leading principal components and centered on the solution. Without noise the parallel machine settles onto the period-2 pair $\mathbf{s}^{A}$ and $\mathbf{s}^{B}$ (large red points); under the noisy operating dynamics the visited states scatter around these configurations (small red points), while the sequential machine and PIMI converge to
    the solution at the center.}
    \label{fig:period2_landscape}
\end{figure}


\subsection{Lyapunov analysis of period-2 oscillations}
\label{sec:period-2-lyapunov}

\begin{figure}[htbp]
    \centering
    \includegraphics[width=1\textwidth]{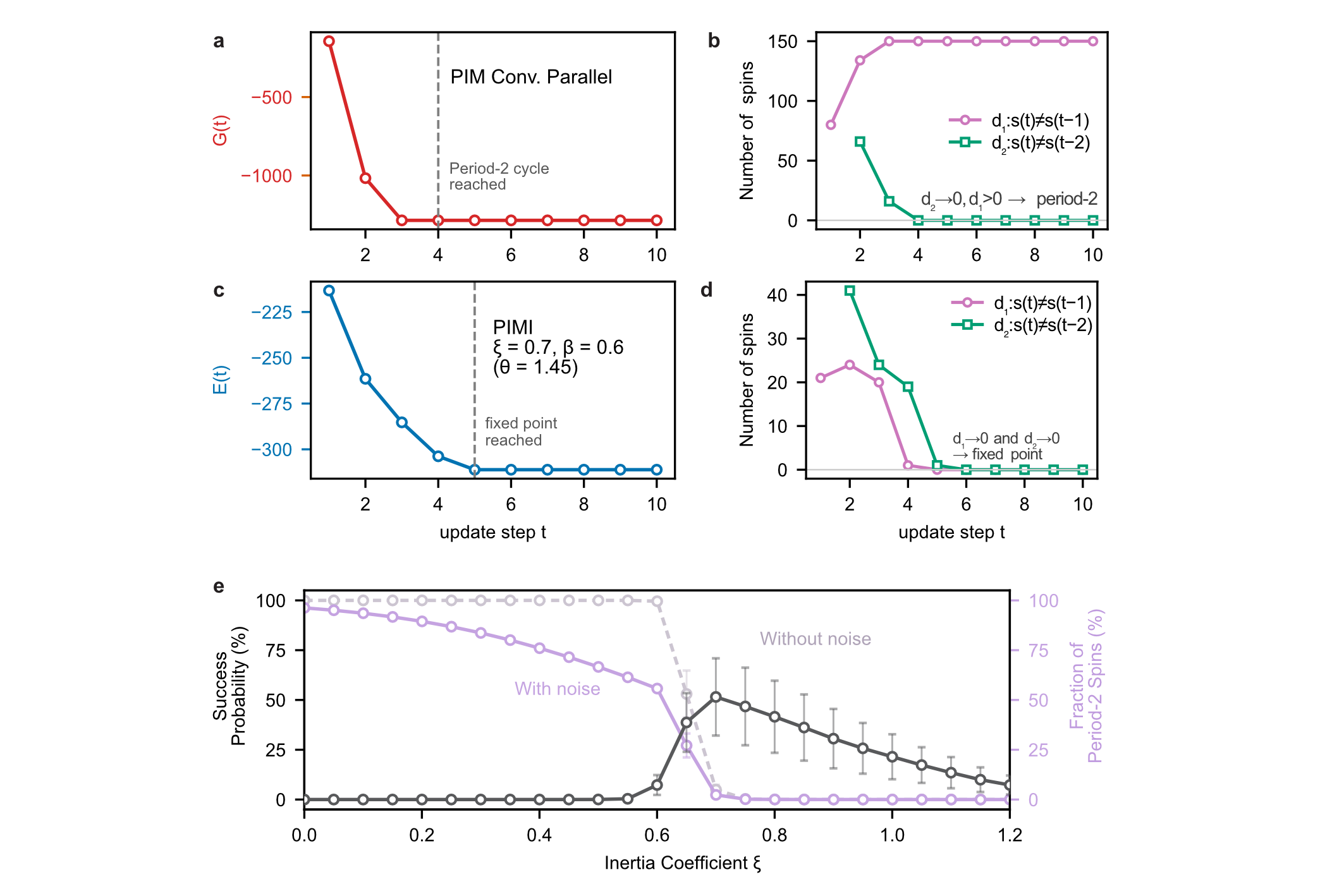}
    \caption{
    \textbf{Lyapunov analysis of period-2 oscillations.} 
    Results for a Max-Cut instance at $N=150$ with $50\%$ edge density.
    \panel{a,b} Conventional PIM with fully parallel updates. \panel{a} The two-state energy $G(t)=-\mathbf{s}(t-1)^{\mathsf T}J\,\mathbf{s}(t)$ decreases monotonically to a plateau; the dashed line marks where the period-2 cycle is reached.
    \panel{b} The number of spins that change between update steps, $d_1(t)$
    ($s_i(t)\neq s_i(t-1)$) and $d_2(t)$ ($s_i(t)\neq s_i(t-2)$): $d_1$ remains large while $d_2\to0$, the signature of a period-2 cycle. 
    \panel{c,d} PIMI with $\xi=0.7$, $\beta=0.6$ ($\theta=1.45$) and no noise, for which the effective coupling is $W=J+\theta I$ and the dynamics converge to a fixed point rather than a period-2 cycle.
    \panel{c} The matched energy $E(t)=-\mathbf{s}(t-1)^{\mathsf T}(J+\theta I)\,\mathbf{s}(t)$ decreases monotonically to a plateau (dashed line, fixed point reached). 
    \panel{d} Both $d_1(t)$ and $d_2(t)$ fall to zero, confirming convergence to a fixed point rather than a period-2 cycle. 
    \panel{e} Success probability (left axis, black) and the fraction of spins in period-2 oscillation (right axis) versus the inertia coefficient $\xi$, for the same Max-Cut instances ($N=150$ spins, $100$ instances, $256$ independent anneals each, $100N$ steps per anneal, fixed noise schedule). 
    For each anneal the period-2 fraction is measured over the final $1000$ steps as the fraction of spins satisfying $s_i(t-2)=s_i(t)\neq s_i(t-1)$, then averaged over anneals and instances. The \emph{with-noise} curve (purple) uses the operating schedule, whereas the \emph{without-noise} curve (gray, dashed) switches the noise off for those final $1000$ steps, allowing the dynamics to relax to their underlying deterministic local attractor; this curve therefore reports whether the attractor is a period-2 cycle or a fixed point. 
    The attractor dynamics change from a period-2 cycle to a fixed point near $\xi\approx0.65$ (gray), and the success probability --- the fraction of anneals reaching the reference target energy state --- increases rapidly at that point and is maximized at $\xi^\star\approx0.70$.
    }
    \label{fig:period2_lyapunov}
\end{figure}

\paragraph{Conventional parallel updates.}
Consider the deterministic (noise-free) parallel dynamics
$s_i(t+1)=\mathrm{sgn}\,I_i(t)$ with $\mathbf{I}(t)=J\mathbf{s}(t)$ and $J=J^{\mathsf T}$.
Since the Ising energy $\Ham$ is not monotone under fully parallel updates, it cannot characterize their convergence directly. Following Goles~et~al.~\cite{Goles1985}, we instead define a two-state energy that depends on two consecutive configurations,
\begin{equation}
G(t) = -\,\mathbf{s}(t-1)^{\mathsf T} J\, \mathbf{s}(t),
\label{eq:goles}
\end{equation}
which acts as a Lyapunov function for the parallel updates. Its change over one update step is
\begin{equation}
G(t+1)-G(t) = -\sum_i I_i(t)\big(s_i(t+1)-s_i(t-1)\big).
\label{eq:goles-monotone}
\end{equation}
When noise is absent, the update $s_i(t+1)=\mathrm{sgn}\,I_i(t)$ aligns each spin with its own field, so that $s_i(t+1)\,I_i(t)=|I_i(t)|\ge s_i(t-1)\,I_i(t)$. Each term $I_i(t)\big(s_i(t+1)-s_i(t-1)\big)$ is therefore non-negative, and hence
\begin{equation}
G(t+1)-G(t) \le 0 :
\label{eq:goles-nonincreasing}
\end{equation}
$G$ can never increase.

With only $2^N$ possible configurations, $G$ takes values in a discrete set, and since it is non-increasing and bounded below, it must settle onto a constant plateau after finitely many steps.
Once $G$ stops changing, Eq.~\eqref{eq:goles-monotone} requires every term in the sum to be zero, which forces
\begin{equation}
s_i(t+1)=s_i(t-1)
\label{eq:period2-condition}
\end{equation}
for every spin with $I_i(t)\neq 0$; spins with zero field are fixed by a tie-breaking convention, under which the period-2 conclusion still holds~\cite{Goles1981}.
The state therefore repeats every second step, and the attractors of the parallel dynamics are restricted to fixed points ($\mathbf{s}(t+1)=\mathbf{s}(t)$) and period-2 cycles ($\mathbf{s}(t+1)=\mathbf{s}(t-1)\neq\mathbf{s}(t)$, cycles of period three or greater are excluded) \cite{Goles1985}.

We illustrate this on a Max-Cut instance at $N=150$ with $50\%$ edge density
(Fig.~\ref{fig:period2_lyapunov}(a,b)): $G(t)$ decreases monotonically to a
plateau (a). To identify the resulting attractor, we count the number of spins that change between two update steps,
\begin{equation}
d_1(t)=\big|\{i:s_i(t)\ne s_i(t-1)\}\big|,
\qquad
d_2(t)=\big|\{i:s_i(t)\ne s_i(t-2)\}\big|,
\label{eq:hamming}
\end{equation}
where $d_1(t)$ compares the configuration with the preceding step and $d_2(t)$ with the step two earlier. 
Panel (b) shows that $d_1$ remaining large while $d_2\to 0$: the configuration keeps changing at every step, yet returns to itself every second step, which is consistent with the period-2 cycle prediction.

\paragraph{The PIMI update.}

The bare energy \eqref{eq:goles} does not apply for PIMI, whose update
$s_i(t+1)=\mathrm{sgn}\!\big[\tanh(\beta I_i(t))+\xi s_i(t)\big]$ (without noise) is not the sign of the $J$-field. 
Here we try to transform the PIMI update in a form that allows us to apply the same Lyapunov argument.
For $\xi\in(0,1)$,
\begin{equation}
    \mathrm{sgn}\!\big[\tanh(\beta I_i)+\xi s_i\big]
    = \mathrm{sgn}\!\big[I_i + \theta\, s_i\big],
    \qquad \theta = \frac{\mathrm{atanh}(\xi)}{\beta},
\label{eq:reduction}
\end{equation}
so PIMI's deterministic dynamics is itself a symmetric threshold network with
effective coupling $W'=J+\theta I$: the $\tanh$ and the inertia term are absorbed into the diagonal shift $\theta$. 

The same argument then applies to the \emph{matched} energy
\begin{equation}
    E(t) = -\,\mathbf{s}(t-1)^{\mathsf T}(J+\theta I)\,\mathbf{s}(t).
\label{eq:matched}
\end{equation}
By \eqref{eq:reduction}, the PIMI update is the sign of the effective field
$I'_i(t)=I_i(t)+\theta s_i(t)=\big[(J+\theta I)\,\mathbf{s}(t)\big]_i$, so that
$s_i(t+1)\,I'_i(t)=|I'_i(t)|\ge s_i(t-1)\,I'_i(t)$ and
\begin{equation}
E(t+1)-E(t) = -\sum_i I'_i(t)\big(s_i(t+1)-s_i(t-1)\big) \le 0.
\label{eq:matched-monotone}
\end{equation}
$E$ is therefore non-increasing, and PIMI likewise admits only fixed points or period-2 cycles. 
The diagonal shift removes these cycles through the inertia term, which rewards consecutive configurations for agreeing and so penalizes the reversals that sustain the oscillation. 
We show this by contradiction: we suppose a period-2 cycle does exist and ask what it would require. 
The assumption leads to a condition that becomes harder to
satisfy as $\theta$ grows.

Suppose two distinct configurations $\mathbf{s}^{A}$ and $\mathbf{s}^{B}$ formed a period-2 cycle, each obtained from the other in a single update,
$\mathbf{s}^{A}=\mathrm{sgn}(W\mathbf{s}^{B})$ and
$\mathbf{s}^{B}=\mathrm{sgn}(W\mathbf{s}^{A})$.
Since $\mathbf{s}^{B}=\mathrm{sgn}(W\mathbf{s}^{A})$, the state $\mathbf{s}^{B}$ maximizes the overlap $\mathbf{x}^{\mathsf T}(W\mathbf{s}^{A})$ over all $\pm1$ configurations $\mathbf{x}$, and in particular yields an overlap no smaller than $\mathbf{s}^{A}$ itself, so
$(\mathbf{s}^{B})^{\mathsf T}W\mathbf{s}^{A}\ge(\mathbf{s}^{A})^{\mathsf T}W\mathbf{s}^{A}$.
With $W=W^{\mathsf T}$ this reads
$(\mathbf{s}^{A})^{\mathsf T}W\mathbf{s}^{B}\ge(\mathbf{s}^{A})^{\mathsf T}W\mathbf{s}^{A}$, and the same reasoning applied to $\mathbf{s}^{A}=\mathrm{sgn}(W\mathbf{s}^{B})$
gives $(\mathbf{s}^{A})^{\mathsf T}W\mathbf{s}^{B}\ge(\mathbf{s}^{B})^{\mathsf T}W\mathbf{s}^{B}$.
This is the same property that made $E$ non-increasing in \eqref{eq:matched-monotone}.

Substituting these into the expansion
\begin{equation}
(\mathbf{s}^{A}-\mathbf{s}^{B})^{\mathsf T}W(\mathbf{s}^{A}-\mathbf{s}^{B})
= (\mathbf{s}^{A})^{\mathsf T}W\mathbf{s}^{A}
  +(\mathbf{s}^{B})^{\mathsf T}W\mathbf{s}^{B}
  -2\,(\mathbf{s}^{A})^{\mathsf T}W\mathbf{s}^{B}\le 0 .
\label{eq:no-2cycle}
\end{equation}
Writing $\boldsymbol{\delta}=\mathbf{s}^{A}-\mathbf{s}^{B}$ and separating the
diagonal term, \eqref{eq:no-2cycle} becomes
\begin{equation}
\boldsymbol{\delta}^{\mathsf T} J\, \boldsymbol{\delta}
\;\le\; -\,\theta\,\|\boldsymbol{\delta}\|^{2}.
\label{eq:2cycle-condition}
\end{equation}
A period-2 cycle can therefore persist only if the two configurations differ along a
direction for which $\boldsymbol{\delta}^{\mathsf T}J\boldsymbol{\delta}$ is sufficiently negative, and increasing $\theta$ steadily narrows the set of directions that qualify. 
In particular, once $\theta$ exceeds $|\lambda_{\min}(J)|$ no direction satisfies \eqref{eq:2cycle-condition} and every period-2 cycle is removed, which is the case in which $W$ is positive definite. 
This bound is conservative, since it must exclude all cycles simultaneously, and it is not reached at the inertia values used here; the suppression achieved in practice is characterized empirically below.

Here, we show that the fixed points mentioned above correspond to good solutions of the original Ising problem, rather than to configurations stabilized only by the inclusion of inertia.
A fixed point of the PIMI update satisfies $s_i=\mathrm{sgn}(I_i+\theta s_i)$, which is equivalent to
\begin{equation}
    s_i\,I_i \ge -\theta \quad\text{for all } i.
    \label{eq:fp-condition}
\end{equation}
This is weaker than the condition $s_i\,I_i\ge 0$ that defines an Ising local minimum: the inertia lets a spin stay fixed even when it is misaligned with its field by up to $\theta$, so at a given $\theta$ a PIMI fixed point is not necessarily a local minimum of the original Hamiltonian.
Because $\theta=\mathrm{atanh}(\xi)/\beta$ decreases as $\beta$ increases, this tolerance narrows over the course of the trial. The fixed points sustained by the deterministic dynamics therefore approach the single-spin-stability condition of the original Ising Hamiltonian as $\theta$ decreases.
In practice, we tune the hyperparameters --- the inertia $\xi$ together with the $\beta$ and noise schedules --- so that the PIMI dynamics reach the low-energy configurations of the original Ising Hamiltonian, as reflected in the success probabilities on the Max-Cut and SK-1 benchmarks in our main results.

Fig.~\ref{fig:period2_lyapunov}(c,d) illustrates this behavior for $\xi=0.7$ and $\beta=0.6$, corresponding to $\theta\approx1.45$: the matched energy $E(t)$ decreases monotonically to a plateau in panel c, while both $d_1$ and $d_2$ approach zero in panel d, showing that the dynamics converge to a fixed point rather than a period-2 cycle.
In our implementation, $\xi$ is treated as a tunable parameter and optimized for solution quality rather than fixed, since the operating machine also injects noise.

\subsection{PIMI implementation with stochasticity}
\label{sec:period-2-operating}


The reduction to $W=J+\theta I$ is used above as an analytical tool to understand the stability of the PIMI dynamics. 
This deterministic dynamics is not the regime in which PIMI operates --- our implementation injects time-dependent noise during operation. 
Fig.~\ref{fig:period2_lyapunov}(e) sweeps the inertia $\xi$ and reports, for each value, the fraction of spins in period-2 oscillation (the spins satisfying $s_i(t-2)=s_i(t)\neq s_i(t-1)$) measured over the final $1000$ update steps of each anneal and averaged over $100$ instances and $256$ anneals per instance. 
The \emph{with-noise} curve uses the full operating schedule of our PIMI implementation. The \emph{without-noise} curve switches the noise off for those final $1000$ steps, allowing the dynamics to relax to their underlying local attractors; by the analysis above this attractor is either a fixed point or a period-2 cycle, so this curve reports directly which of the two the machine has settled into.

The without-noise curve confirms the deterministic prediction: it remains near $100\%$ at small $\xi$ and drops toward zero around $\xi\approx0.65$. 
Below this value the underlying attractor is a period-2 cycle; above it, the attractor becomes a fixed point. 
The crossover occurs at a considerably smaller inertia than the bound above would require, indicating that destabilizing the period-2 cycles actually reached under the operating schedule does not demand the worst-case condition.
With noise, the period-2 fraction is lower at every $\xi$ and falls off smoothly: the noise itself partially alleviates the oscillation even at small inertia, and the same transition toward a non-oscillating regime occurs.
Noise maintains exploration and helps trajectories escape deterministic attractors, since reaching low-energy solutions also requires the noise to drive the system and explore the spin configuration space.

The success probability provides a complementary, performance-based view of how $\xi$ influences the system's behavior. 
It stays near zero for all $\xi$ below the collapse of the without-noise curve, then increases significantly as the underlying attractor becomes a fixed point, since only then can the machine settle onto a solution rather than being drawn back into oscillation. It peaks at an intermediate $\xi^\star\approx0.7$ and declines for larger $\xi$, since excessive inertia over-damps the dynamics, freezing spins and suppressing the noise-driven exploration needed to reach low-energy states.

\clearpage
\section{Effect of the noise distribution on PIMI performance}
\label{sec:noise_distribution}

PIMI uses additive Gaussian noise to maintain stochastic exploration during fully parallel updates. To understand the importance of this choice, consider the deterministic part of the update argument,
\begin{equation}
    a_i(t)
    =
    \tanh\!\bigl(\beta(t)I_i(t)\bigr)
    +\xi s_i(t),
\end{equation}
such that the complete PIMI update can be written as
\begin{equation}
    s_i(t+1)
    =
    \operatorname{sign}
    \left[
        a_i(t)+\eta(t)z_i(t)
    \right],
    \qquad
    z_i(t)\sim\mathcal{N}(0,1).
\end{equation}

The magnitude $|a_i(t)|$ represents the deterministic update margin: a noise realization must exceed this margin in the opposite direction to change the deterministic update outcome. For Gaussian noise, the probability of doing so is
\begin{equation}
    P_{\mathrm{G},i}(t)
    =
    \Phi\!\left(
        -\frac{|a_i(t)|}{\eta(t)}
    \right),
\end{equation}
where $\Phi$ is the cumulative distribution function of the standard normal distribution. Because a Gaussian distribution has unbounded support, this probability remains nonzero for every finite update margin. Gaussian noise can therefore occasionally perturb spins even when the deterministic local field and inertia term strongly favor the current update direction.

This behavior differs qualitatively from bounded uniform noise. For a uniform perturbation supported on $[-A(t),A(t)]$, the corresponding probability is
\begin{equation}
    P_{\mathrm{U},i}(t)
    =
    \begin{cases}
    \displaystyle
    \frac{1}{2}
    \left(
        1-\frac{|a_i(t)|}{A(t)}
    \right),
    & |a_i(t)|<A(t), \\[8pt]
    0,
    & |a_i(t)|\geq A(t).
    \end{cases}
\end{equation}
Once the deterministic margin exceeds the largest available uniform perturbation, the noise cannot alter the update outcome. Consequently, spins can become effectively locked by the deterministic local field and inertia term, even when the resulting configuration is not the desired low-energy solution. Gaussian noise retains rare larger perturbations that can unlock such spins and help the dynamics escape locally stable configurations.

\begin{figure}[htbp]
    \centering
    \includegraphics[width=0.85\textwidth]{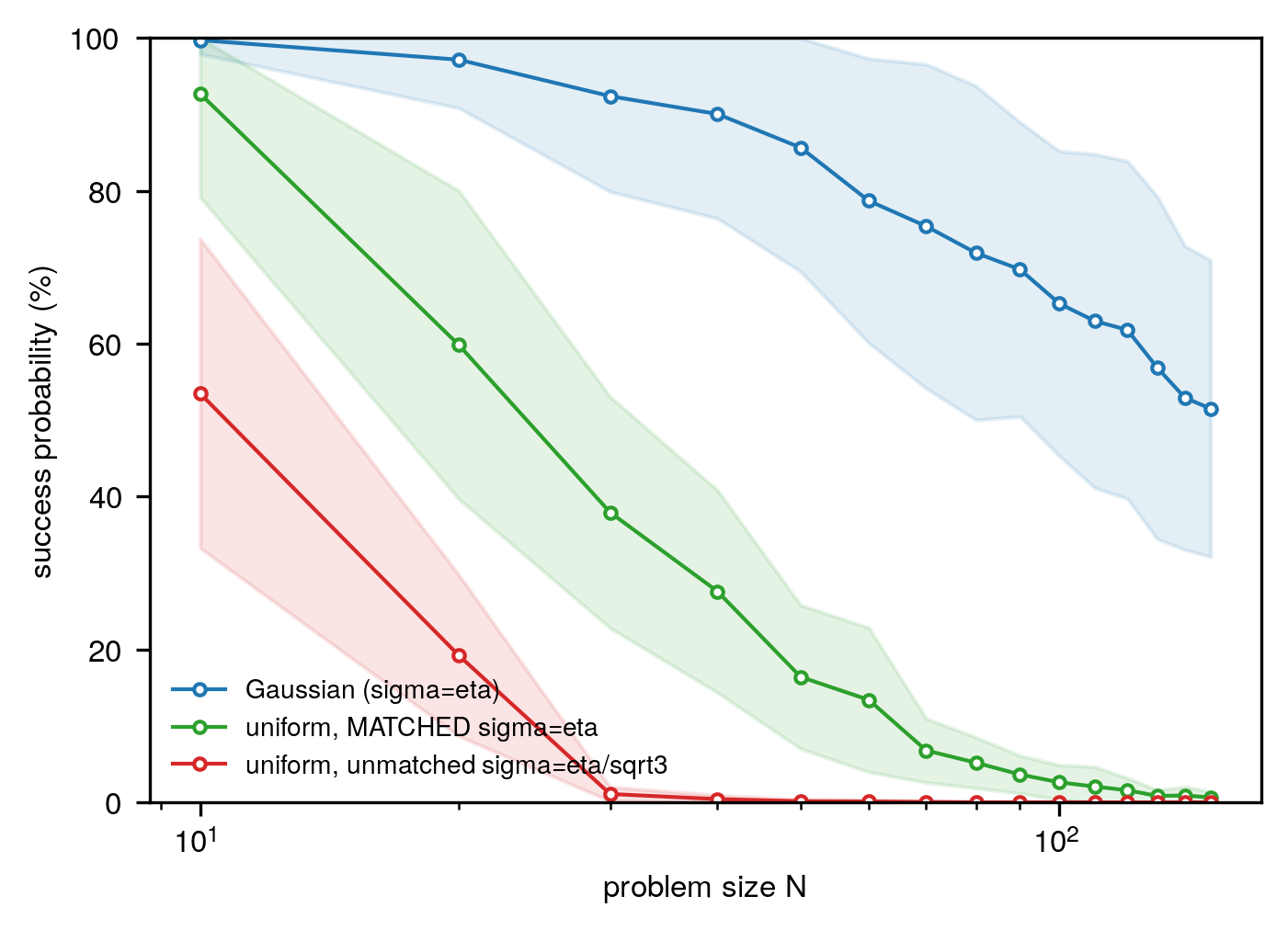}
    \caption{
    \textbf{Gaussian noise improves the success probability of PIMI.}
    Success probability is shown as a function of problem size $N$ for Gaussian noise with standard deviation $\sigma=\eta$, variance-matched uniform noise with $\sigma=\eta$, and unscaled uniform noise with $\sigma=\eta/\sqrt{3}$.
    All other PIMI parameters and experimental conditions are held fixed.
    Lines and markers show success probabilities averaged over the ensemble of problem instances, and shaded regions indicate instance-to-instance variation.
    }
    \label{fig:noise_distribution}
\end{figure}

Figure~\ref{fig:noise_distribution} shows that Gaussian noise produces substantially higher success probabilities than uniform noise, with the difference becoming increasingly pronounced at larger problem sizes. Directly replacing the Gaussian samples with $\eta(t)U(-1,1)$ also reduces the standard deviation of the injected noise to $\eta(t)/\sqrt{3}$. We therefore include a rescaled uniform distribution, $\sqrt{3}\eta(t)U(-1,1)$, having the same standard deviation as the Gaussian noise. Although this rescaling improves performance relative to the unscaled uniform case, it does not reproduce the success probability obtained with Gaussian noise.

The comparison therefore indicates that the advantage of Gaussian noise is associated with its distributional shape and, in particular, its unbounded tails. Rare larger perturbations allow the PIMI dynamics to overcome update margins that cannot be overcome by bounded uniform noise. This property becomes increasingly important as the problem size grows and the dynamics encounter a larger number of locally stable configurations. We therefore use Gaussian noise throughout the PIMI experiments and hardware implementations reported in this work. In the FPGA implementation, Gaussian samples are generated offline and stored in on-board memory, so this choice does not require an online Gaussian random-number generator in the solver datapath.

\clearpage

\section{Clock cycles per sweep explanation}
\label{sec:CC_per_sweep}

In a conventional PIM with sequential updates, spins must be updated one at a time.
For each spin update, the corresponding local field is computed as a vector–vector accumulation between one row of the coupling matrix and the current spin vector.
Because the spin configuration changes after each update, this accumulation must complete and the updated spin value must be written back before the local field for the next spin can be evaluated.
As a result, $N$ such vector–vector accumulations are performed serially within each \stepname.

Each vector–vector accumulation requires summing $N$ products, which is implemented using a reduction tree of depth $\log_2 N$.
Consequently, the total clock cycles per \stepname\ scale as $\mathcal{O}(N \log_2 N)$ for conventional sequential PIMs.

In PIMI, the spin vector is explicitly duplicated in memory so that each row of the coupling matrix can access the same spin values simultaneously.
This removes the update dependency present in sequential PIMs and enables all local fields to be computed in a single matrix–vector pass.
As a result, all $N^2$ scalar multiplications are performed in parallel, and the $N$ products contributing to each local field are accumulated using $N$ independent adder trees operating in parallel.
The latency of each accumulation is therefore determined by the depth of the adder tree, which scales as $\log_2 N$, rather than by the number of spins.
Consequently, the clock cycles per \stepname\ scale as $\mathcal{O}(\log_2 N)$.

\clearpage

\section{Estimation of optimal clock cycles to solution}
\label{sec:CCTS_estimation}


To quantify the computational effort required to reach a target success confidence, we combine the hardware clock-cycle cost of a single \runname{} with the number of independent \runname{s} required to obtain at least one successful outcome with that confidence. We define the resulting metric as the \emph{clock cycles to solution} (CCTS).

\subsection{Analysis parameters}

Unless otherwise stated, for each problem size \(N\), we evaluate \(M=100\) problem instances. For each instance, we perform \(B=256\) independent \runname{s}. Each \runname{} is simulated for up to \(s_{\max}(N)\) \stepname{s}, where \(s_{\max}(N)=100N\) for \(N\le 100\) and \(s_{\max}(N)=N^2\) for \(N>100\). Since \(J\in\mathbb{Z}^{N\times N}\) and \(s\in\{-1,+1\}^{N}\), every attainable energy \(H(s)\) is an integer, and any two distinct energy levels are separated by at least unity. A \runname{} is considered successful if it reaches an energy at or below \(\bar{H}_j+10^{-6}\), where \(\bar{H}_j\) is the reference-state energy of instance \(j\) (see \ref{sec:mc_gen} and ~\ref{sec:sk_gen} for how reference energies are generated), the small offset guarding against floating-point round-off.

\subsection{Instance-level success probabilities}

For each problem instance \(j\), we perform \(B\) independent \runname{s} and record the energy trajectory \(H_{i,j}(t)\) of \runname{} \(i\) as a function of \stepname{} \(t\), where \(i=1,\dots,B\). 
For a given number of update steps per trial, $T_{\mathrm{steps}}$, we evaluate the best energy reached within the trial, $\min_{t\le T_{\mathrm{steps}}} H_{i,j}(t).$
A \runname{} is declared successful if this quantity reaches the reference-state energy \(\bar{H}_j\) of instance \(j\), that is,
\begin{equation}
X_{i,j}(T_{\mathrm{steps}})=
\begin{cases}
1, & \text{if } \min_{t\le T_{\mathrm{steps}}} H_{i,j}(t)\le \bar{H}_j+10^{-6},\\
0, & \text{otherwise}.
\end{cases}
\label{eq:success_var}
\end{equation}
The empirical success probability for instance \(j\) is then
\begin{equation}
p_j(T_{\mathrm{steps}})=\frac{1}{B}\sum_{i=1}^{B} X_{i,j}(T_{\mathrm{steps}}).
\label{eq:success_prob_instance}
\end{equation}

\subsection{Number of \runname{s} required to reach the target success probability}

For each problem size \(N\), we next compute the mean success probability across the ensemble of \(M\) instances:
\begin{equation}
\bar{p}(T_{\mathrm{steps}};N)=\frac{1}{M}\sum_{j=1}^{M} p_j(T_{\mathrm{steps}}),
\label{eq:success_prob_mean}
\end{equation}
where \(p_j(T_{\mathrm{steps}})\) is the empirical single-\runname{} success probability for instance \(j\). From this mean success probability, we estimate the number of independent \runname{s} required to obtain at least one successful outcome with probability \(1-\epsilon\) as
\begin{equation}
n_{\mathrm{\runname}}(T_{\mathrm{steps}};N)=\frac{\log \epsilon}{\log\!\bigl(1-\bar{p}(T_{\mathrm{steps}};N)\bigr)}.
\label{eq:nrun}
\end{equation}
Here, \(\epsilon\) is the target failure probability, so that \(1-\epsilon\) is the desired overall probability of success. We use a target failure probability \(\epsilon=10^{-3}\), corresponding to a target success confidence of \(1-\epsilon=99.9\%\).

Following established time-to-solution benchmarking conventions, Eq.~\eqref{eq:nrun} is evaluated using the ensemble-averaged success probability $\bar{p}$ defined in Eq.~\eqref{eq:success_prob_mean}.
This provides a compact problem-size-level summary that is applied consistently to both solvers.
Because the transformation from success probability to the required number of trials is nonlinear, the resulting ensemble-level CCTS is not generally identical to the mean of CCTS values evaluated separately for individual instances.
We therefore complement the ensemble-level results with the paired instance-by-instance comparisons reported in \fig{max_cut}{e} and \fig{sk1}{e}.

\subsection{Clock cycles per \stepname}

For a problem of size $N$, the number of FPGA clock cycles required per \stepname, denoted $C_{\mathrm{step}}(N)$, is obtained from fitted, cycle-accurate timing models calibrated to the synthesized FPGA implementation. The estimated relationships are:
\begin{align}
    C_{\mathrm{seq}} \simeq N \log_2(N) + 8N + 4.67, \quad
    C_{\mathrm{para}} \simeq 1.1 \log_2(N) + 7, \quad
    C_{\mathrm{PIMI}} \simeq 1.1 \log_2(N) + 8.6.
    \label{eq:clock_cycles}
\end{align}
For PIMI and conventional PIMs, these fitted relations capture the dependence of $C_{\mathrm{step}}(N)$ on problem size and are used directly to convert \stepname\ counts into clock cycles in all timing analyses.
These cycle counts are obtained directly from the synthesized FPGA design and are used as fixed conversion factors in post-processing.

Note that in Eq.~\ref{eq:clock_cycles}, the sequential-update cost $C_{\mathrm{seq}}$ is reported per sweep; when computing CCTS for conventional PIMs, we therefore convert this quantity to a per-step cost by dividing by the problem size N, i.e., $C_{\mathrm{seq\;step}} = C_{\mathrm{seq\;sweep}}/N$.
The update cost for PIMI, $C_{\mathrm{PIMI}}$, was taken directly from Eq.~\ref{eq:clock_cycles}, since each PIMI update step corresponds to a full sweep in which all spins are updated simultaneously.

\subsection{Clock cycles to solution}

The hardware cost of a single \runname{} with \(T_{\mathrm{steps}}\) \stepname{s} is
\begin{equation}
C_{\mathrm{\runname}}(T_{\mathrm{steps}};N)=T_{\mathrm{steps}}\,C_{\mathrm{step}}(N),
\label{eq:crun}
\end{equation}
where \(C_{\mathrm{step}}(N)\) is the number of clock cycles required per \stepname{} for a problem of size \(N\). This quantity depends on the algorithm and is evaluated separately for the conventional PIM and the proposed PIMI.

Combining Eqs.~\eqref{eq:nrun} and \eqref{eq:crun}, the clock cycles to solution are given by
\begin{equation}
\mathrm{CCTS}(T_{\mathrm{steps}};N)
=
n_{\mathrm{\runname}}(T_{\mathrm{steps}};N)\,C_{\mathrm{\runname}}(T_{\mathrm{steps}};N)
=
n_{\mathrm{\runname}}(T_{\mathrm{steps}},N)\,T_{\mathrm{steps}}\,C_{\mathrm{step}}(N),
\label{eq:ccts_def}
\end{equation}
or equivalently,
\begin{equation}
\mathrm{CCTS}(T_{\mathrm{steps}};N)
=
\frac{T_{\mathrm{steps}}\,C_{\mathrm{step}}(N)\,\log \epsilon}
{\log\!\bigl(1-\bar{p}(T_{\mathrm{steps}};N)\bigr)}.
\label{eq:ccts_expanded}
\end{equation}
This metric estimates the clock-cycle cost required to obtain at least one successful solution with cumulative probability $1-\epsilon$, under the ensemble-averaged success-probability convention.

\subsection{Conversion to wall-clock time}

Clock cycles are converted to wall-clock time using the measured FPGA clock frequency $f_{\mathrm{clk}}$.
The wall-clock time to solution is given by
\begin{equation}
T_{\mathrm{wall}}
=
\frac{\mathrm{CCTS}}{f_{\mathrm{clk}}}.
\end{equation}

\subsection{Optimal step budget}

For a fixed problem size \(N\), the number of \stepname{s} allocated to each \runname{}, denoted by \(T_{\mathrm{steps}}\), introduces a tradeoff between per-\runname{} cost and per-\runname{} success probability. When \(T_{\mathrm{steps}}\) is too small, the probability of success in a single \runname{} remains low, so many repeated \runname{s} are required. When \(T_{\mathrm{steps}}\) is too large, the success probability may improve, but each \runname{} becomes more expensive in clock cycles. The optimal number of \stepname{s} per \runname{} is therefore the value of \(T_{\mathrm{steps}}\) that minimizes the total expected cost of obtaining a successful solution.

In practice, for each problem size \(N\), we evaluate \(\mathrm{CCTS}(T_{\mathrm{steps}};N)\) over all candidate values of \(T_{\mathrm{steps}}\),
\begin{equation}
T_{\mathrm{steps}} \in \{10,20,\dots,T_{\mathrm{steps},\max}(N)\},
\label{eq:s_budget_range}
\end{equation}
or over the sampled subset of \stepname{} counts used in the numerical analysis, and define the optimal \stepname{} count as
\begin{equation}
T_{\mathrm{steps}}^{*}(N)=\arg\min_{T_{\mathrm{steps}}\le T_{\mathrm{steps},\max}(N)} \mathrm{CCTS}(T_{\mathrm{steps}};N).
\label{eq:s_opt}
\end{equation}
This quantity identifies the most efficient \runname{} length for a given problem size and architecture.

\subsection{Optimal clock cycles to solution}

Once the optimal number of \stepname{s} per \runname{}, \(T_{\mathrm{steps}}^{*}(N)\), has been identified, the corresponding optimal clock cycles to solution are defined as
\begin{equation}
\mathrm{CCTS}^{*}(N)=\mathrm{CCTS}\!\left(T_{\mathrm{steps}}^{*}(N);N\right).
\label{eq:ccts_opt}
\end{equation}
Equivalently,
\begin{equation}
\mathrm{CCTS}^{*}(N)
=
\min_{T_{\mathrm{steps}}\le T_{\mathrm{steps},\max}(N)} \mathrm{CCTS}(T_{\mathrm{steps}};N).
\label{eq:ccts_opt_min}
\end{equation}

Thus, $\mathrm{CCTS}^{*}(N)$ represents the minimum CCTS required to reach the target success probability $1-\epsilon$ under the ensemble-averaged convention, after optimizing the number of \stepname{s} per \runname{}..

\subsection{Speedup metrics}

To compare architectures, we compute the ratio of the optimal CCTS values,
\begin{equation}
S_{\mathrm{CCTS}}(N)=
\frac{\mathrm{CCTS}^{*}_{\mathrm{PIM}}(N)}
{\mathrm{CCTS}^{*}_{\mathrm{PIMI}}(N)},
\label{eq:speedup_ccts}
\end{equation}
which measures the end-to-end hardware speedup achieved by PIMI after accounting for both convergence behavior and architecture-specific per \stepname{} execution cost.

\newpage
\subsection{Extended figures for the Max-cut problem}

\begin{figure}[htbp]
    \centering
    \includegraphics[width=1\textwidth]{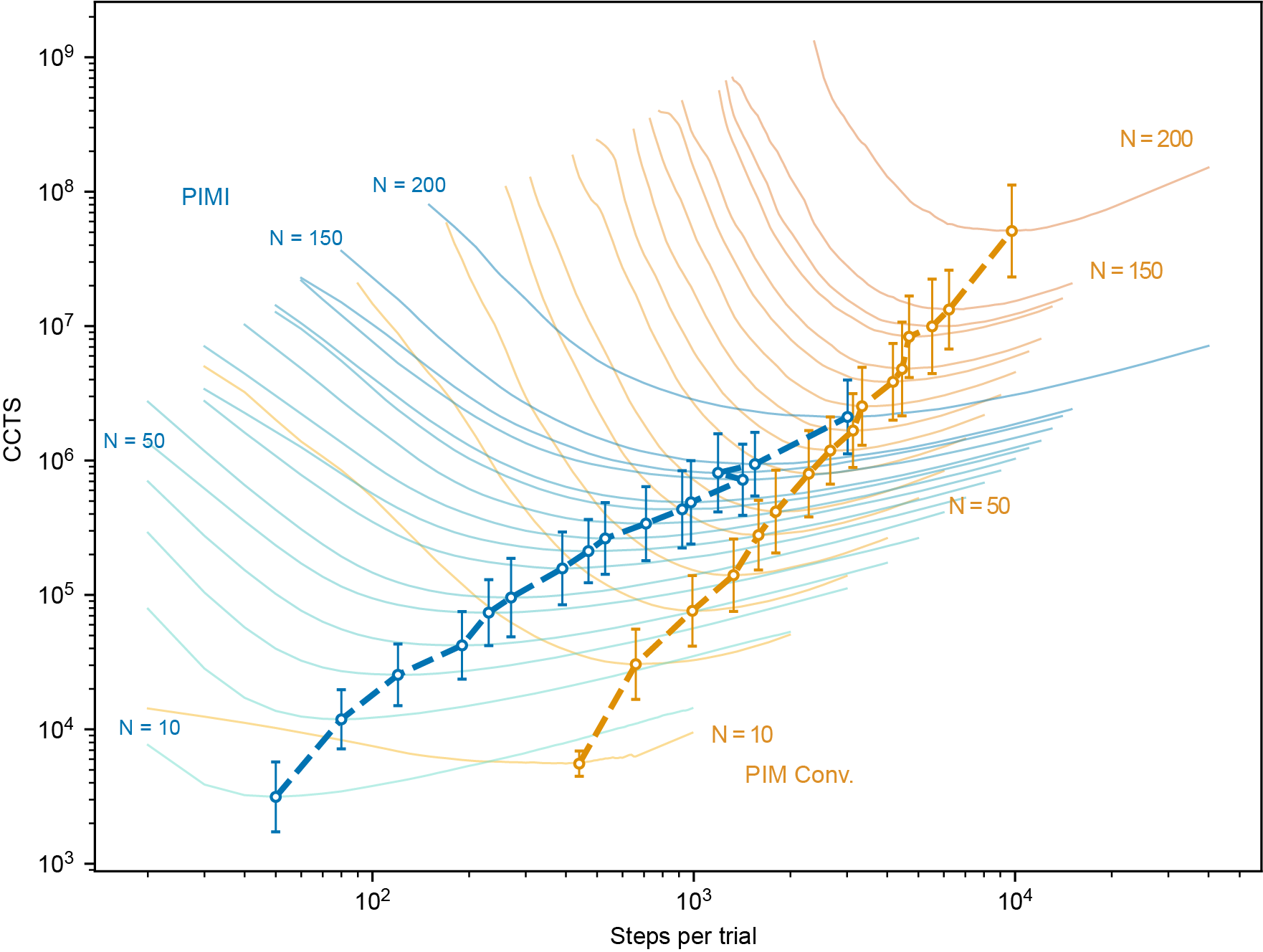}
    \caption{\textbf{Clock cycles to solution (CCTS) as a function of the step budget per trial for the max-cut problem for all problem sizes studied in this work ($N=10\ldots$, 150 and 200). }
    This is an extension to \fig{max_cut}{b}.
    Blue curves correspond to probabilistic Ising machine with inertia (PIMI), whereas orange curves correspond to the conventional probabilistic Ising machine with sequential updates (PIM Conv.).
    For each problem size, the results were collected over 100 instances. For every instance and number of steps per trial, the success probability was estimated from 256 independent trials, and these values were then averaged across instances.
    The mean success probability was converted into the number of independent trials required to reach the target cumulative success probability using Eq.~\ref{eq:nrun}; multiplying this by the steps per trial and the FPGA clock-cycle cost per step yielded the CCTS (Eq.~\ref{eq:ccts_expanded}).
    For each problem size, CCTS exhibits a minimum at an intermediate step budget, reflecting the trade-off between steps per trial and success probability. 
    Dashed lines connect the measured optima, and faint curves show the full landscapes for different $N$.}
    \label{fig:landscape_all}
\end{figure}

\begin{figure}[htbp]
    \centering
    \includegraphics[width=1\textwidth]{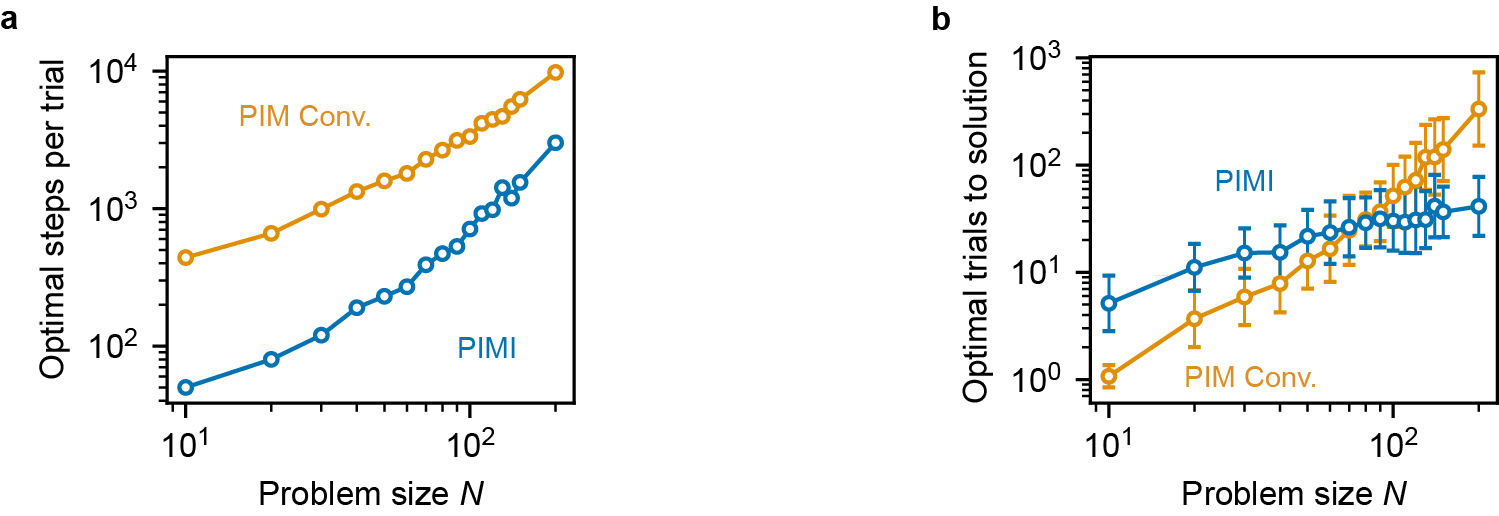}
    \caption{\textbf{Optimal steps per trial and corresponding trials to solution for the max-cut problem different problem sizes}.
    (a) Optimal number of steps per trial, $T_\mathrm{steps}^*(N)$, as a function of problem size $N$ for conventional PIM and PIMI (Eq.~\ref{eq:s_opt}). Across all tested sizes, PIMI attains its minimum CCTS at a smaller step budget than conventional PIM. 
    (b) Corresponding optimal number of trials to solution, evaluated at $T_\mathrm{steps}^*(N)$ for each problem size using Eq.~\ref{eq:nrun}. Error bars denote standard deviation across instances of the same problem size.}
    \label{fig:data_at_optimal}
\end{figure}

\begin{figure}[htbp]
    \centering
    \includegraphics[width=1\textwidth]{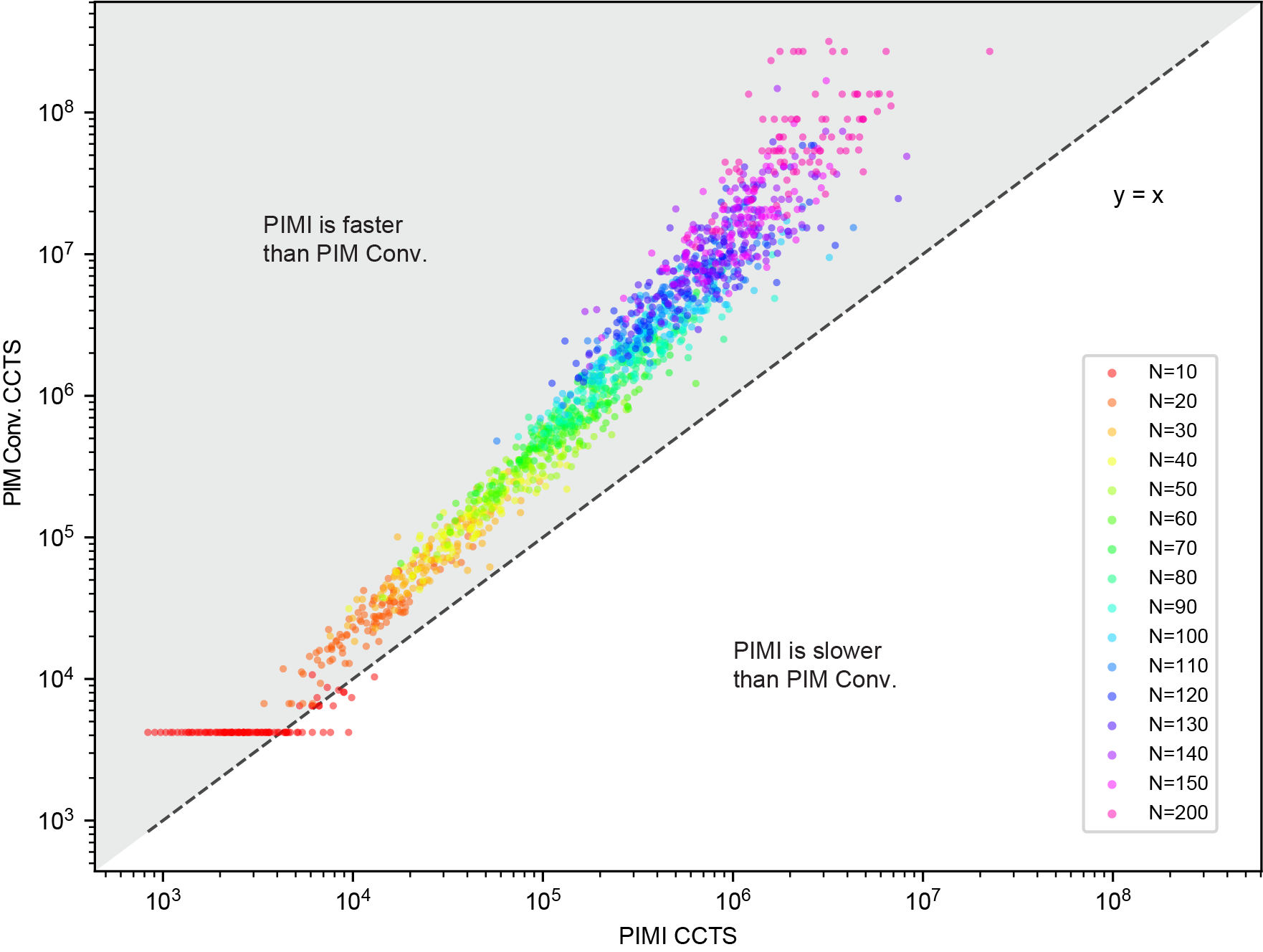}
    \caption{\textbf{Instance-by-instance comparison of the optimal CCTS obtained with PIMI and with the conventional PIM for the max-cut problem for all problem sizes studied in this work ($N=10\ldots$, 150 and 200), with 100 instances shown for each problem size. }
    This is an extension to \fig{max_cut}{e}.
    Each point represents one instance and is colored by problem size. Points above the dashed \(y=x\) line correspond to instances for which PIMI required fewer clock cycles to solution than the conventional PIM. }
    \label{fig:scatter_full}
\end{figure}

\newpage

\subsection{Extended figures for the SK-1 model}
\begin{figure}[htbp]
    \centering
    \includegraphics[width=1\textwidth]{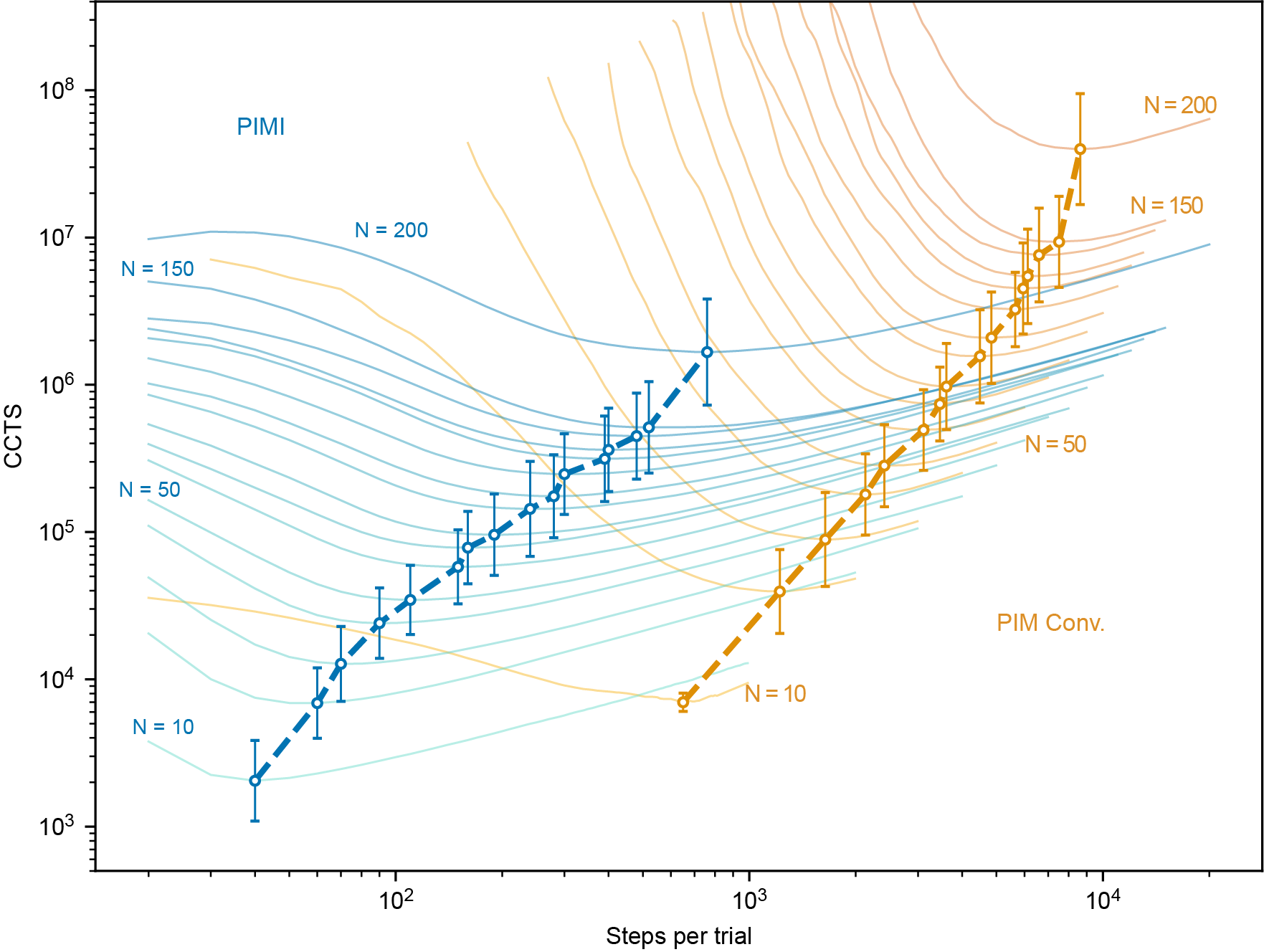}
    \caption{\textbf{Clock cycles to solution (CCTS) as a function of the step budget per trial for the sk-1 problem for all problem sizes studied in this work ($N=10\ldots$, 150 and 200). }
    This is an extension to \fig{sk1}{b}. Blue curves correspond to probabilistic Ising machine with inertia (PIMI), whereas orange curves correspond to the conventional probabilistic Ising machine with sequential updates (PIM Conv.).
    For each problem size, the results were collected over 100 instances. For every instance and number of steps per trial, the success probability was estimated from 256 independent trials, and these values were then averaged across instances.
    The mean success probability was converted into the number of independent trials required to reach the target cumulative success probability using Eq.~\ref{eq:nrun}; multiplying this by the steps per trial and the FPGA clock-cycle cost per step yielded the CCTS (Eq.~\ref{eq:ccts_expanded}).
    For each problem size, CCTS exhibits a minimum at an intermediate step budget, reflecting the trade-off between steps per trial and success probability. 
    Dashed lines connect the measured optima, and faint curves show the full landscapes for different $N$.}
    \label{fig:landscape_all_sk}
\end{figure}

\begin{figure}[htbp]
    \centering
    \includegraphics[width=1\textwidth]{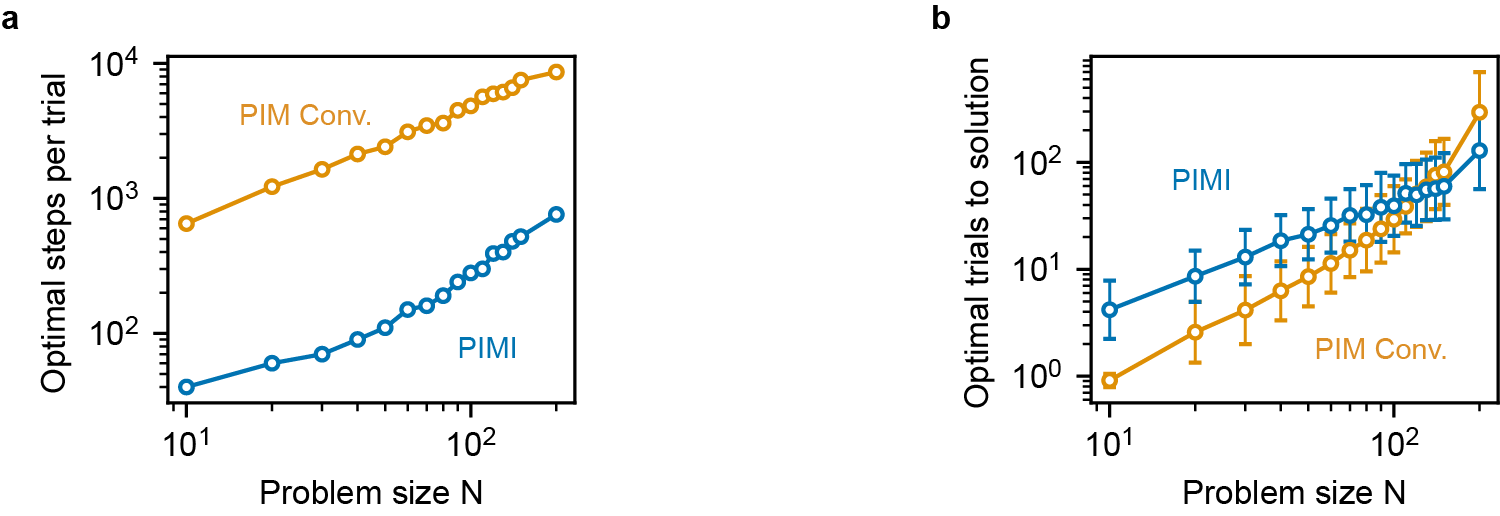}
    \caption{\textbf{Optimal steps per trial and corresponding trials to solution for the sk-1 model with different problem sizes}.
    (a) Optimal number of steps per trial, $T_\mathrm{steps}^*(N)$, as a function of problem size $N$ for conventional PIM and PIMI (Eq.~\ref{eq:s_opt}). Across all tested sizes, PIMI attains its minimum CCTS at a smaller step budget than conventional PIM. 
    (b) Corresponding optimal number of trials to solution, evaluated at $T_\mathrm{steps}^*(N)$ for each problem size using Eq.~\ref{eq:nrun}. Error bars denote standard deviation across instances of the same problem size.}
    \label{fig:data_at_optimal_sk}
\end{figure}

\begin{figure}[htbp]
    \centering
    \includegraphics[width=1\textwidth]{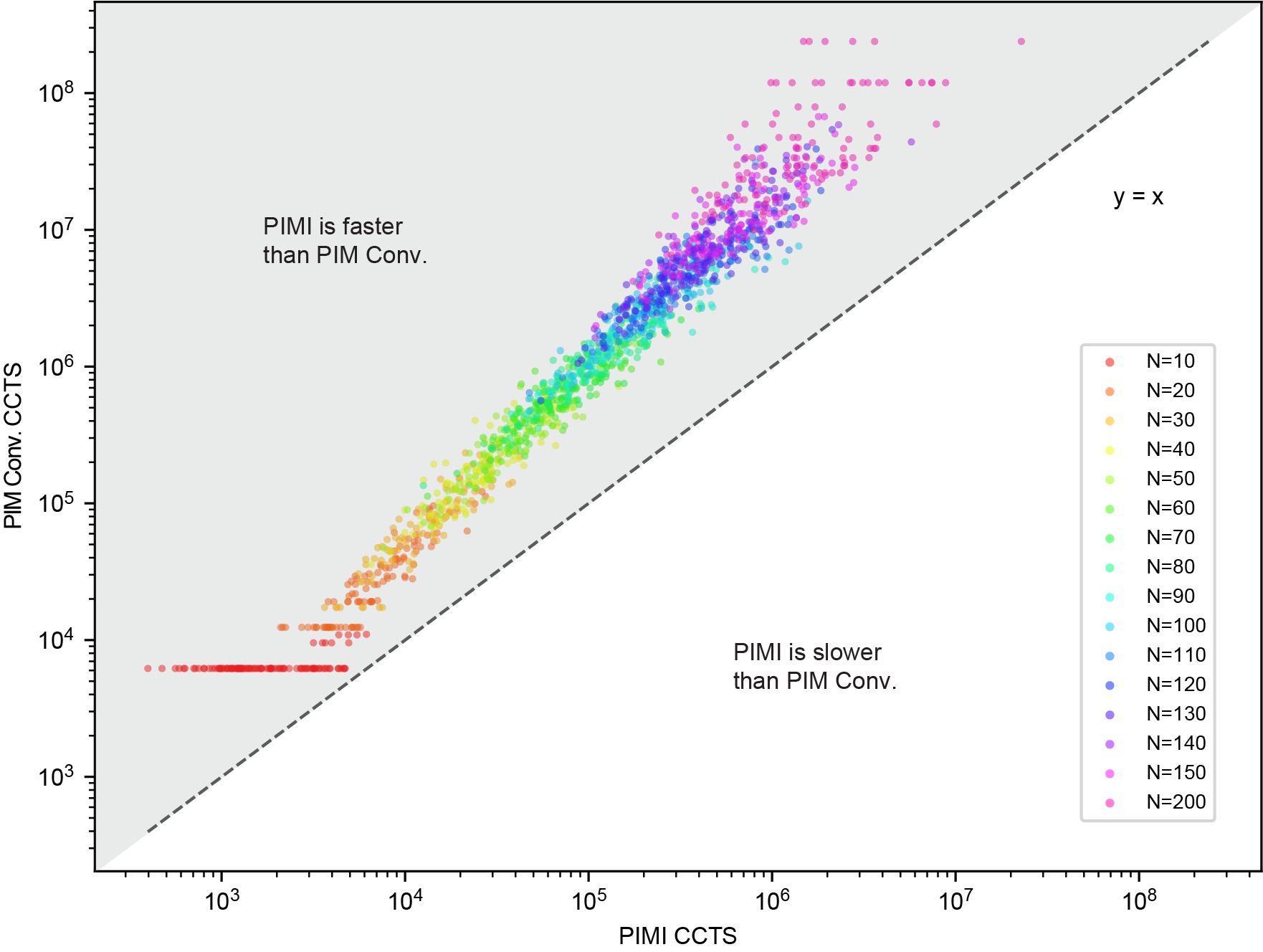}
    \caption{\textbf{Instance-by-instance comparison of the optimal CCTS obtained with PIMI and with the conventional PIM for the sk-1 model for all problem sizes studied in this work ($N=10\ldots$, 150 and 200), with 100 instances shown for each problem size. }
    This is an extension to \fig{sk1}{e}.
    Each point represents one instance and is colored by problem size. Points above the dashed \(y=x\) line correspond to instances for which PIMI required fewer clock cycles to solution than the conventional PIM. }
    \label{fig:scatter_full_sk}
\end{figure}

\clearpage

\section{Quantization scheme}
\label{sec:quant_sch}

\subsection{Numerical representations}

All experiments use fixed-point arithmetic to ensure consistency between software simulation and FPGA execution.  
The Python simulator and FPGA implementation are explicitly calibrated to use identical word lengths, scaling, truncation, and saturation behavior, so that numerical effects observed in simulation faithfully reflect hardware behavior.

\paragraph{Calibration between software and hardware.}
In software, fixed-point arithmetic is emulated using a uniform quantize--dequantize model with truncation toward zero and saturation at the representable range.  
On the FPGA, the same behavior is implemented natively using Vitis HLS fixed-point types ($\texttt{ap\_fixed}$ and $\texttt{ap\_int}$) with truncation and saturation enabled.
The two representations are matched exactly in total bit width, number of integer bits, fractional precision, rounding mode, and overflow handling.

\paragraph{Max--Cut and SK-1 benchmarks.}
For the Max--Cut and SK-1 problem classes, the Ising coupling matrix $J$ and all intermediate quantities in the solver datapath are represented using a low-precision fixed-point format.
Specifically, all such values use
\[
    \texttt{ap\_fixed<8,3,AP\_TRN\_ZERO,AP\_SAT>},
\]
corresponding to a signed 8-bit fixed-point representation with 3 integer bits (including sign) and truncation toward zero with saturation on overflow.
Although this normalization is defined at the algorithmic level, in the hardware implementation it is applied as a post-multiplication scaling of the accumulated interaction terms, avoiding additional quantization of the stored couplings and preserving fixed-point dynamic range.

Spin states are stored separately using a compact integer representation,
\[
    \texttt{ap\_int<2>},
\]
reflecting their binary nature.

\paragraph{MIMO detection benchmarks.}
For the MIMO detection experiments, higher numerical precision is required to accommodate denser couplings and accumulated interaction terms.
Accordingly, the Ising coupling matrix $J$, local fields $h$, noise samples, and all intermediate arithmetic values are represented using
\[
    \texttt{ap\_fixed<16,4,AP\_TRN\_ZERO,AP\_SAT>},
\]
i.e., a signed 16-bit fixed-point format with 4 integer bits (including sign) and 12 fractional bits, again using truncation toward zero and saturation on overflow.
Spin states are stored using the same \texttt{ap\_int<2>} representation as in the Max--Cut and SK-1 benchmarks.

\paragraph{Consistency of numerical behavior.}
All arithmetic operations—including accumulation, stochastic updates, and noise injection—use the same fixed-point format.
This calibrated quantization scheme ensures that differences between software and FPGA results arise from algorithmic or architectural effects, rather than mismatches in numerical representation.

For MIMO detection, synthesized results for resource usage with respect to bitwidth can be found in \ref{sec:resource_v_bw}, Tables~\ref{tab:bitwidth_8x8}~\&~\ref{tab:bitwidth_16x16}.

\subsection{Tanh lookup-table approximation}

To reduce computational cost and ensure deterministic fixed-point behavior, the $\tanh(\cdot)$ activation in the Ising update rule is realized as a small lookup table (LUT), used identically in software simulation and on the FPGA. The number of levels $L$ is fixed at design time, so the table entries and comparison thresholds are precomputed offline and the LUT incurs no runtime cost that scales with $L$. The LUT stores $L$ evenly spaced output values in $[-1,1]$; for $L=4$ these are $\{-1,-\tfrac{1}{3},\tfrac{1}{3},1\}$. For an input $x$, the quantized output is obtained by rounding $\tanh(x)$ to the nearest stored value.

The input thresholds separating adjacent output levels are placed at $x=\tanh^{-1}\!\big(\tfrac{1}{2}(\ell_k+\ell_{k+1})\big)$, so that each transition occurs where $\tanh$ reaches the midpoint between the two neighboring levels. The thresholds are therefore non-uniformly spaced, denser where $\tanh$ is steep and sparser where it saturates. 
For $L=4$, the transitions lie at $x=\{-0.805,\,0,\,0.805\}$, and the saturation of $\tanh$ at large $|x|$ is reproduced by clamping to the endpoint levels $\pm 1$. The mapping thus follows the shape of $\tanh$ while reducing to a fixed $L$-level LUT with $L-1$ precomputed thresholds.
Details about the implementation can be found in Algorithm~\ref{alg:tanh_lut}.

For MIMO detection, resource usage versus tanh levels is reported in \ref{sec:resource_v_tanh}, Tables~\ref{tab:tanh_8x8}~\&~\ref{tab:tanh_16x16}.

\begin{algorithm}[h!]
\caption{$L$-level $\tanh$ approximation}
\label{alg:tanh_lut}
\KwData{Input value $x$; number of levels $L$ (fixed at design time)}
\KwResult{Quantized output $\tilde{x}$}

\tcp{Precomputed once, offline}
Define $L$ uniformly spaced output levels $\ell_0,\dots,\ell_{L-1}$ in $[-1,1]$ \tcp*[r]{$L=4$: $\ell=\{-1,-\tfrac13,\tfrac13,1\}$}
Define $L-1$ input thresholds $t_k = \tanh^{-1}\!\big(\tfrac{1}{2}(\ell_k+\ell_{k+1})\big)$\;
\tcp{$L=4$: \ $x<-0.805\to-1$, \ $[-0.805,0)\to-\tfrac13$, \ $[0,0.805)\to\tfrac13$, \ $x\ge0.805\to1$}

\tcp{Per-sample lookup: threshold comparison only}
\eIf{$x < t_0$}{
  $\tilde{x} \gets \ell_0$\;
}{
  \eIf{$x \ge t_{L-2}$}{
    $\tilde{x} \gets \ell_{L-1}$\;
  }{
    Find interval $k$ such that $x \in [t_{k-1}, t_k)$\;
    $\tilde{x} \gets \ell_k$\;
  }
}
\Return $\tilde{x}$\;
\end{algorithm}

\clearpage

\section{Hardware-software numerical equivalence and scalability limits}
\label{sec:hw_equiv}

The Python-based emulation is configured to match the hardware at the numerical level, including fixed-point precision, \(\tanh\) look-up table discretization, noise statistics, and update schedules.
All arithmetic operations follow the same quantization and saturation rules as those used in the synthesized FPGA design, ensuring consistent numerical dynamics between software and hardware executions.

For the Max-Cut and Sherrington--Kirkpatrick-1 (SK-1) benchmarks, PIMI employs a fully unrolled update architecture in which the local-field computation is carried out as a ternary popcount and requires no multipliers.
While this design enables fast convergence, it still incurs substantial logic and routing consumption that grows with problem size.
Using a Xilinx Alveo U55C FPGA, we were able to map the fully unrolled PIMI architecture for dense coupling matrices up to problem sizes of \(N = 64\), beyond which on-chip routing and logic constraints become prohibitive.
Experimental results for problem sizes that can be mapped to hardware (\(N \le 60\)) are therefore obtained from direct execution on an FPGA.
For larger problem sizes (\(N > 70\)), results are obtained using a Python-based software emulation that is numerically calibrated to match the FPGA behavior.

For MIMO, the matrix-vector multiplication is partially unrolled and therefore all problem sizes are implementable on FPGAs.

These limitations are architectural rather than algorithmic and do not affect the validity of the PIMI dynamics.
Hardware-calibrated software emulation consequently enables systematic exploration of larger problem sizes while preserving numerically accurate and hardware-faithful behavior.

\clearpage
\section{Max-cut performance at different quantization and tanh LUT levels}
\label{sec:mc_quant_tanh_map}

\begin{figure}[htbp]
    \centering
    \includegraphics[width=0.8\textwidth]{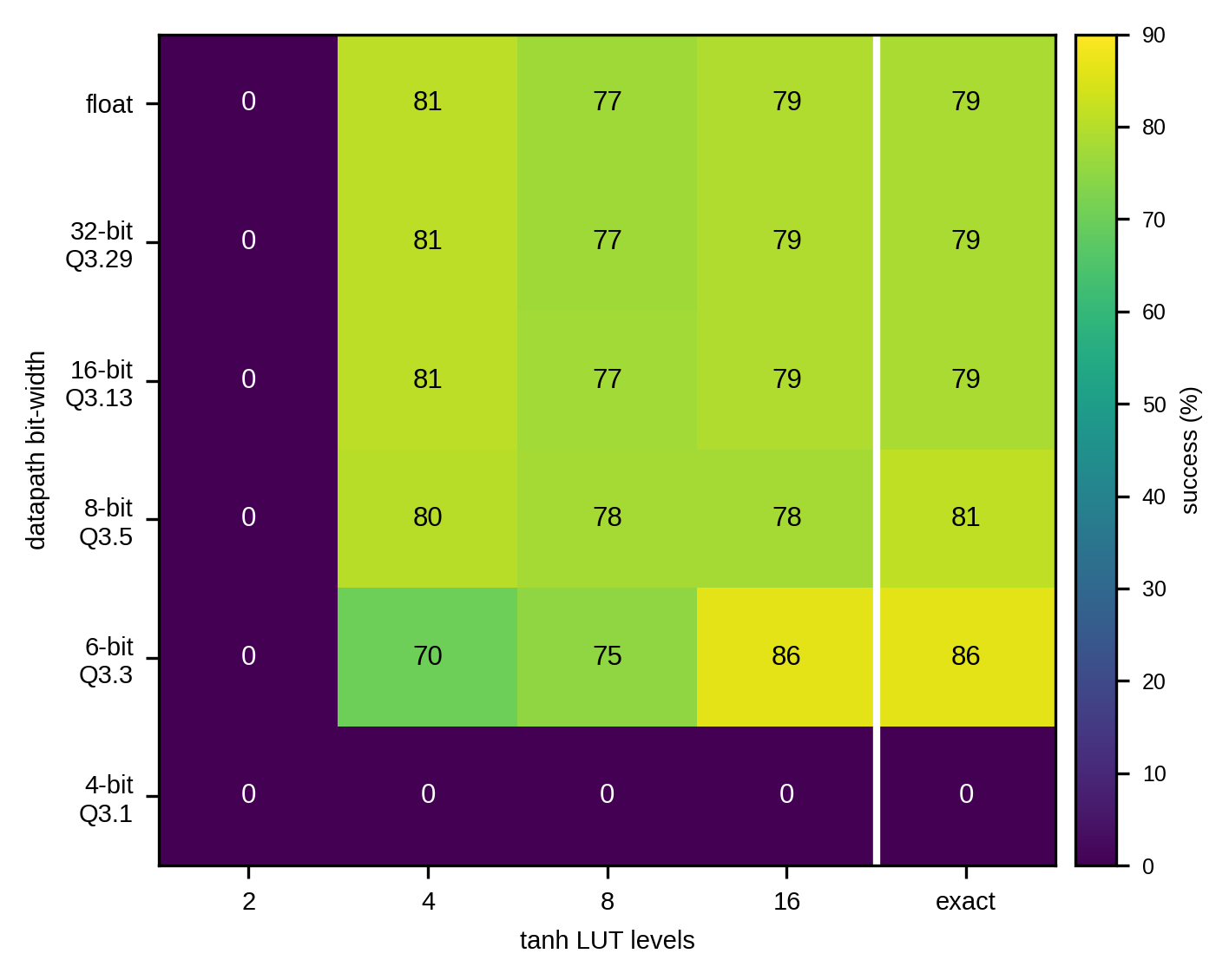}
    \caption{\textbf{Joint sweep of datapath bitwidth and $\tanh$ look-up-table
    resolution for Max-Cut at $N=60$.} 
    Success probability (\%) for every combination of fixed-point datapath format (rows) and number of levels in the $\tanh$ look-up table (columns). 
    Fixed-point formats are labeled in Q$m.n$ notation, $m$ integer bits (including one bit for sign) and $n$ fractional bits. 
    The \texttt{float} row and the \texttt{exact} column denote full floating-point arithmetic and an unquantized $\tanh$ evaluation, respectively, and together define the full-precision baseline. 
    Results are averaged over 100 instances with 256 \runname{s} each.}
    \label{fig:codesign_grid}
\end{figure}

\fig{codesign_grid} shows the success probability over the full grid of datapath bitwidths and $\tanh$ look-up-table resolutions, from which the operating point used in the hardware implementation was selected. 
The grid exhibits two sharp failure boundaries. First, a two-level $\tanh$ look-up table fails for every datapath width, including full floating-point arithmetic, since with only two output levels the approximation reduces to a sign function. 
Second, datapath widths of four bits and below fail regardless of the $\tanh$ resolution, as the remaining fractional precision is insufficient to represent the accumulated local fields. 
Above both boundaries the success probability forms a broad plateau between approximately $70\%$ and $86\%$ and is largely insensitive to further increases in either parameter: the $8$-bit, four-level configuration reaches $80\%$, matching the $81\%$ obtained with full floating-point arithmetic and an unquantized $\tanh$. 
We therefore adopt this configuration for the hardware implementation, as it is the least costly point on the plateau. The residual variation within the plateau, including the $86\%$ observed at six bits with sixteen or more levels, is comparable to the spread across instances.

\clearpage

\section{PIMI FPGA implementation for Max-cut and SK-1}
\label{sec:PIMI_MC_SK}

To accelerate combinatorial optimization on Ising formulations such as Max-Cut and Sherrington–Kirkpatrick-1 (SK-1), we employ the Probabilistic Ising machine with inertia (PIMI), which augments conventional probabilistic Ising dynamics with \emph{fully parallel} spin updates. Each iteration of PIMI consists of (i) evaluating all effective fields in a single matrix–vector multiply (MVM) and (ii) updating all spins simultaneously using a momentum-augmented stochastic rule. The FPGA architecture directly employs this two-stage structure and implements both stages as a deeply unrolled computation scheme.

\subsection{Update schedule}

For Max--Cut and SK-1 benchmarks, we use a deterministic \schedulename\ tailored to the PIMI update rule and fully parallel spin updates.
The schedule jointly controls the inverse temperature $\beta(t)$, the stochastic noise amplitude $\eta(t)$, and the self-alignment parameter $\xi$, which biases spins toward their previous values.
In practice, the shapes of the \schedulename were selected, and the corresponding schedule parameters were extensively tuned through software-based parameter search to obtain the best achievable performance and resource efficiency under fixed-point arithmetic and fully unrolled updates. The same tuning protocol was used across methods, following established benchmarking best practices to ensure a fair comparison \cite{Bartz-Beielstein2020}.

\paragraph{Inverse-temperature schedule.}
In both Max--Cut and SK-1 experiments, the inverse temperature follows a monotonic schedule given by
\begin{equation}
    \beta(t) = \beta_{\mathrm{scale}} \,
    \tanh\!\bigl(\beta_{\mathrm{init}} + \Delta\beta \, t\bigr),
    \qquad t = 0,1,\ldots,T_{\mathrm{steps}}-1.
\end{equation}
This choice produces a rapid initial increase in the local-field gain followed by gradual saturation.
Because the stochasticity of the PIMI update depends jointly on $\beta(t)$, $\eta(t)$, and the fixed self-alignment term $\xi$, $\beta(t)$ should not be interpreted as controlling the effective temperature independently of the other update parameters.

\paragraph{Noise schedule.}
To maintain stochasticity during the early and intermediate phases of the \runname, additive noise is injected into each spin update with an amplitude that depends on the instantaneous inverse temperature:
\begin{equation}
    \eta(t) = \sqrt{\beta(t)/5}.
\end{equation}
Accordingly, the absolute noise amplitude increases together with $\beta(t)$ rather than decreasing over the course of the trial.
This coupled schedule was selected empirically to maintain stochastic transitions as the local-field response becomes stronger.
The noise amplitude grows sublinearly with $\beta(t)$: in the approximately linear regime, $\tanh(\beta I_i)\approx\beta I_i$, whereas $\eta(t)\propto\sqrt{\beta(t)}$.
Consequently, the local-field contribution generally becomes stronger relative to the injected noise as $\beta(t)$ increases, while the increasing absolute noise amplitude preserves stochastic exploration.
Both quantities gradually saturate at later update steps because $\beta(t)$ itself saturates.

\paragraph{Self-alignment parameter.}
A constant self-alignment term $\xi$ is included in the PIMI update to bias each spin toward its previous value and suppress oscillatory behavior under fully parallel updates.
For Max--Cut, we use $\xi = 0.7$ for all problem sizes.
For SK-1, we use $\xi = 0.5$ for all problem sizes.

\subsection{Matrix, state, and schedule storage}
Each instance is defined by its coupling matrix $J$ and initial spin configuration, which are streamed from the host to the accelerator.  
The nonlinear term $\tanh(\cdot)$ is implemented using a compact 4-level LUT, and both the \schedulename\ $\beta(t)$ and noise samples $\mathcal{N}_i(t)$ are pre-generated offline and pre-loaded onto the FPGA as on-board read-only data.
The accelerator operates in a continuous manner: new $J$–state pairs enter as earlier instances advance, and completed results stream out without reprogramming the FPGA.

Spin variables are stored using \verb|ap_int<2>|, a two-bit signed integer representation sufficient for encoding $\{-1,+1\}$.  
The $J$ matrix and all intermediate values (including the effective fields, tanh outputs, and noise samples) use the fixed-point type \verb|ap_fixed<8,3,AP_TRN_ZERO,AP_SAT>|, an 8-bit representation with 3 integer bits (including sign) and 5 fractional bits, employing truncation-to-zero and saturating overflow.
Internal representations with wider bitwidth are employed during the MVM accumulation to maintain numerical integrity across the $N$ multiply–accumulate terms.

\subsection{Fully unrolled MVM and activation pipeline}

Since every spin is $\pm 1$ and every coupling is $+1$, $-1$, or $0$ (absent edges, as in Max-Cut where $J=-A$), the local field $I_i(t)=\sum_j J_{ij}s_j$ requires no multiplication. We unroll the row and column dimensions to evaluate all $N$ terms in parallel, each taking the value $+1$, $-1$, or $0$ according to whether the edge is present with agreeing signs, present with disagreeing signs, or absent. Summing these ternary terms in a compressor tree reduces the field to a popcount that uses only LUT and carry logic rather than DSP slices, with the terms zero-padded to the next power of two to balance the tree.

Immediately downstream, a fully unrolled activation stage performs the PIMI update rule: each spin independently reads its $I_i(t)$, applies the LUT-based $\tanh(\beta(t) I_i)$, adds the self-alignment and noise terms, and computes the sign. Because both stages are spatially unrolled, the architecture produces one entire updated spin vector in a single pass.

\subsection{Output handling and solution selection}
All intermediate spin states $\mathbf{s}(t)$ generated during the \schedulename\ are returned to the host processor. The host evaluates the Ising energy for each state and selects the lowest-energy configuration as the final solution. This enables solution extraction even when the best state appears before the final iteration.

\subsection{Kernel execution}
Only one PIMI kernel is active on the FPGA at a time, and the kernel performs one \runname\ per invocation. Each \runname\ consists of a pre-defined number of sequential update steps; within each step, the MVM and activation computations are fully unrolled in hardware. 

\subsection{Pseudocode}

\begin{algorithm}[h!]
\caption{Fully unrolled PIMI update for Max-Cut/ SK1 Problems}
\label{alg:PIMI_single}
\KwData{Couplings $J_{ij}\in\{+1,-1,0\}$, fields $h_i$, initial spins $s_i(0)$}
\KwResult{Final spin configuration $s_i(T_{\mathrm{steps}})$}

\textbf{Indices:}
$i,j$ = spin indices; \quad $t$ = \stepname\;

\For{$t \gets 0$ \KwTo $T_{\mathrm{steps}}-1$}{

  \ForEach{spin $i$ \tcp*[r]{HLS: fully unrolled over $i$}}{

    $u_i \gets 0$ \tcp*[r]{ternary-term accumulator}

    \ForEach{spin $j$ \tcp*[r]{HLS: fully unrolled over $j$}}{
      $\mathrm{term}_{ij} \gets
        \begin{cases}
          +1 & J_{ij}\neq 0 \ \text{and}\ \mathrm{sign}(J_{ij})=s_j(t)\\
          -1 & J_{ij}\neq 0 \ \text{and}\ \mathrm{sign}(J_{ij})\neq s_j(t)\\
          \ \ 0 & J_{ij}=0
        \end{cases}$\;
      $u_i \gets u_i + \mathrm{term}_{ij}$ \tcp*[r]{compressor (adder) tree}
    }

    $I_i(t) \gets u_i + h_i$\;
  }

  \ForEach{spin $i$ \tcp*[r]{HLS: fully unrolled}}{

    $\tilde{s}_{i}(t) \gets
        \tanh\!\bigl(\beta(t)\, I_{i}(t)\bigr)$ \tcp*[r]{4-level tanh LUT}

    $\delta_{i}(t) \gets
        \tilde{s}_{i}(t)
        + \xi\, s_{i}(t)
        + \eta\, \mathcal{N}(0,1)$\;

    $s_{i}(t+1) \gets
        \mathrm{sign}\!\bigl(\delta_{i}(t)\bigr)$\;
  }
}
\end{algorithm}

\clearpage
\section{Conventional PIM FPGA implementation for Max-cut and SK-1}
\label{sec:CPIM_MC_SK}

As a comparison, we implement a conventional probabilistic Ising machine (PIM) on the same FPGA platform and using the same fixed-point arithmetic, memory organization, and tanh approximation as the PIMI design. 
The key differences between the two architectures lie in the algorithmic parallelism (e.g. pipelining and unrolling) and the update dynamics.

\subsection{Update schedule}

For both Max--Cut and Sherrington--Kirkpatrick (SK--1) benchmarks, the conventional PIM employs the same shape of \schedulename, consisting of a fixed inverse temperature, and a decaying noise amplitude. 
Similar to PIMI, the shapes of the \schedulename{} were selected, and the corresponding schedule parameters were extensively tuned through software-based parameter search to obtain the best achievable performance and resource efficiency under fixed-point arithmetic and fully unrolled updates. The same tuning protocol was used across methods, following established benchmarking best practices to ensure a fair comparison \cite{Bartz-Beielstein2020}.

The inverse temperature is held constant throughout the \runname,
\begin{equation}
    \beta(t) = \beta_{\mathrm{scale}},
\end{equation}
with $\beta_{\mathrm{scale}} = 0.2$. Stochastic exploration is introduced via additive noise whose amplitude decays with annealing step $t$ according to
\begin{equation}
    \eta(t) =
    \max\!\left(
        \frac{\eta_{\mathrm{scale}}}{\sqrt{t+1}},
        \eta_{\mathrm{floor}}
    \right),
    \qquad t = 0,1,\ldots,T_{\mathrm{steps}}-1,
\end{equation}
where $\eta_{\mathrm{scale}}$ and $\eta_{\mathrm{floor}}$ are determined with a parameter search that gives the highest success probability.

\subsection{Update structure and hardware mapping}

In the conventional PIM, spins are updated sequentially (one at each step). As illustrated in Algorithm~\ref{alg:pim_maxcut}, the outer loop over spin index $i$ is \emph{pipelined}, while the inner loop over $j$ is fully unrolled to compute the local field $I_i(t)$ via a dot product. This results in one spin update being completed per pipeline initiation interval, in contrast to the fully parallel updates employed by PIMI.

Because only a single spin is updated at a time, the coupling matrix $J$ and current spin state $\mathbf{s}(t)$ are accessed in a broadcast fashion, and the effective field, activation, and stochastic update are computed for one spin per cycle. The tanh nonlinearity is implemented using the same 4-level LUT as in the PIMI design, and the same fixed-point data types are used throughout.

\clearpage
\subsection{Pseudocode}

\begin{algorithm}[h!]
\caption{Conventional PIM update for Max-Cut / SK-1}
\label{alg:pim_maxcut}
\KwData{Couplings $J_{ij}\in\{+1,-1,0\}$, fields $h_i$, initial spins $s_i(0)$}
\KwResult{Final spin configuration $s_i(T_{\mathrm{steps}})$}

\textbf{Indices:}
$i,j$ = spin indices;\quad $t$ = \stepname\;

\For{$t \gets 0$ \KwTo $T_{\mathrm{steps}}-1$}{
    $i \gets t \bmod N$  \tcp*[r]{current spin index to update}

    $u_i(t) \gets 0$ \tcp*[r]{ternary-term accumulator}

    \ForEach{spin $j$ \tcp*[r]{HLS: fully unrolled over $j$}}{
      $\mathrm{term}_{ij} \gets
        \begin{cases}
          +1 & J_{ij}\neq 0 \ \text{and}\ \mathrm{sign}(J_{ij})=s_j(t)\\
          -1 & J_{ij}\neq 0 \ \text{and}\ \mathrm{sign}(J_{ij})\neq s_j(t)\\
          \ \ 0 & J_{ij}=0
        \end{cases}$\;
      $u_i(t) \gets u_i(t) + \mathrm{term}_{ij}$ \tcp*[r]{compressor (adder) tree}
    }

    $I_{i}(t) \gets u_{i}(t)$\;

    $\tilde{s}_{i}(t) \gets
        \tanh\!\bigl(\beta(t)\, I_{i}(t)\bigr)$\ \tcp*[r]{4-level tanh LUT}
    $\delta_{i}(t) \gets
        \tilde{s}_{i}(t)
        + \eta\, U(-1,1)$\;
    $s_{i}(t+1) \gets
        \mathrm{sign}\!\bigl(\delta_{i}(t)\bigr)$\;
  }

\end{algorithm}

\clearpage

\section{PIMI FPGA implementation for MIMO detection}
\label{sec:PIMI_MIMO}

The PIMI framework is also applied to the MIMO detection problem, with the same update rule and FPGA pipeline structure used for the Max-Cut and SK-1 benchmarks. 
The key differences are the required numerical precision, the multi-\runname\ parallelism used to achieve high throughput, and the use of a MIMO-specific \schedulename. 
In addition, unlike the combinatorial benchmarks where all intermediate states are retained, the MIMO detector uses only the final spin configuration as the detected symbol vector.

\subsection{Update schedule for MIMO detection}

For MIMO detection, the PIMI solver employs a fixed \schedulename\ that prioritizes numerical stability and high-throughput execution under fully parallel updates.
Unlike combinatorial benchmarks, the MIMO problem structure is determined by the channel realization, and reliable detection is achieved using a constant inverse temperature combined with a controlled noise schedule.
The shapes of the \schedulename{} were selected, and the corresponding schedule parameters were extensively tuned through software-based parameter search to obtain the best achievable performance and resource efficiency under fixed-point arithmetic and fully unrolled updates. The same tuning protocol was used across methods, following established benchmarking best practices to ensure a fair comparison \cite{Bartz-Beielstein2020}.
All \schedulename\ parameters for MIMO detection are fixed across channel realizations and SNRs.

\paragraph{Inverse temperature.}
For all MIMO experiments, the inverse temperature is held constant throughout the \runname,
\begin{equation}
    \beta(t) = \beta_{\mathrm{scale}},
    \qquad
    t = 0,1,\ldots,T_{\mathrm{steps}}-1,
\end{equation}
with $\beta_{\mathrm{scale}} = 1$.
This choice simplifies hardware control and avoids sensitivity to schedule
shape, while providing sufficient nonlinear gain for reliable symbol
detection across signal-to-noise ratios.

\paragraph{Noise schedule.}
Stochastic noise is injected to improve robustness against local minima.
The noise amplitude follows a linearly decreasing schedule of the form
\begin{equation}
    \eta(t)
    = \sqrt{\frac{1}{5\,\gamma(t)}},
\end{equation}
where
\begin{equation}
    \gamma(t)
    = \gamma_{\mathrm{init}}
      + (\gamma_{\mathrm{final}}-\gamma_{\mathrm{init}})
        \frac{t}{T_{\mathrm{steps}}-1}.
\end{equation}
This schedule provides high stochasticity during early iterations and gradually suppresses noise as the detector stabilizes, while keeping the inverse temperature itself fixed.

\paragraph{Self-alignment term.}
A constant self-alignment parameter $\xi = 2$ is used for all MIMO experiments.

\subsection{Problem-specific coupling matrix and precision}

In the MIMO detection problem, the Ising model contains both a coupling matrix $J$ and the external field $h$, both derived from DI-MIMO. 
The entries of $J$ and $h$ in the MIMO setting require a substantially wider dynamic range. 
To accommodate this, both $J$ and $h$ are supplied to the accelerator as part of the instance stream and stored on-chip in an extended fixed-point format \verb|ap_fixed<16,4,AP_TRN_ZERO,AP_SAT>|, which allocates four integer bits (including sign) and twelve fractional bits. All intermediate quantities in the MIMO update pipeline also use this 16-bit representation to preserve numerical fidelity across the wider dynamic range required by the channel-induced couplings.

\subsection{Grouped, parallel field computation}

To increase throughput, the MIMO implementation evaluates multiple \runname* in parallel. 
The total of $M$ \runname* is partitioned into groups, with each group containing 4 independent \runname* initialized from different spin configurations. 
The FPGA kernel processes one such group at a time: within the MVM stage, the four \runname* in the group are evaluated concurrently, while the outer loop sequentially advances across all groups.
This is accomplished through a pipelined computation scheme in which the outer loops over spins and \runname\ indices advance each cycle (initiation interval of one), while the inner multiply–accumulate loop is unrolled to sustain independent MAC operations per cycle. 

After the grouped MVM computation, the activation stage (tanh LUT, self-alignment term, and noise injection) proceeds identically for the four replicas using the same logic and dataflow as in the Max-Cut and SK-1 solver. 
No additional modification is required beyond the factor-of-four unrolling. The final spin configuration $\mathbf{s}(N)$ from each replica is used directly as the detected symbol vector candidate.

\subsection{Kernel execution}
A single PIMI kernel is deployed on the FPGA, and each run of the kernel processes all $M$ \runname* by sweeping over \emph{groups of 4 \runname*} (in parallel). 
The \schedulename\ and noise are pre-loaded on board, while $J$ and $h$ for each MIMO instance are streamed from the host. The kernel operates continuously, processing one instance after another without idling.
At the end of each instance, the FPGA returns the final spin states for all \runname*, and the host selects the configuration with the lowest Ising energy as the final detected symbol vector.

\subsection{Host-side task scheduling}
To maintain continuous throughput, a lightweight task scheduler operates on the host computer. The scheduler has $K$ workers, each responsible for driving one PIMI kernel on the FPGA. 
Incoming MIMO detection instances are inserted into a global queue and dispatched to workers in a round-robin fashion, ensuring balanced load distribution and preventing kernel idle time. 
Each worker streams the corresponding $J$ matrix, $h$ vector and initial spin states to the device, receives the final spin configuration after the \runname\ completes, and reports the resulting detected symbol vector to the upper-layer receiver pipeline.  
Details about the scheduler can be found in \ref{sec:host_scheduler}

\subsection{Pseudocode}

\begin{algorithm}[h!]
\caption{PIMI implementation for MIMO detection}
\label{alg:PIMI_mimo}

\KwData{
  Couplings $J_{ij}$; local fields $h_i$; \\
  initial spins $s_{i,a}(0)$ for all spins $i$ and \runname* $a$
}
\KwResult{Final spin states $s_{i,a}(T_{\mathrm{steps}})$}


\textbf{Indices:}
$i,j$ = spin indices;\quad
$a$ = \runname\ index;\quad
$g$ = group index;\quad
$t$ = \stepname.\;

\For{$t \gets 0$ \KwTo $T_{\mathrm{steps}}-1$}{
  \ForEach{group $g$}{
    \ForEach{spin $i$ \tcp*[r]{HLS: pipelined, II = 1}}{
      \ForEach{\runname\ $a$ in group $g$ \tcp*[r]{HLS: pipelined, II = 1}}{

        $u_{i,a} \gets 0$ \tcp*[r]{accumulator for dot product}

        \ForEach{spin $j$ \tcp*[r]{HLS: unrolled, factor = 4}}{
          $u_{i,a} \gets u_{i,a} + J_{ij}\, s_{j,a}(t)$\;
        }

        $I_{i,a}(t) \gets u_{i,a} + h_i$\;
      }
    }
  }

  \ForEach{\runname\ $a$ \tcp*[r]{HLS: pipelined, II = 1}}{
    \ForEach{spin $i$ \tcp*[r]{HLS: unrolled, factor = 4}}{
      $\tilde{s}_{i,a}(t) \gets
        \tanh\!\bigl(\beta(t)\, I_{i,a}(t)\bigr)$\ \tcp*[r]{4-level tanh LUT}
      $\delta_{i,a}(t) \gets
        \tilde{s}_{i,a}(t)
        + \xi\, s_{i,a}(t)
        + \eta\, \mathcal{N}(0,1)$\;
      $s_{i,a}(t+1) \gets
        \mathrm{sign}\!\bigl(\delta_{i,a}(t)\bigr)$\;
    }
  }
}
\end{algorithm}

\clearpage

\section{Conventional PIM FPGA implementation for MIMO detection}
\label{sec:CPIM_MIMO}

For comparison, we implement a conventional probabilistic Ising machine (PIM) for MIMO detection on the same FPGA platform and using the same numerical representations, LUT-based activation, and host–device dataflow as the PIMI implementation. The key differences are the update ordering, the pipelining strategy, and the \schedulename.

\subsection{Update schedule}

The shapes of the \schedulename{} were selected, and the corresponding schedule parameters were extensively tuned through software-based parameter search to obtain the best achievable performance and resource efficiency under fixed-point arithmetic and fully unrolled updates. The same tuning protocol was used across methods, following established benchmarking best practices to ensure a fair comparison \cite{Bartz-Beielstein2020}.
For this implementation, we employ a fixed inverse temperature combined with a decaying noise amplitude. The inverse temperature is held constant throughout the \runname,
\begin{equation}
    \beta(t) = 1,
    \qquad
    t = 0,1,\ldots,T_{\mathrm{steps}}-1.
\end{equation}

Stochastic exploration is introduced through an additive noise term whose amplitude decays with iteration count according to
\begin{equation}
    \eta(t) =
    \frac{1}{\sqrt{(t+1)/5}},
\end{equation}
which provides strong randomness during early iterations and gradually suppresses noise as the \runname\ progresses. This schedule avoids explicit temperature ramping and simplifies hardware control, while maintaining sufficient stochasticity for reliable convergence under fixed-point arithmetic and sequential spin updates.

\subsection{Sequential update structure and pipelining}

In the conventional PIM, spins are updated sequentially rather than in parallel. As shown in Algorithm~\ref{alg:pim_mimo}, each update step corresponds to a single spin index $i = t \bmod N$, which needs to be updated across all \runname* before advancing to the next spin. 
For a given spin $i$, the local field $I_{i,a}(t)$ is computed for all \runname* $a$ using a grouped matrix–vector multiply. The inner loop over spin index $j$ is fully unrolled to compute the dot product, while the outer loops over \runname\ index and \runname\ group are pipelined with an initiation interval of one. This enables concurrent processing of multiple \runname* while maintaining the sequential update semantics of conventional PIM.

Following field computation, the activation stage updates the current spin using the same 4-level $\tanh$ LUT as in PIMI, but without a self-alignment term. Because only one spin is updated per step, the overall update latency scales linearly with the number of spins $N$, in contrast to the fully parallel PIMI architecture.

\subsection{Kernel execution and output handling}

A single conventional PIM kernel is deployed on the FPGA. At each update step $t$, the kernel updates a single spin index $i = t \bmod N$ while processing \emph{16 \runname* in parallel}. For the selected spin $i$, the local field $I_{i,a}(t)$ is computed concurrently across the 16 \runname* using grouped, pipelined matrix--vector multiplication, and the subsequent activation and stochastic update are performed in parallel across all \runname*. Over $T_{\mathrm{steps}}$ update steps, the kernel therefore executes $T_{\mathrm{steps}}$ single-spin updates (equivalently $T_{\mathrm{steps}}/N$ full sweeps). The coupling matrix $J$ and local field vector $h$ are streamed from the host, while the \schedulename\ parameters and noise tables are stored on chip.

At the end of each instance, the FPGA returns the final spin configuration for each \runname. The host evaluates the corresponding Ising energies and selects the lowest-energy configuration as the detected symbol vector, identical to the PIMI-based MIMO workflow.

\subsection{Pseudocode}

\begin{algorithm}[h]
\caption{Conventional PIM implementation for MIMO detection}
\label{alg:pim_mimo}

\KwData{
  Couplings $J_{ij}$; local fields $h_i$; \\
  initial spins $s_{i,a}(0)$ for all spins $i$ and \runname* $a$
}
\KwResult{Final spin states $s_{i,a}(T_{\mathrm{steps}})$}


\textbf{Indices:}
$i,j$ = spin indices;\quad
$a$ = anneal index;\quad
$g$ = group index;\quad
$t$ = update step (with $i = t \bmod N$).\;

\For{$t \gets 0$ \KwTo $T_{\mathrm{steps}}-1$}{
  $i \gets t \bmod N$  \tcp*[r]{current spin index to update}

  \ForEach{group $g$}{
    \ForEach{\runname\ $a$ in group $g$ \tcp*[r]{HLS: pipelined, II = 1}}{

      $u_{i,a}(t) \gets 0$ \tcp*[r]{accumulator for dot product}

      \ForEach{spin $j$ \tcp*[r]{HLS: fully unrolled}}{
        $u_{i,a}(t) \gets u_{i,a}(t) + J_{ij}\, s_{j,a}(t)$\;
      }

      $I_{i,a}(t) \gets u_{i,a} + h_i$\;
    }
  }

  \ForEach{\runname\ $a$ \tcp*[r]{HLS: fully unrolled}}{
    $\tilde{s}_{i,a}(t) \gets
        \tanh\!\bigl(\beta(t)\, I_{i,a}(t)\bigr)$\ \tcp*[r]{4-level tanh LUT}
    $\delta_{i,a}(t) \gets
        \tilde{s}_{i,a}(t)
        + \eta\, U(-1,1)$\;
    $s_{i,a}(t+1) \gets
        \mathrm{sign}\!\bigl(\delta_{i,a}(t)\bigr)$\;
  }
}
\end{algorithm}

\clearpage

\section{MIMO problem definition}
\label{sec:MIMO_def}

Multiple-Input Multiple-Output (MIMO) refers to a wireless communication technique where several antennas are used at both the transmitter and the receiver \cite{Tse2005, Yang2015}. 
By sending and receiving multiple signals at the same time, MIMO creates parallel spatial paths between the transmitter and receiver. These extra paths allow the system to carry more data in the same bandwidth and make the link more reliable in the presence of noise and interference. For example, in an $N_t \times N_r$ MIMO system, $N_t$ signals are transmitted simultaneously to a base station equipped with $N_r$ receive antennas. 
MIMO communication operates in two directions: \emph{downlink}, from the base station to the users, where the main operation is \emph{precoding}; and \emph{uplink}, from the users to the base station, where the main task is \emph{detection} of the transmitted symbols. 
In this work, we focus on the uplink scenario: given the received signal at the \textbf{base station} and an estimate of the channel, the task is to infer the transmitted symbol vector from all users. 

\subsection{Uplink MIMO detection}
In an uplink MIMO system, $N_t$ symbols are transmitted to a base station with $N_r$ antennas. The transmitted symbol vector is denoted by $\tilde{x}\in\Phi^{N_t}$, where $\Phi$ is a quadrature amplitude modulation (QAM) constellation \cite{Tse2005, Proakis2008}. 
QAM encodes information in both amplitude and phase of complex symbols, with typical choices including 4-QAM, 16-QAM, and 64-QAM. 
The propagation environment between transmit and receive antennas are represented by the complex channel matrix $\tilde{H}\in\mathbb{C}^{N_r\times N_t}$, whose entries capture the gain and phase shift of each transmit–receive path. The corresponding received signal vector at the base station is $\tilde{y}\in\mathbb{C}^{N_r}$. We use the tilde notation to emphasize that $H$, $x$, and $y$ are complex-valued; later we will convert them into an equivalent real-valued form for algorithmic convenience. 
The baseband input–output relation is
\begin{equation}
    \tilde{y} \;=\; \tilde{H}\,\tilde{x} \;+\; n,
    \label{eq:complex_model}
\end{equation}
where $n$ is additive white Gaussian noise (AWGN) \cite{Shannon1948, Shannon1949}. We model this noise as
\[
n \sim \mathcal{CN}(0,\sigma^2 I_{N_r}),
\]
meaning that each entry of $n$ is an independent circularly symmetric complex Gaussian random variable with zero mean and variance $\sigma^2$ per real dimension. Here, $\mathcal{C}$ indicates the complex domain, $\mathcal{N}$ denotes a Gaussian distribution, and $\sigma^2$ controls the noise power.

\subsection{Maximum-Likelihood (ML) detection}

Maximum likelihood (ML) detection is a standard principle in statistical inference: among all possible candidates, choose the one that maximizes the probability of producing the observed data. In the context of MIMO detection, this means selecting the transmitted symbol vector that is most likely to have generated the received signal given the channel \cite{Tse2005, Verdu1998}. Mathematically, given $\tilde{y}$ and $\tilde{H}$, ML detection seeks the symbol vector $\tilde{x}$ that minimizes the Euclidean distance between the received signal and all candidate transmit vectors:
\begin{equation}
    \hat{x}
    \;=\;
    \arg\min_{\tilde{u}\in\Phi^{N_t}}
    \;\bigl\|\tilde{y}-\tilde{H}\,\tilde{u}\bigr\|^2.
    \label{eq:ml_complex}
\end{equation}
The search space grows exponentially with $N_t$ (and $\log_2 M$), which is why ML detection is computationally challenging in large systems.

\subsection{Real-Valued equivalent}
\label{sec:real_valued_equivalent}

For algorithmic convenience, the complex system in \eqref{eq:complex_model} is often recast into a real-valued form by stacking real and imaginary parts \cite{Tse2005}:
\begin{equation}
    H \;=\;
    \begin{bmatrix}
        \Re(\tilde{H}) & -\Im(\tilde{H}) \\
        \Im(\tilde{H}) & \phantom{-}\Re(\tilde{H})
    \end{bmatrix},\quad
    y \;=\;
    \begin{bmatrix}
        \Re(\tilde{y}) \\ \Im(\tilde{y})
    \end{bmatrix},\quad
    x \;=\;
    \begin{bmatrix}
        \Re(\tilde{x}) \\ \Im(\tilde{x})
    \end{bmatrix}.
    \label{eq:real_transform}
\end{equation}
Let $\Re(\Phi)$ and $\Im(\Phi)$ denote the real and imaginary symbol alphabets. The ML problem becomes
\begin{equation}
    \hat{x}
    \;=\;
    \arg\min_{u \in \bigl[\Re(\Phi)^{N_t},\,\Im(\Phi)^{N_t}\bigr]}
    \;\bigl\| y - H\,u \bigr\|^2,
    \label{eq:ml_real}
\end{equation}
and $\hat{x}$ is mapped back to $\hat{\tilde{x}}$ via the inverse of \eqref{eq:real_transform}.

\subsection{Uplink MIMO instance generation}
\label{sec:gen_instances}

\paragraph{System sizes and constellations.}
Our focus is on the \emph{large MIMO} regime, where the number of antennas is sufficiently high that conventional detectors face a steep performance–complexity trade-off.  We consider square MIMO systems with
\[
(N_t,N_r)\in\{(4,4),(8,8),(16,16)\},
\]
and evaluate QAM constellations with $M\!\in\!\{4,16,64\}$. Transmitted vectors $\tilde{x}\in\Phi^{N_t}$ are drawn uniformly at random from the chosen constellation.

\paragraph{Channel generation.}
For each problem instance, we generate an i.i.d.\ Rayleigh fading channel \cite{Tse2005}:  
\begin{equation}
    \tilde{H} \;=\; \tfrac{1}{\sqrt{2}}\bigl(H_\mathrm{R}+jH_\mathrm{I}\bigr),
    \quad H_\mathrm{R},H_\mathrm{I}\stackrel{\text{i.i.d.}}{\sim}\mathcal{N}(0,1)^{N_r\times N_t},
\end{equation}
so that each entry of $\tilde{H}$ is distributed as $\mathcal{CN}(0,1)$. Here, $\mathcal{CN}(\mu,\sigma^2)$ denotes a circularly symmetric complex Gaussian distribution with mean $\mu$ and variance $\sigma^2$.

\paragraph{Noise power from $E_b/N_0$.}
To control the noise level in our simulations, we specify the energy-per-bit to noise spectral density ratio, $E_b/N_0$, which is a standard way of defining signal-to-noise ratio (SNR) in digital communications. Here, $E_b$ is the average energy per information bit, and $N_0$ is the one-sided noise power spectral density.  

Unless otherwise stated, $E_b/N_0$ denotes the linear-scale quantity, while values expressed in decibels are explicitly labeled as dB. The target $E_b/N_0$ is typically specified in decibels and converted as
\[
\frac{E_b}{N_0} = 10^{\tfrac{(E_b/N_0)_{\mathrm{dB}}}{10}}.
\]

Since each QAM symbol carries $b=\log_2 M$ bits, the corresponding energy-per-symbol to noise ratio is
\[
\frac{E_s}{N_0} = b \cdot \frac{E_b}{N_0},
\]
where $E_s$ is the average energy of a transmitted symbol.  

Given a noiseless receive vector $y_\mathrm{rec}=\tilde{H}\tilde{x}$, we compute the empirical average received power per antenna as
\[
E_y \;\approx\; \frac{1}{N_r}\sum_{i=1}^{N_r}\!\bigl| (y_\mathrm{rec})_i \bigr|^2.
\]
We then set the complex noise variance so that the effective received SNR per antenna matches $E_s/N_0$:
\begin{equation}
    \sigma^2 = \frac{E_y}{E_s/N_0}
             = \frac{E_y}{\,b \cdot (E_b/N_0)}.
\end{equation}
Finally, we generate additive white Gaussian noise (AWGN) samples as
\[
n = \sqrt{\tfrac{\sigma^2}{2}}\,(w_\mathrm{R}+j w_\mathrm{I}), 
\quad w_\mathrm{R},w_\mathrm{I}\sim\mathcal{N}(0,1)^{N_r}.
\]

\subsection{Baseline: linear MMSE detector}
\label{sec:mmse_baseline}

Let $(E_b/N_0)_{\mathrm{dB}}$ be the target energy-per-bit to noise ratio in decibels and
\[
\frac{E_b}{N_0} = 10^{(E_b/N_0)_{\mathrm{dB}}/10}
\]
its linear-scale value. With $b=\log_2 M$ bits per QAM symbol, the symbol SNR is
\[
\frac{E_s}{N_0} \;=\; b \cdot \frac{E_b}{N_0}.
\]
The linear MMSE filter is then
\begin{equation}
    \mathbf{W}_\text{MMSE}
    \;=\;
    \Bigl(\tilde{H}^\mathrm{H}\tilde{H}
    \;+\; \tfrac{1}{E_s/N_0}\,I_{N_t}\Bigr)^{-1}\tilde{H}^\mathrm{H}
    \;=\;
    \Bigl(\tilde{H}^\mathrm{H}\tilde{H}
    \;+\; \tfrac{1}{\,b\,(E_b/N_0)}\,I_{N_t}\Bigr)^{-1}\tilde{H}^\mathrm{H}.
    \label{eq:mmse_filter_ebn0_non_norm}
\end{equation}
The linear estimate is obtained as $\tilde{z}=\mathbf{W}_\text{MMSE}\tilde{y}$ and then mapped to the nearest constellation points through hard slicing:
\begin{equation}
    \hat{\tilde{x}}_{\text{MMSE}} \;=\; \mathcal{Q}_\Phi(\tilde{z}).
\end{equation}
Here, $\mathcal{Q}_\Phi(\cdot)$ refers to hard slicing, which maps each component of $\tilde{z}$ to its closest symbol in the constellation $\Phi$, effectively making an independent maximum-likelihood decision for each transmitted symbol.

\subsection{Performance metric: bit-error rate (BER)}
\label{sec:ber_metric}

Each QAM symbol $\tilde{x}_i\in\Phi$ carries $b=\log_2 M$ information bits via a fixed Gray-coded mapping
$\mathcal{M}:\{0,1\}^b \leftrightarrow \Phi$.
Accordingly, the transmitted symbol vector $\tilde{x}\in\Phi^{N_t}$ corresponds to a binary information vector
$\mathbf{b}\in\{0,1\}^{bN_t}$.
At the receiver, the detected symbol vector $\hat{\tilde{x}}$ is mapped back to a bit vector
$\hat{\mathbf{b}}$ using the same inverse mapping $\mathcal{M}^{-1}$.

Performance is quantified using the bit-error rate (BER), defined as the fraction of incorrectly detected information bits.
For a given detection outcome, the BER is computed as
\begin{equation}
    \mathrm{BER}
    \;=\;
    \frac{1}{bN_t}
    \sum_{k=1}^{bN_t}
    \mathbb{I}\!\left[b_k \neq \hat{b}_k\right],
    \label{eq:ber_def}
\end{equation}
where $\mathbb{I}[\cdot]$ denotes the indicator function.

For a fixed signal-to-noise ratio (SNR), the reported BER is obtained by averaging \eqref{eq:ber_def}
over multiple independent realizations of the channel $\tilde{H}$, transmitted symbol vector $\tilde{x}$,
and noise vector $n$.
This Monte Carlo averaging yields an empirical estimate of the expected bit-error probability,
\begin{equation}
    \overline{\mathrm{BER}}
    \;=\;
    \mathbb{E}_{\tilde{H},\,\tilde{x},\,n}
    \!\left[
        \frac{1}{bN_t}
        \sum_{k=1}^{bN_t}
        \mathbb{I}\!\left[b_k \neq \hat{b}_k\right]
    \right].
    \label{eq:ber_expectation}
\end{equation}

The BER is reported as a function of $E_b/N_0$ and serves as a direct measure of end-to-end detection accuracy
under QAM signaling in the presence of fading and additive noise.

\clearpage

\section{Delta-Ising MIMO (DI-MIMO) formulation}
\label{sec:DI_MIMO}

The central challenge of maximum-likelihood (ML) MIMO detection lies in the exponential search over all possible transmit symbol vectors. 
Prior Ising-machine-based approaches map the ML-MIMO problem directly into an Ising model, but struggle to sustain performance for higher-order modulations (e.g., $16$-QAM, $64$-QAM). 
To address this, Singh et al. introduced a perturbation-based formulation called \emph{Delta-Ising MIMO} (DI-MIMO) \cite{Singh2022}. The key idea is to first obtain a low-complexity estimate of the transmit vector (e.g., using MMSE), and then formulate an Ising problem that searches only for the \emph{correction} to this estimate, rather than the full symbol vector itself.

\subsection{From ML-MIMO to perturbation formulation}
Recall the real-valued ML-MIMO detection problem from \ref{sec:real_valued_equivalent}
\begin{equation}
    \hat{x}
    \;=\;\arg\min_{u \in [\Re(\Phi)^{N_t},\,\Im(\Phi)^{N_t}]}
    \|y - H u\|^2,
    \label{eq:ml_mimo_real}
\end{equation}
where $H \in \mathbb{R}^{2N_r \times 2N_t}$ is the real-valued channel matrix, 
$y \in \mathbb{R}^{2N_r}$ is the received signal, 
$u \in [\Re(\Phi)^{N_t},\,\Im(\Phi)^{N_t}] \subset \mathbb{R}^{2N_t}$ is the transmitted symbol vector in real-valued form, 
and $\hat{x} \in \mathbb{R}^{2N_t}$ is the estimated result from ML. 
Here, $\Phi$ denotes the $M$-QAM constellation. 

DI-MIMO begins with a polynomial-time detector, such as MMSE (see \ref{sec:mmse_baseline} for details), and denotes the corresponding real-valued estimate as $x_m \in \mathbb{R}^{2N_t}$. 
Instead of directly solving for $u$, a perturbation variable is defined as
\begin{equation}
    d = u - x_m,
    \label{eq:d_def}
\end{equation}
with $d \in \mathbb{Z}^{2N_t}$ representing the correction to the MMSE solution. 
The detection task is thus reframed as
\begin{equation}
    \hat{d} 
    \;=\;
    \arg\min_{d \in \mathcal{D}^{2N_t}} 
    \| y - H(x_m + d) \|^2,
    \label{eq:ml_d}
\end{equation}
where $\mathcal{D}$ denotes a finite integer-valued correction set. 
For example, in the case of 4-QAM, one may choose $\mathcal{D} = \{-2,0,2\}$ to cover all symbol differences, while higher-order constellations require larger sets.

This formulation searches only within a small neighborhood around $x_m$, significantly reducing the effective search space at moderate-to-high SNR.

\subsection{Spin representation and Ising mapping}

The correction vector $d$ can be expressed in terms of spin variables that take values in $\{-1,1\}$. 
Each scalar correction $c$ is represented as a weighted sum of a small number of spins. 
For instance, when $c \in \{-2,0,2\}$, two spin variables $s_1, s_2 \in \{-1,1\}$ are sufficient, with 
$c = s_1 + s_2$. 
When $c \in \{-4,-2,0,2,4\}$, three spins are required, and $c$ can be written as 
$c = 2s_1 + s_2 + s_3$.

Extending this construction to the full correction vector gives
\begin{equation}
    d = T s,
    \label{eq:d_transform}
\end{equation}
where $T$ is a transform matrix determined by the choice of $\mathcal{D}$, and $s\in\{-1,1\}^{K}$ is the corresponding spin vector.
\begin{equation}
T = 
\begin{cases}
    [\, I_{2N_t}, \, I_{2N_t} \,], 
    & \mathcal{D} = \{-2,0,2\}, \\[6pt]
    [\, 2I_{2N_t}, \, I_{2N_t}, \, I_{2N_t} \,], 
    & \mathcal{D} = \{-4,-2,0,2,4\},
\end{cases}
\label{eq:T_def}
\end{equation}
Thus, $K=4N_t$ for $\mathcal{D}=\{-2,0,2\}$ and $K=6N_t$ for $\mathcal{D}=\{-4,-2,0,2,4\}$.
Throughout this work, we use $\mathcal{D}=\{-2,0,2\}$, and hence $K=4N_t$.

Substituting \eqref{eq:d_transform} into \eqref{eq:ml_d} yields the Ising form:
\begin{equation}
    \hat{s}
    \;=\;
    \arg\min_{s \in \{-1,1\}^{K}}
    -h^\top s - s^\top J s,
    \label{eq:ising_form}
\end{equation}
where
\begin{equation}
    J = -\text{zeroDiag}(T^\top H^\top H T), 
    \qquad
    h = 2 (y - H x_m)^\top H T,
    \label{eq:ising_params}
\end{equation}
where $\text{zeroDiag}(\cdot)$ sets the diagonal of a matrix to zero, ensuring compatibility with Ising solvers.

\textbf{NOTE}: Auxiliary spins are not used, as the implementation supports a separate input for the local field vector $h$, and the high-level synthesis tool exhibits better resource utilization when the coupling matrix J is square with power-of-two dimensions.

\subsection{Reconstruction of the transmitted signal}
\label{sec:reconstruction}

Solving the Ising problem \eqref{eq:ising_form} yields an estimated spin configuration
$\hat{s}\in\{-1,1\}^{K}$, from which the corresponding correction vector is reconstructed as
\begin{equation}
    \hat{d} = T \hat{s},
\end{equation}
using the same transform matrix $T$ defined in \eqref{eq:T_def}.
The refined real-valued estimate of the transmitted signal is then obtained by
\begin{equation}
    \hat{x} = x_m + \hat{d},
    \label{eq:x_recon}
\end{equation}
where $x_m$ is the initial MMSE estimate.

The reconstructed vector $\hat{x}\in\mathbb{R}^{2N_t}$ is mapped back to the complex domain
via the inverse of the real-valued transformation in \eqref{eq:real_transform}, yielding
$\hat{\tilde{x}}\in\mathbb{C}^{N_t}$.
Finally, each component of $\hat{\tilde{x}}$ is projected onto the nearest constellation
point in $\Phi$ through hard slicing,
\begin{equation}
    \hat{\tilde{x}} \;=\; \mathcal{Q}_\Phi(\hat{\tilde{x}}),
\end{equation}
producing the detected transmit symbol vector.
This symbol estimate is subsequently used for bit reconstruction and BER evaluation
as described in \ref{sec:ber_metric}.

\clearpage
\section{MIMO computational cost breakdown}
\label{sec:MIMO_cost_breakdown}

The end-to-end MIMO detection workflow consists of host-side preprocessing, FPGA-based PIMI solver execution, and host-side postprocessing. 
Among these components, the overall runtime and throughput are dominated by the PIMI solver kernels executing on the FPGA.

The DI-MIMO formulation on the host involves linear-algebra operations on the received signal and channel matrix, followed by the construction of Ising parameters $(J,h)$ for each MIMO instance.
The PIMI solver constitutes the dominant computational workload.
For each MIMO instance, a solver kernel performs a fixed number of \runname*, each consisting of a fixed number of fully parallel spin-update steps.
As a result, the solver runtime scales linearly with the number of \runname* and \stepname*, and inversely with the number of instantiated PIMI kernels on the FPGA.
Because kernels operate independently and access data through HBM, overall throughput scales approximately linearly with kernel count until memory bandwidth becomes limiting.

After FPGA execution, candidate spin configurations are returned to the host for energy evaluation and selection of the minimum-energy state.
This postprocessing step involves evaluating the Ising energy for each candidate and performing a minimum reduction across \runname*.
Its computational complexity scales linearly with the number of returned candidates and is negligible compared to the iterative ising solving performed on the FPGA.

Overall, the end-to-end latency and throughput of the workflow are governed by the PIMI solver execution, while host-side formulation, scheduling, and energy evaluation contribute only minor overhead.

\section{Throughput requirement for LTE/5G systems}
\label{sec:throughput}

Orthogonal Frequency-Division Multiplexing (OFDM) is the multicarrier modulation scheme used in modern cellular systems, including LTE and 5G.
An OFDM transmission is organized into subcarriers in frequency and OFDM symbols in time; in an uplink MIMO receiver, each \emph{data-bearing OFDM symbol} must be detected across all allocated subcarriers.
Under the DI--MIMO formulation, one Ising ``instance'' corresponds to detecting one such OFDM data symbol.
Therefore, the required Ising-solver throughput is determined by the \emph{data-symbol rate} of the wireless system.

\paragraph{LTE (10~MHz).}
According to \cite{singh2025thesis}, for a typical LTE deployment operating within a 10~MHz bandwidth using 15~kHz subcarrier spacing (SCS), the throughput requirement is
\[
    \Theta_{\mathrm{req}}^{\mathrm{LTE}}
    = 8400\ \text{instances/ms.}
\]

\paragraph{5G NR (50~MHz).}
For 5G New Radio (NR), we consider a 50~MHz bandwidth configuration using 15~kHz SCS.
According to the 5G NR resource-grid specification (3GPP TS 38.104) \cite{3gpp38104}, a 50~MHz carrier at 15~kHz SCS contains $N_{\mathrm{PRB}} = 270$ physical resource blocks. Each PRB spans 12 subcarriers, resulting in a total of $270 \times 12 = 3,240$ active subcarriers.
With 15~kHz SCS, the frame structure provides 1 slot per millisecond.
Following the methodology in \cite{Singh2025mmgap}, we assume 11 data-bearing symbols per slot after accounting for DMRS and control overhead.
This results in a real-time throughput requirement of:
\[
    \Theta_{\mathrm{req}}^{\mathrm{5G}}
    = 3240 \times 11 = 35,640\ \text{instances/ms.}
\]

\clearpage

\section{Host-side multi-kernel scheduler}
\label{sec:host_scheduler}

To maximize device utilization, the host launches one worker thread per FPGA kernel (compute unit) and dynamically assigns work using an atomic batch counter.
This scheduling strategy is used for both PIMI and conventional PIM implementations.
Each worker repeatedly acquires the next unprocessed batch, transfers the corresponding input buffers to the device, launches the kernel, reads back results, and records profiling information.
The main thread waits until all batches are completed. 
Details about the implementations is listed in Algorithm \ref{alg:host_scheduler}.

\begin{algorithm}[h!]
\caption{Host-side multi-kernel scheduler for batched PIMI execution}
\label{alg:host_scheduler}
\KwData{
Batched inputs $\{J_b, S_{\mathrm{init},b}, h_b\}_{b=1}^{B}$, number of kernels $K$
}
\KwResult{
Batched outputs $\{S_{\mathrm{final},b}\}_{b=1}^{B}$
}

\textbf{Notation:} $b$ denotes the batch index and $B$ the total number of batches\;

Initialize atomic counter $\texttt{next\_batch} \gets 0$\;
Initialize atomic counter $\texttt{completed\_batches} \gets 0$\;

\For{$k \gets 1$ \KwTo $K$ \KwInParallel}{
    Launch worker thread associated with kernel $k$\;
    
    \While{true}{
        Atomically acquire next batch index $b \gets \texttt{next\_batch.fetch\_add}(1)$\;
        \If{$b \ge B$}{
            \textbf{terminate worker}\;
        }
        
        Allocate device buffers for $J_b$, $S_{\mathrm{init},b}$, $h_b$, and $S_{\mathrm{final},b}$\;
        Configure kernel arguments for kernel $k$\;
        
        Transfer input buffers to device memory\;
        Execute PIMI kernel for batch $b$\;
        Transfer output buffer $S_{\mathrm{final},b}$ back to host\;
        
        Record kernel execution latency for profiling\;
        Atomically increment $\texttt{completed\_batches}$\;
    }
}

Block until $\texttt{completed\_batches} = B$\;
\Return $\{S_{\mathrm{final},b}\}_{b=1}^{B}$\;
\end{algorithm}

\clearpage

\section{Throughput measurement for MIMO detection}
\label{sec:MIMO_throughput_measure}

Throughput is defined as the number of complete DI--MIMO detection instances solved per millisecond. Each instance corresponds to one uplink MIMO detection problem with fixed channel $H$, received vector $y$, as well as a pre-defined number of \runname* and \stepname*.

\subsection{Measurement procedure}

\begin{enumerate}
    \item \textbf{FPGA programming (not timed).}  
    The FPGA bitstream is programmed onto the device, and $K$ independent solver kernels are instantiated.  
    This programming time is measured separately and excluded from the throughput measurement.
    
    \item \textbf{Instance preparation and batching (not timed).}  
    A total of $N_{\mathrm{inst}} = 100{,}000$ DI--MIMO problem instances are first loaded into host memory from disk.  
    The instances are then grouped into fixed-size batches, and all host-side preprocessing is completed before timing begins, ensuring that the measured runtime reflects only the solver’s execution and the required host--device transfers.

    \item \textbf{Scheduling.}  
    The host launches $K$ worker threads, each responsible for driving one solver kernel.  
    A shared counter distributes batches dynamically so that all kernels remain fully utilized.  
    For each batch, a worker transfers the problem data to the FPGA, triggers the solving kernel, and reads back the final spin configurations.  
    This process continues until all instances have been processed.
    
    \item \textbf{Timer start.}  
    A global timer is started immediately before launching the multi-kernel execution.  
    The timing window therefore captures:
    \begin{itemize}
        \item transfers of batched MIMO instances from host memory to the FPGA,
        \item all computation performed by the FPGA solver kernels,
        \item transfers of the resulting spin configurations back to the host, and
        \item any scheduling or dispatch overhead during kernel execution.
    \end{itemize}
    
    \vspace{1mm}

    The detailed runtime profiling in \cite{singh2025thesis} (Chapter 6.6) has shown that the solving kernel dominates the computation time, whereas subsequent energy-evaluation steps contribute negligibly to the overall latency. \textbf{We therefore \emph{did not} include these post-processing operations in the throughput measurement.}

    \item \textbf{Timer stop.}  
    The timer is stopped once the final batch of results has been received from the FPGA.  
    The resulting elapsed time,
    \[
        T_{\mathrm{run}},
    \]
    captures the full end-to-end time of FPGA-based solving and host--device communication.

    \item \textbf{Throughput extraction.}  
    The MIMO detection throughput is computed as
    \[
        \Theta_{\mathrm{MIMO}}
        = \frac{N_{\mathrm{inst}}}{T_{\mathrm{run}}}
        \qquad (\text{instances/s}),
    \]
    with a corresponding per-instance latency
    \[
        t_{\mathrm{inst}} = \frac{T_{\mathrm{run}}}{N_{\mathrm{inst}}}.
    \]
\end{enumerate}

Measured latencies for $8\times8$ and $16\times16$ MIMO detection with 16-QAM using PIMI and conventional PIMs are reported in ~\ref{sec:MIMO_latency_table}, Tables~\ref{tab:mimo_PIMI_results_8x8}--\ref{tab:mimo_conv_results_16x16_sweeps}.

\clearpage

\section{Selection of parameters in MIMO}
\label{sec:slt_anneal_params}

\begin{figure}[htbp]
    \centering
    \includegraphics[width=0.9\textwidth]{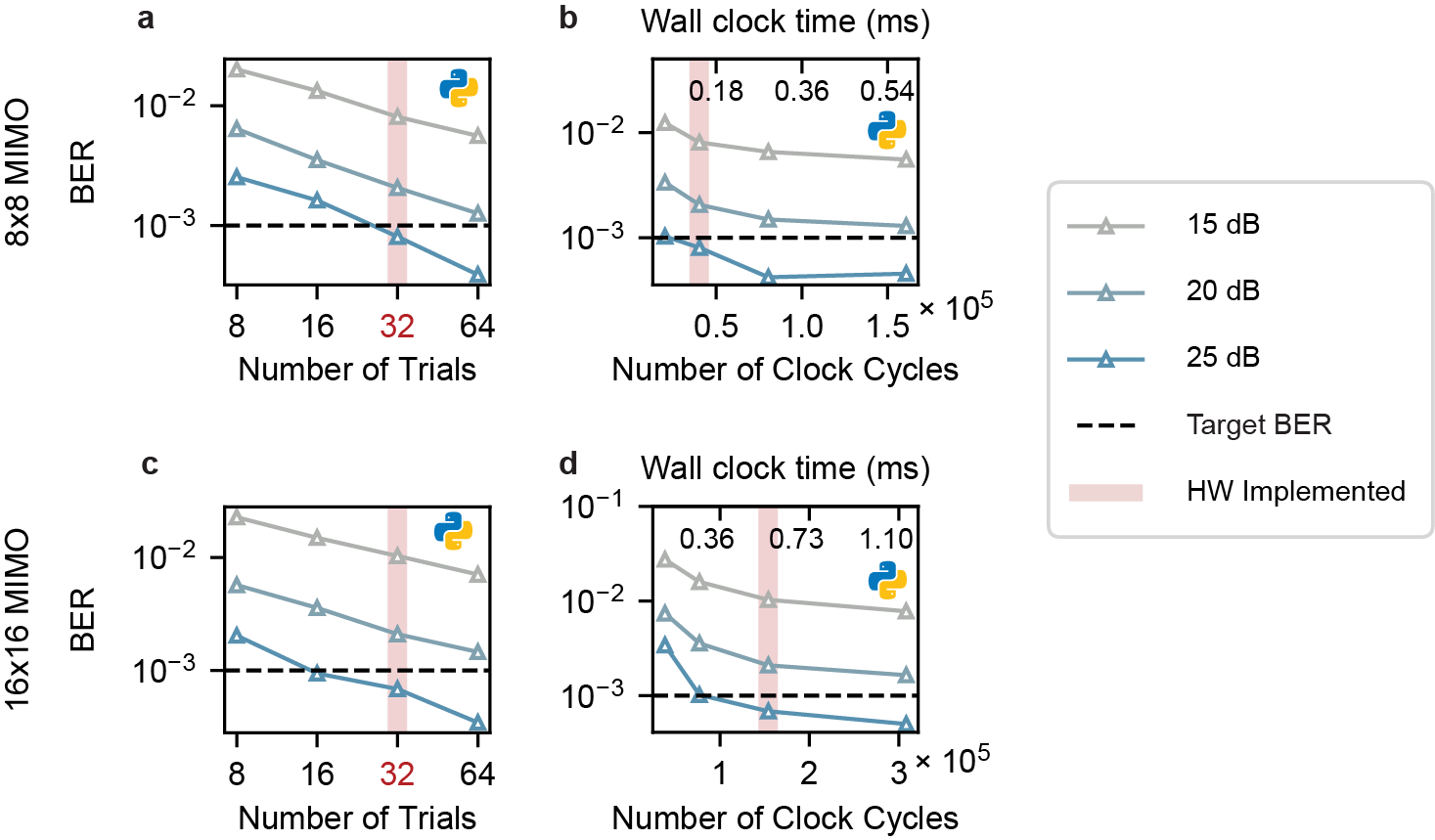}
    \caption{\textbf{Parameter selection for FPGA implementation.} BER performance and hardware timing for 16-QAM modulation. \panel{a\&c} BER versus number of trials for $8\times8$ and $16\times16$ MIMO systems at SNR levels of 15, 20, and 25~dB. The dashed line indicates the target BER of $10^{-3}$. \panel{b\&d} BER versus clock cycles and wall-clock time for each system size. The pink shaded region marks the selected hardware configuration (32 trials; 32 annealing steps for $8\times8$ MIMO, 64 steps for $16\times16$ MIMO), representing a tradeoff between accuracy and latency.}
    \label{fig:supp_step_anneal}
\end{figure}

Fig.~\ref{fig:supp_step_anneal} illustrates how the updating parameters used in the FPGA implementation (\ref{tab:BER_SNR_steps})—namely the number of \runname* and the number of \stepname*—were selected for $16$-QAM modulation.
The goal of this study is to determine parameter values that achieve the target bit error rate (BER) while remaining compatible with FPGA timing constraints.

Panels~\textbf{a} and~\textbf{c} show the BER as a function of the number of \runname* for $8\times8$ and $16\times16$ MIMO systems, respectively, evaluated at multiple signal-to-noise ratio (SNR) levels.
For both system sizes, increasing the number of \runname* improves solution quality, with progressively smaller gains at larger \runname\ counts.
Based on these curves, a value of 32 \runname* was selected as the operating point for hardware implementation, as it reliably achieves the target BER of $10^{-3}$ under typical SNR conditions while avoiding unnecessary repetitions.

Panels~\textbf{b} and~\textbf{d} illustrate the BER as a function of the number of clock cycles elapsed for each instance and the corresponding wall clock time.
The total number of clock cycles is obtained from high-level synthesis (HLS) reports of the PIMI kernel, which account for all \stepname* executed on the FPGA.
Using the FPGA clock frequency, the cycle counts are converted to wall-clock execution time, as indicated on the upper axes.
For the FPGA implementation, 32 \stepname* are used for the $8\times8$ MIMO system, while 64 annealing steps are used for the $16\times16$ MIMO system.
This mapping allows the selected parameters to be directly evaluated against real-time latency constraints.

Together, these results define the configuration used in the FPGA experiments reported in the main text.
The selected number of \runname* and \stepname* represent a practical tradeoff between solution accuracy and hardware execution time, and are fixed across experiments unless otherwise stated.

\clearpage
\section{Hardware--software co-design for MIMO detection}
\label{sec:MIMO_co_design}

\begin{figure}[htbp]
    \centering
    \includegraphics[width=\textwidth]{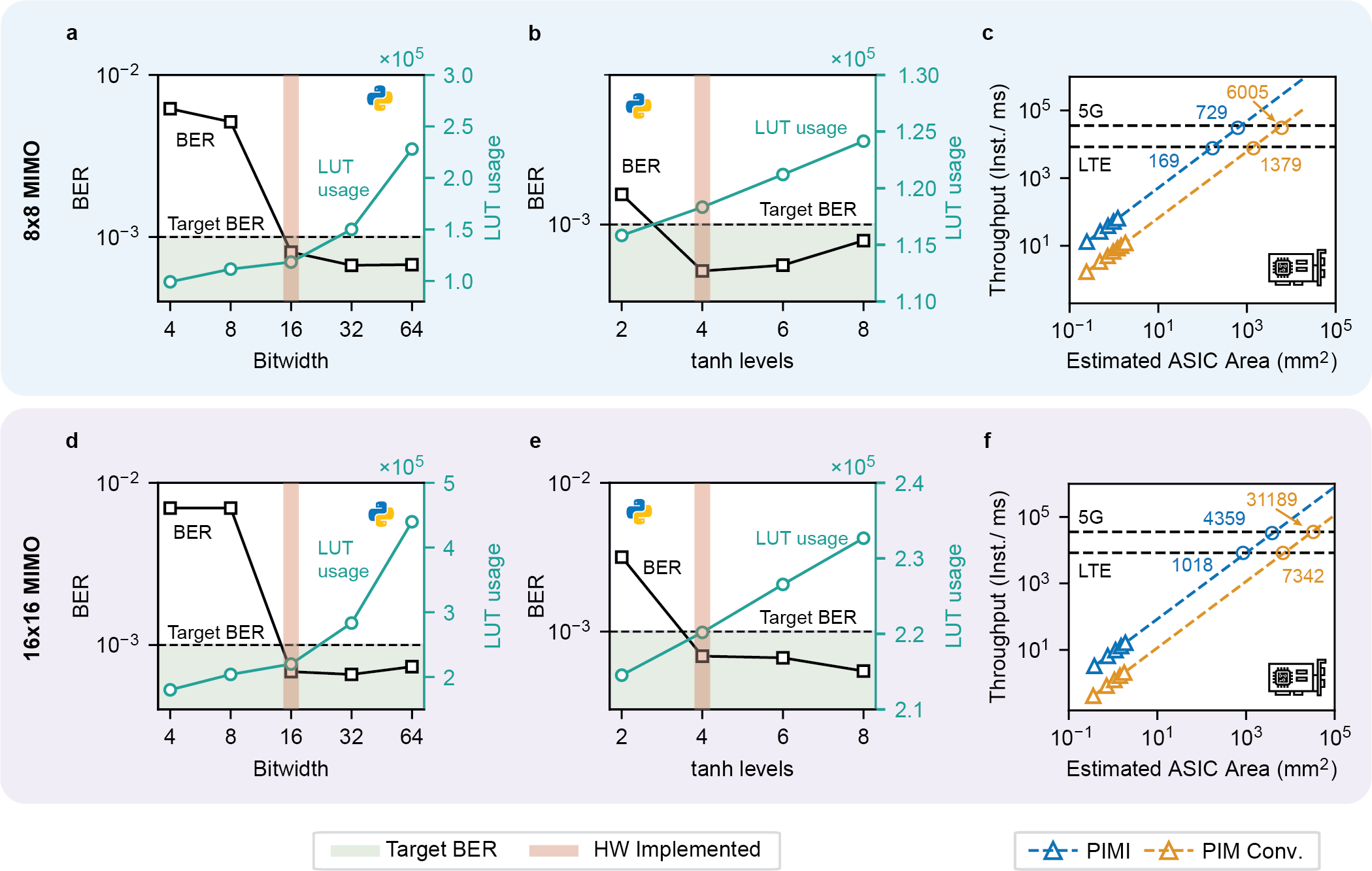}
    \caption{
    {Hardware--software co-design of PIMI enables efficient resource usage to meet BER and throughput targets for MIMO detection.}
    \panel{a--c} Results for an $8\times8$ MIMO system with 16-QAM modulation; 
    \panel{d--f} corresponding results for a $16\times16$ MIMO system with 16-QAM modulation.
    \panel{a\&d} BER (left axis) and look-up-table (LUT) usage (right axis) as a function of quantization bitwidth.
    BER improves with increasing precision and reaches the target value of $10^{-3}$ at 16 bit, beyond which additional precision yields negligible accuracy gains at increased hardware cost.
    \panel{b\&e} BER and LUT usage as a function of the number of quantization levels in the $\tanh$ look-up table (LUT).
    Four LUT levels are sufficient to reach the target BER.
    \panel{c\&f} Measured FPGA solver throughput versus projected aggregate logic area for PIMI and conventional probabilistic Ising machine (PIM) implementations.
    All BER results in this figure are estimated  at an SNR of 25~dB.
    Dashed horizontal lines indicate throughput targets for LTE (10~MHz) and 5G (50~MHz) systems.
    Shaded regions denote the parameter choices selected for hardware implementation, corresponding to the minimum LUT usage that achieves the target BER.
    Data in \textbf{a,b,d,e} are obtained from software-level simulations. In \textbf{c,f}, throughput is measured from FPGA implementations, while aggregate logic area is projected using the synthesis-based procedure described in \ref{sec:ASIC_area_est}.
    }
    \label{fig:MIMO_co_design}
\end{figure}

\fig{MIMO_co_design} presents the hardware--software co-design decisions obtained from Python-based simulations, which determine the implementation parameters of PIMI for $8\times8$ (panels~{a -- c}) and $16\times16$ (panels~{d -- f}) MIMO detection with 16-QAM modulation.
All bit error rate (BER) results in this figure are evaluated at a fixed $\mathrm{E_b}/\mathrm{N_0} = 25$~dB, corresponding to a realistic high-SNR operating regime in practical MIMO receivers, where quantization and algorithmic effects dominate over channel noise. 
Panels~a and~d show the BER (left axis) and FPGA look-up-table (LUT) usage (right axis) as a function of quantization bitwidth. 
For both $8\times8$ and $16\times16$ MIMO systems, BER improves with increasing precision and meets the target value of $10^{-3}$ at 16-bit quantization. 
Meanwhile, LUT usage increases with respect to bitwidth.
Since FPGA resource efficiency is maximized for power-of-two bitwidths (\ref{sec:resource_v_bw} includes data for non-power-of-two bitwidths), we implement a 16-bit fixed-point representation, which constitutes the lowest hardware-efficient precision that satisfies the BER requirement.
Panels~b and~e show the BER as a function of the number of quantization levels in the $\tanh$ look-up table (more details in \ref{sec:resource_v_tanh}). 
The BER decreases with increasing LUT resolution and reaches the target BER of $10^{-3}$ at 4 quantization levels. 
In contrast, LUT usage increases with the number of $\tanh$ levels. Balancing BER performance and hardware cost, we therefore employ a 4-level $\tanh$ look-up table in the hardware implementation to minimize per-kernel resource usage.

\fig{MIMO_co_design}{c} and~f show the MIMO detection throughput versus estimated ASIC chip area. 
Throughput reported in these panels was measured in real time on an AMD Xilinx U55C FPGA by scaling the number of parallel Ising solver kernels (raw latency measurements for 100,000 instances can be found in \ref{sec:MIMO_latency_table}).
Vitis HLS post-synthesis netlists were mapped to equivalent gate counts using Yosys \cite{Wolf2013}, and silicon area was estimated using the 7-nm ASAP7 process design kit \cite{Clark2016} (see \ref{sec:ASIC_area_est} for estimation details and mapping breakdown).
Both PIMI and conventional PIM implementations were optimized for minimal hardware cost.
For $8 \times 8$ MIMO, PIMI requires an ASIC of $729\,\si{mm^2}$ ($\sim2.7\,\si{cm} \times 2.7\,\si{cm}$) to satisfy the 5G throughput requirement (5G New Radio, 50~MHz with 15~kHz subcarrier spacing, see \ref{sec:throughput} for details), compared to $6005\,\si{mm^2}$ for conventional PIM, corresponding to an area reduction of $\sim 8\times$. 
For $16 \times 16$ MIMO, PIMI achieves the 5G target throughput with $4359\,\si{mm^2}$ ($\sim 6.6\,\si{cm} \times 6.6\,\si{cm}$), compared to $31{,}189\,\si{mm^2}$ for conventional PIM, with an area reduction of $\sim 7\times$. 
These results show that PIMI improves BER while also providing the hardware efficiency needed to meet next-generation wireless standards.

\begin{figure}[htbp]
    \centering
    \includegraphics[width=\textwidth]{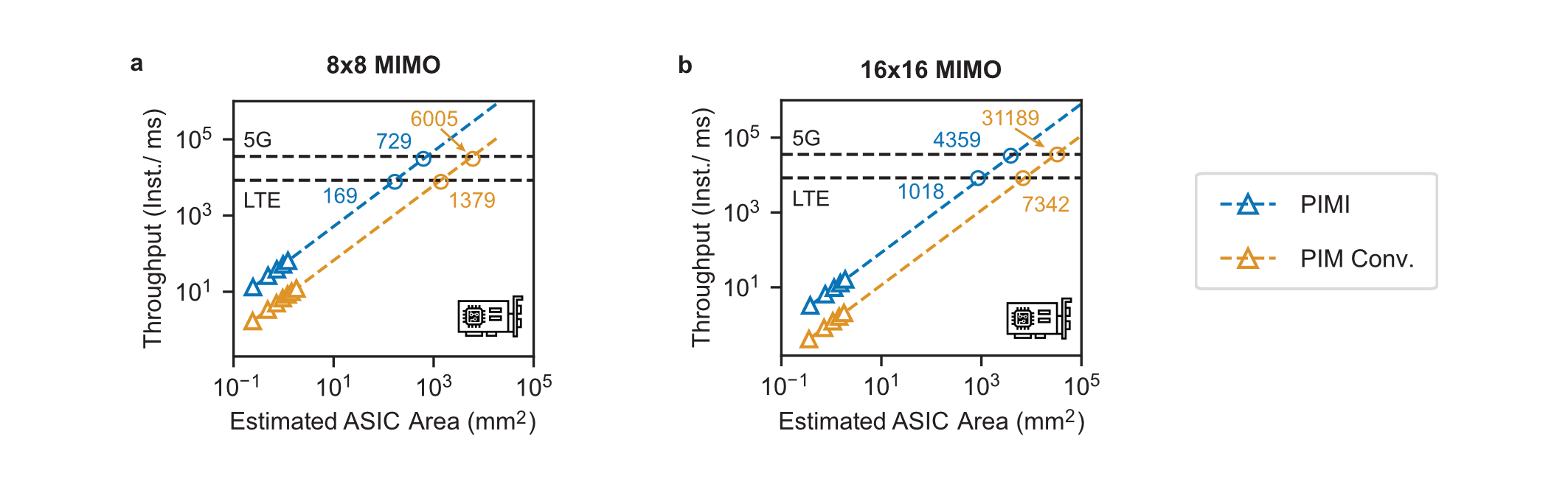}
    \caption{
    \textbf{Hardware--software co-design of PIMI enables efficient resource usage to meet BER and throughput targets for MIMO detection.}
    Throughput versus estimated aggregated ASIC area for PIMI and conventional probabilistic Ising machine (PIM) implementations, derived from FPGA resource utilization reports.
    All BER results in this figure are estimated  at an SNR of 25~dB.
    Dashed horizontal lines indicate throughput targets for LTE (10~MHz) and 5G (50~MHz) systems.
    Throughput data is measured from FPGA implementations.
    }
    \label{fig:MIMO_chip_area}
\end{figure}

\clearpage

\section{FPGA latency measurement results}
\label{sec:MIMO_latency_table}

\begin{table}[h!]
\centering
\begin{tabular}{lr}
\toprule
\multicolumn{2}{c}{\textbf{Experiment Configuration for a Single Kernel}} \\
\midrule
Problem             & 8$\times$8 MIMO PIMI \\
QAM                 & 16-QAM \\
Total \runname*       & 32 \\
Update steps     & 32 \\
Total spin updates  & 32{,}768\\
Trials in parallel & 4 \\
Batch size          & 500 \\
Total instances     & 100{,}000 \\
Designed Frequency (MHz)     & 274 \\
Designed Clock Cycles        & 40{,}202 \\
Designed Theoretical Time ($\mu$s) & 134.01 \\
Timing based on Frequency ($\mu$s) & 146.72 \\
BRAM (total = 4032) & 92 \\
DSP (total = 9024)  & 0 \\
FF (total = 2,607,360) & 44{,}325 \\
LUT (total = 1,303,680) & 118{,}270 \\
LUT after placement (total = 1,303,680) & 28{,}764 \\
\midrule
\end{tabular}

\vspace{0.5em}

\begin{tabular}{lrrrrr}
\toprule
Number of Kernels & 2 & 4 & 6 & 8 & 10 \\
\midrule
Measured avg. instance time ($\mu$s) & 146.773 & 146.852 & 146.787 & 146.792 & 146.845 \\
Measured avg. batch time (500 inst., ms) & 73.387 & 73.426 & 73.394 & 73.396 & 73.432 \\
Measured total kernel time (100k inst., ms) & 7{,}338.65 & 3{,}671.3 & 2{,}446.45 & 1{,}834.9 & 1{,}468.45 \\
Measured total run time (100k inst., ms) & 7{,}584 & 3{,}809 & 2{,}590 & 1{,}929 & 1{,}541 \\
\bottomrule
\end{tabular}
\caption{Performance and resource results for 8$\times$8 MIMO PIMI.}
\label{tab:mimo_PIMI_results_8x8}
\end{table}

\begin{table}[h!]
\centering
\begin{tabular}{lr}
\toprule
\multicolumn{2}{c}{\textbf{Experiment Configuration for a Single Kernel}} \\
\midrule
Problem             & 8$\times$8 MIMO PIM Conv \\
QAM                 & 16-QAM \\
Total \runname*       & 32 \\
Update steps     & 1024 \\
Total spin updates  & 32{,}768\\
Trials in parallel & 16 \\
Batch size          & 500 \\
Total instances     & 100{,}000 \\
Designed Frequency (MHz)     & 247.2 \\
Designed Clock Cycles        & 299{,}299 \\
Designed Theoretical Time ($\mu$s) & 997.66 \\
Timing based on Frequency ($\mu$s) & 1210.76 \\
BRAM (total = 4032) & 96 \\
DSP (total = 9024)  & 32 \\
FF (total = 2,607,360) & 35{,}855 \\
LUT (total = 1,303,680) & 102{,}314 \\
LUT after placement (total = 1,303,680) & 21{,}079 \\
\midrule
\end{tabular}

\vspace{0.5em}

\begin{tabular}{lrrrrrrrr}
\toprule
Number of Kernels & 2 & 4 & 6 & 8 & 10 & 12 & 15 \\
\midrule
Measured avg. instance time ($\mu$s) & 1174.27 & 1174.28 & 1174.28 & 1174.30 & 1174.30 & 1174.34 & 1174.30 \\
Measured avg. batch time (500 inst., ms) & 587.13 & 587.14 & 587.14 & 587.10 & 587.20 & 587.17 & 587.00 \\
Measured total kernel time (100k inst., ms) & 58{,}715 & 29{,}357.5 & 19{,}571.7 & 14{,}678.8 & 11{,}743 & 9{,}785.8 & 7{,}828.7 \\
Measured total run time (100k inst., ms) & 58{,}995 & 29{,}522 & 20{,}062 & 14{,}780 & 11{,}832 & 10{,}060 & 8{,}279 \\
\bottomrule
\end{tabular}
\caption{Performance and resource results for 8$\times$8 MIMO PIM Conv.}
\label{tab:mimo_conv_results_8x8}
\end{table}

\begin{table}[h!]
\centering
\begin{tabular}{lr}
\toprule
\multicolumn{2}{c}{\textbf{Experiment Configuration for a Single Kernel}} \\
\midrule
Problem             & 16$\times$16 MIMO PIMI \\
QAM                 & 16-QAM \\
Total \runname*       & 32 \\
Update steps     & 64 \\
Total spin updates  & 131{,}072 \\
Trials in parallel & 4 \\
Batch size          & 100 \\
Total instances     & 100{,}000 \\
Designed Frequency (MHz)     & 258.1 \\
Designed Clock Cycles        & 154{,}074 \\
Designed Single Instance time ($\mu$s) & 596.95 \\
BRAM (total = 4032) & 196 \\
DSP (total = 9024)  & 0 \\
FF (total = 2,607,360) & 76{,}975 \\
LUT (total = 1,303,680) & 220{,}148 \\
LUT after placement (total = 1,303,680) & 71{,}650 \\
\midrule
\midrule
\end{tabular}

\vspace{0.5em}

\begin{tabular}{lrrrrr}
\toprule
Number of Kernels & 2 & 4 & 6 & 8 & 10 \\
\midrule
Measured avg. instance time ($\mu$s) & 600.2 & 600.2 & 600.2 & 600.4 & 600.3 \\
Measured avg. batch time (100 inst., ms) & 60.02 & 60.02 & 60.02 & 60.04 & 60.03 \\
Measured total kernel time (100k inst., ms) & 30{,}010 & 15{,}005 & 10{,}003 & 7{,}505 & 6{,}003 \\
Measured total run time (100k inst., ms) & 30{,}620 & 15{,}343 & 10{,}270 & 7{,}738 & 6{,}169 \\
\bottomrule
\end{tabular}
\caption{Performance and resource results for 16$\times$16 MIMO PIMI.}
\end{table}

\begin{table}[h!]
\centering
\begin{tabular}{lr}
\toprule
\multicolumn{2}{c}{\textbf{Experiment Configuration for a Single Kernel}} \\
\midrule
Problem             & 16$\times$16 MIMO Conv \\
QAM                 & 16-QAM \\
Total \runname*       & 32 \\
Update steps     & 4096 \\
Total spin updates  & 131{,}072 \\
Trials in parallel & 16 \\
Batch size          & 100 \\
Total instances     & 100{,}000 \\
Designed Frequency (MHz)     & 267.1 \\
Designed Clock Cycles        & 1{,}325{,}443 \\
Designed Single Instance time ($\mu$s) & 4418.14 \\
BRAM (total = 4032) & 128 \\
DSP (total = 9024)  & 64 \\
FF (total = 2,607,360) & 51{,}706 \\
LUT (total = 1,303,680) & 148{,}470 \\
LUT after placement (total = 1,303,680) & 23{,}333 \\
\midrule
\end{tabular}

\vspace{0.5em}

\begin{tabular}{lrrrrr}
\toprule
Number of Kernels & 2 & 4 & 6 & 8 & 10 \\
\midrule
Measured avg. instance time ($\mu$s) & 4826.62 & 4826.61 & 4826.72 & 4826.64 & 4826.71 \\
Measured avg. batch time (100 inst., ms) & 482.66 & 482.66 & 482.67 & 482.66 & 482.67 \\
Measured total kernel time (100k inst., ms) & 242{,}047 & 121{,}123 & 80{,}928 & 60{,}537 & 48{,}454 \\
Measured total run time (100k inst., ms) & 242{,}761 & 121{,}498 & 81{,}193 & 60{,}722 & 48{,}617 \\
\bottomrule
\end{tabular}
\caption{Performance and resource results for 16$\times$16 MIMO Conv (sweeps).}
\label{tab:mimo_conv_results_16x16_sweeps}
\end{table}

\section{ASIC area estimate from HLS design}
\label{sec:ASIC_area_est}


We estimate aggregated ASIC silicon area by converting post-synthesis logic into NAND2-equivalent gate counts and projecting area using the 7-nm ASAP7 CMOS process \cite{Clark2016}.
RTL (register-transfer level) generated by Vitis HLS was synthesized using Yosys\cite{Wolf2013} with SystemVerilog support. FPGA-specific IP blocks (e.g., memories and AXI interfaces) were treated as black boxes. 
The design hierarchy was flattened and normalized using standard Yosys passes, and the resulting netlist was mapped with ABC to a restricted set of Boolean primitives (NAND, AND, OR, NOR, NOT, XOR/XNOR, and multiplexers), eliminating FPGA-specific optimizations.

Each logic primitive was converted to a NAND2-equivalent gate count using fixed weights derived from standard CMOS decompositions (e.g., NAND: 1.0, NOT: 0.5; details can be found in the following tables). All flip-flops were modeled as 6 NAND2 equivalents. The total NAND2-equivalent gate count was obtained by summing the weighted contributions reported by Yosys.
To account for non-functional overheads not captured in logic synthesis, including clock-tree synthesis, buffering, and routing overhead, \textbf{a uniform 50\% area margin was applied}. Silicon area was then projected using the ASAP7 predictive process design kit, assuming a NAND2 standard-cell area of 0.027~$\mu$m$^{2}$. 
This methodology is intended for relative architectural comparison rather than absolute tape-out prediction. Because the same synthesis and estimation flow is applied consistently across all evaluated designs, the resulting area trends reliably reflect relative hardware efficiency.

Estimated ASIC areas (with 7-nm process) for $8\times8$ and $16\times16$ MIMO detection with 16-QAM using PIMI and conventional PIMs are reported in Tables~\ref{tab:asic_8x8_PIMI}--\ref{tab:asic_16x16_cpim}.

\begin{table}[h!]
\centering
\begin{tabular}{l l r r r}
\toprule
Class & Primitive & Count & NAND2/each & NAND2 total \\
\midrule
Gate & \texttt{\_MUX\_} & 264,174 & 4.00 & 1,056,696 \\
Gate & \texttt{\_AND\_} & 126,759 & 1.50 & 190,138 \\
Gate & \texttt{\_OR\_} & 71,816 & 2.00 & 143,632 \\
Gate & \texttt{\_XOR\_} & 25,974 & 4.00 & 103,896 \\
Gate & \texttt{\_XNOR\_} & 22,345 & 4.00 & 89,380 \\
Gate & \texttt{\_NOR\_} & 31,791 & 2.50 & 79,477 \\
Gate & \texttt{\_NOT\_} & 22,883 & 0.50 & 11,441 \\
Gate & \texttt{\_NAND\_} & 4,278 & 1.00 & 4,278 \\
FF & \texttt{\_DFFE\_PP\_} & 205,172 & 6.00 & 1,231,032 \\
FF & \texttt{\_DFF\_P\_} & 8,847 & 6.00 & 53,082 \\
FF & \texttt{\_SDFFCE\_PP0P\_} & 4,579 & 6.00 & 27,474 \\
FF & \texttt{\_SDFFE\_PP0P\_} & 3,723 & 6.00 & 22,338 \\
FF & \texttt{\_SDFF\_PP0\_} & 431 & 6.00 & 2,586 \\
FF & \texttt{\_SDFFCE\_PN0P\_} & 32 & 6.00 & 192 \\
FF & \texttt{\_SDFFE\_PP1P\_} & 26 & 6.00 & 156 \\
FF & \texttt{\_SDFF\_PP1\_} & 18 & 6.00 & 108 \\
FF & \texttt{\_DFFE\_PN\_} & 12 & 6.00 & 72 \\
FF & \texttt{\_SDFFE\_PP0N\_} & 2 & 6.00 & 12 \\
FF & \texttt{\_SDFFCE\_PP1P\_} & 1 & 6.00 & 6 \\
\midrule
\textbf{Total (raw)} &  &  &  & \textbf{3,015,997} \\
\textbf{+50\% overhead} &  &  &  & \textbf{4,523,996} \\
\textbf{Area (ASAP7)} &  &  &  & \textbf{122,148 µm²} \\
\textbf{Area (ASAP7)} &  &  &  & \textbf{0.122148 mm²} \\
\bottomrule
\end{tabular}
\caption{NAND2-equivalent breakdown and ASAP7 area for one PIMI kernel for 8x8 MIMO (assumptions: 1 FF = 6 NAND2, +50\% overhead, 0.027 $\mu$m$^2$/NAND2).}
\label{tab:asic_8x8_PIMI}
\end{table}

\begin{table}[h!]
\centering
\begin{tabular}{l l r r r}
\toprule
Class & Primitive & Count & NAND2/each & NAND2 total \\
\midrule
Gate & \texttt{\_MUX\_} & 286,373 & 4.00 & 1,145,492 \\
Gate & \texttt{\_AND\_} & 91,426 & 1.50 & 137,139 \\
Gate & \texttt{\_NOR\_} & 31,785 & 2.50 & 79,462 \\
Gate & \texttt{\_XOR\_} & 18,424 & 4.00 & 73,696 \\
Gate & \texttt{\_OR\_} & 35,313 & 2.00 & 70,626 \\
Gate & \texttt{\_XNOR\_} & 14,857 & 4.00 & 59,428 \\
Gate & \texttt{\_NOT\_} & 17,643 & 0.50 & 8,821 \\
Gate & \texttt{\_NAND\_} & 3,184 & 1.00 & 3,184 \\
FF & \texttt{\_DFFE\_PP\_} & 221,724 & 6.00 & 1,330,344 \\
FF & \texttt{\_SDFFE\_PP0P\_} & 3,740 & 6.00 & 22,440 \\
FF & \texttt{\_DFF\_P\_} & 3,593 & 6.00 & 21,558 \\
FF & \texttt{\_SDFFCE\_PP0P\_} & 2,300 & 6.00 & 13,800 \\
FF & \texttt{\_SDFF\_PP0\_} & 627 & 6.00 & 3,762 \\
FF & \texttt{\_SDFFE\_PP1P\_} & 49 & 6.00 & 294 \\
FF & \texttt{\_SDFFCE\_PN0P\_} & 32 & 6.00 & 192 \\
FF & \texttt{\_SDFF\_PP1\_} & 19 & 6.00 & 114 \\
FF & \texttt{\_DFFE\_PN\_} & 12 & 6.00 & 72 \\
FF & \texttt{\_SDFFE\_PP0N\_} & 2 & 6.00 & 12 \\
FF & \texttt{\_SDFFCE\_PP1P\_} & 1 & 6.00 & 6 \\
\midrule
\textbf{Total (raw)} &  &  &  & \textbf{2,970,443} \\
\textbf{+50\% overhead} &  &  &  & \textbf{4,455,664} \\
\textbf{Area (ASAP7)} &  &  &  & \textbf{120,303 µm²} \\
\textbf{Area (ASAP7)} &  &  &  & \textbf{0.120303 mm²} \\
\bottomrule
\end{tabular}
\caption{NAND2-equivalent breakdown and ASAP7 area for one conventional PIM kernel for 8x8 MIMO (assumptions: 1 FF = 6 NAND2, +50\% overhead, 0.027 $\mu$m$^2$/NAND2).}
\label{tab:asic_8x8_cpim}
\end{table}

\begin{table}[h!]
\centering
\begin{tabular}{l l r r r}
\toprule
Class & Primitive & Count & NAND2/each & NAND2 total \\
\midrule
Gate & \texttt{\_MUX\_} & 361,611 & 4.00 & 1,446,444 \\
Gate & \texttt{\_AND\_} & 277,659 & 1.50 & 416,488 \\
Gate & \texttt{\_OR\_} & 135,748 & 2.00 & 271,496 \\
Gate & \texttt{\_XOR\_} & 51,240 & 4.00 & 204,960 \\
Gate & \texttt{\_XNOR\_} & 44,107 & 4.00 & 176,428 \\
Gate & \texttt{\_NOR\_} & 62,873 & 2.50 & 157,182 \\
Gate & \texttt{\_NOT\_} & 44,195 & 0.50 & 22,097 \\
Gate & \texttt{\_NAND\_} & 7,613 & 1.00 & 7,613 \\
FF & \texttt{\_DFFE\_PP\_} & 285,256 & 6.00 & 1,711,536 \\
FF & \texttt{\_DFF\_P\_} & 25,790 & 6.00 & 154,740 \\
FF & \texttt{\_SDFFCE\_PP0P\_} & 11,085 & 6.00 & 66,510 \\
FF & \texttt{\_SDFFE\_PP0P\_} & 3,734 & 6.00 & 22,404 \\
FF & \texttt{\_SDFF\_PP0\_} & 489 & 6.00 & 2,934 \\
FF & \texttt{\_SDFFCE\_PN1P\_} & 448 & 6.00 & 2,688 \\
FF & \texttt{\_SDFFCE\_PN0P\_} & 32 & 6.00 & 192 \\
FF & \texttt{\_SDFFE\_PP1P\_} & 26 & 6.00 & 156 \\
FF & \texttt{\_SDFF\_PP1\_} & 24 & 6.00 & 144 \\
FF & \texttt{\_DFFE\_PN\_} & 10 & 6.00 & 60 \\
FF & \texttt{\_SDFFE\_PP0N\_} & 2 & 6.00 & 12 \\
FF & \texttt{\_SDFFCE\_PP1P\_} & 1 & 6.00 & 6 \\
\midrule
\textbf{Total (raw)} &  &  &  & \textbf{4,664,091} \\
\textbf{+50\% overhead} &  &  &  & \textbf{6,996,137} \\
\textbf{Area (ASAP7)} &  &  &  & \textbf{188,896 µm²} \\
\textbf{Area (ASAP7)} &  &  &  & \textbf{0.188896 mm²} \\
\bottomrule
\end{tabular}
\caption{NAND2-equivalent breakdown and ASAP7 area for one PIMI kernel for 16x16 MIMO  (assumptions: 1 FF = 6 NAND2, +50\% overhead, 0.027 $\mu$m$^2$/NAND2).}
\label{tab:asic_16x16_PIMI}
\end{table}

\begin{table}[h!]
\centering
\begin{tabular}{l l r r r}
\toprule
Class & Primitive & Count & NAND2/each & NAND2 total \\
\midrule
Gate & \texttt{\_MUX\_} & 401,720 & 4.00 & 1,606,880 \\
Gate & \texttt{\_AND\_} & 129,672 & 1.50 & 194,508 \\
Gate & \texttt{\_NOR\_} & 58,989 & 2.50 & 147,472 \\
Gate & \texttt{\_XOR\_} & 34,493 & 4.00 & 137,972 \\
Gate & \texttt{\_XNOR\_} & 26,583 & 4.00 & 106,332 \\
Gate & \texttt{\_OR\_} & 42,062 & 2.00 & 84,124 \\
Gate & \texttt{\_NOT\_} & 30,894 & 0.50 & 15,447 \\
Gate & \texttt{\_NAND\_} & 4,191 & 1.00 & 4,191 \\
FF & \texttt{\_DFFE\_PP\_} & 305,862 & 6.00 & 1,835,172 \\
FF & \texttt{\_DFF\_P\_} & 38,665 & 6.00 & 231,990 \\
FF & \texttt{\_SDFFE\_PP0P\_} & 3,746 & 6.00 & 22,476 \\
FF & \texttt{\_SDFFCE\_PP0P\_} & 2,557 & 6.00 & 15,342 \\
FF & \texttt{\_SDFF\_PP0\_} & 661 & 6.00 & 3,966 \\
FF & \texttt{\_SDFFE\_PP1P\_} & 49 & 6.00 & 294 \\
FF & \texttt{\_SDFFCE\_PN0P\_} & 32 & 6.00 & 192 \\
FF & \texttt{\_SDFF\_PP1\_} & 20 & 6.00 & 120 \\
FF & \texttt{\_SDFFCE\_PP1P\_} & 1 & 6.00 & 6 \\
\midrule
\textbf{Total (raw)} &  &  &  & \textbf{4,406,484} \\
\textbf{+50\% overhead} &  &  &  & \textbf{6,609,726} \\
\textbf{Area (ASAP7)} &  &  &  & \textbf{178,463 µm²} \\
\textbf{Area (ASAP7)} &  &  &  & \textbf{0.178463 mm²} \\
\bottomrule
\end{tabular}
\caption{NAND2-equivalent breakdown and ASAP7 area for one conventional PIM kernel for 16x16 MIMO (assumptions: 1 FF = 6 NAND2, +50\% overhead, 0.027 $\mu$m$^2$/NAND2).}
\label{tab:asic_16x16_cpim}
\end{table}

\clearpage

\section{FPGA resource usage vs bitwidth for Max-Cut and SK-1}
\label{sec:resource_v_bw_mc_sk}

\begin{table}[h!]
\centering
\begin{tabular}{lrrrr}
\toprule
Problem & \multicolumn{4}{c}{Max-Cut/ SK-1 PIMI} \\
Problem size $N$ & \multicolumn{4}{c}{60} \\
$\tanh$ levels & \multicolumn{4}{c}{4} \\
\toprule
Bitwidth & 4 & 8 & 16 & 32 \\
\midrule
BRAM (total = 4032) &
30 & 30 & 30 & 30 \\
DSP (total = 9024)  &
0 & 64 & 64 & 128 \\
FF (total = 2,607,360)  &
27{,}935 & 29{,}281 & 31{,}733 & 44{,}895 \\
LUT (total = 1,303,680) &
160{,}196 & 163{,}739 & 170{,}814 & 178{,}653 \\
\bottomrule
\end{tabular}
\caption{PIMI's FPGA resource usage versus datapath bitwidth for Max-Cut/ SK-1 at $N=60$, with the $\tanh$ look-up table fixed at 4 levels.}
\label{tab:bitwidth_maxcut}
\end{table}

\section{FPGA resource usage vs tanh levels for Max-Cut and SK-1}
\label{sec:resource_v_tanh_mc_sk}

\begin{table}[h!]
\centering
\begin{tabular}{lrrrr}
\toprule
Problem & \multicolumn{4}{c}{Max-Cut/ SK-1 PIMI} \\
Problem size $N$ & \multicolumn{4}{c}{60} \\
Bitwidth & \multicolumn{4}{c}{8} \\
\toprule
$\tanh$ levels & 2 & 4 & 6 & 8 \\
\midrule
BRAM (total = 4032) &
30 & 30 & 30 & 30 \\
DSP (total = 9024)  &
64 & 64 & 64 & 64 \\
FF (total = 2,607,360)  &
28{,}269 & 29{,}281 & 29{,}167 & 29{,}488 \\
LUT (total = 1,303,680) &
158{,}365 & 163{,}739 & 166{,}215 & 169{,}804 \\
\bottomrule
\end{tabular}
\caption{PIMI's FPGA resource usage versus the number of $\tanh$ quantization levels for Max-Cut/ SK-1 at $N=60$, with the datapath bitwidth fixed at 8.}
\label{tab:tanh_maxcut}
\end{table}

\clearpage

\section{FPGA resource usage vs bitwidth for MIMO detection}
\label{sec:resource_v_bw}

\begin{table}[h!]
\centering
\begin{tabular}{lrrrrrr}
\toprule
Problem & \multicolumn{6}{c}{8$\times$8 MIMO PIMI} \\
QAM     & \multicolumn{6}{c}{16} \\
Total \runname* & \multicolumn{6}{c}{32} \\
Update steps & \multicolumn{6}{c}{32} \\
Trials in parallel & \multicolumn{6}{c}{4} \\
\toprule
Bitwidth & 4 & 8 & 12 & 16 & 32 & 64 \\
\midrule
BRAM (total = 4032) &
78 & 78 & 92 & 92 & 92 & 108 \\
DSP (total = 9024)  &
0 & 0 & 0 & 0 & 0 & 0 \\
FF (total = 2,607,360)  &
32,622 & 36,524 & 77,382 & 44,325 & 76,509 & 142,562 \\
LUT (total = 1,303,680) &
99,072 & 111,432 & 253,798 & 118,270 & 150,028 & 228,027 \\
\bottomrule
\end{tabular}
\caption{FPGA resource usage versus bitwidth for data representation in 8$\times$8 MIMO PIMI.}
\label{tab:bitwidth_8x8}
\end{table}

\begin{table}[h!]
\centering
\begin{tabular}{lrrrrrr}
\toprule
Problem & \multicolumn{6}{c}{16$\times$16 MIMO PIMI} \\
QAM     & \multicolumn{6}{c}{16} \\
Total \runname* & \multicolumn{6}{c}{32} \\
Update steps & \multicolumn{6}{c}{64} \\
\runname* in parallel & \multicolumn{6}{c}{4} \\
\toprule
Bitwidth & 4 & 8 & 12 & 16 & 32 & 64 \\
\midrule
BRAM (total = 4032) & 
174 & 174 & 196 & 196 & 196 & 204 \\
DSP (total = 9024)  & 
0 & 0 & 0 & 0 & 0 & 0 \\
FF (total = 2,607,360)  & 
53,560 & 61,334 & 124,675 & 76,975 & 141,146 & 269,728 \\
LUT (total = 1,303,680) & 
180,296 & 204,086 & 421,501 & 220,148 & 283,441 & 439,932 \\
\bottomrule
\end{tabular}
\caption{FPGA resource usage versus bitwidth for data representation in 16$\times$16 MIMO PIMI.}
\label{tab:bitwidth_16x16}
\end{table}

\section{FPGA resource usage vs tanh levels for MIMO detection}
\label{sec:resource_v_tanh}

\begin{table}[h!]
\centering
\begin{tabular}{lrrrrrr}
\toprule
Problem & \multicolumn{4}{c}{8$\times$8 MIMO PIMI} \\
QAM     & \multicolumn{4}{c}{16} \\
Total \runname* & \multicolumn{4}{c}{32} \\
Update Steps & \multicolumn{4}{c}{32} \\
Trials in parallel & \multicolumn{4}{c}{4} \\
\toprule
tanh levels & 2 & 4 & 6 & 8 \\
\midrule
BRAM (total = 4032) & 
92 & 92 & 92 & 92  \\
DSP (total = 9024)  & 
0 & 0 & 0 & 0 \\
FF (total = 2,607,360)  & 
44,264 & 44,325 & 44,584 & 45,896 \\
LUT (total = 1,303,680) & 
115,830 & 118,270 & 121,206 & 124,150 \\
\bottomrule
\end{tabular}
\caption{FPGA resource usage versus tanh levels for data representation in 8$\times$8 MIMO PIMI.}
\label{tab:tanh_8x8}
\end{table}

\begin{table}[h!]
\centering
\begin{tabular}{lrrrrrr}
\toprule
Problem & \multicolumn{4}{c}{16$\times$16 MIMO PIMI} \\
QAM     & \multicolumn{4}{c}{16} \\
Total \runname* & \multicolumn{4}{c}{32} \\
Update Steps & \multicolumn{4}{c}{64} \\
Trials in parallel & \multicolumn{4}{c}{4} \\
\toprule
tanh levels & 2 & 4 & 6 & 8 \\
\midrule
BRAM (total = 4032) & 
196 & 196 & 196 & 196  \\
DSP (total = 9024)  & 
0 & 0 & 0 & 0 \\
FF (total = 2,607,360)  & 
76,850 & 76,975 & 77,490 & 79,474 \\
LUT (total = 1,303,680) & 
214,540 & 220,148 & 226,556 & 232,676 \\
\bottomrule
\end{tabular}
\caption{FPGA resource usage versus tanh levels for data representation in 16$\times$16 MIMO PIMI.}
\label{tab:tanh_16x16}
\end{table}

\fi

\end{document}